\documentclass[preprint,12pt]{elsarticle}

\usepackage{graphicx}
\usepackage{amssymb}
\usepackage{upgreek}

\usepackage{lineno}
\usepackage{multirow}
\usepackage{array}

\journal{Nuclear Instruments \& Methods in Physics Research Section A}

\begin{document}
\begin{frontmatter}

\title{ANTARES{\tnoteref{sigle}}:\\the first undersea neutrino telescope}
\tnotetext[sigle]{Astronomy with a Neutrino Telescope and Abyss environmental RESearch}
\author[CPPM]{M.~Ageron}
\author[IFIC]{J.A.~Aguilar}
\author[CPPM]{I.~Al~Samarai}
\author[Colmar]{A.~Albert}
\author[Roma]{F.~Ameli}
\author[Barcelona]{M.~Andr\'e}
\author[Genova]{M.~Anghinolfi}
\author[Erlangen]{G.~Anton}
\author[IRFU/SEDI]{S.~Anvar}
\author[UPV]{M.~Ardid}
% ...

\cortext[tag:corresp]{Corresponding author. Postal address: CE Saclay, B\^at.141, 91191 Gif-sur-Yvette, France}
\cortext[tag:corresp2]{Corresponding author. Postal address: INFN Sezione di Bari, via Amendola 173, 70126 Bari, Italy}
\fntext[tag:1]{\scriptsize{Also at University of Leiden, the Netherlands}}
\fntext[tag:2]{\scriptsize{On leave at DESY, Platanenallee 6, D-15738 Zeuthen, Germany}}
\fntext[tag:4]{\scriptsize{Now at Bergische Universität Wuppertal, Fachbereich C - Mathematik und Naturwissenschaften, 42097 Wuppertal }}
\fntext[tag:3]{\scriptsize{Now at IRFU/DSM/CEA, CE Saclay, 91191 Gif-sur-Yvette, France}}
\fntext[tag:PP]{\scriptsize{Deceased (December 2010).}}

\address[CPPM]{\scriptsize{CPPM, Aix-Marseille Universit\'e, CNRS/IN2P3, Marseille, France}}

\address[IFIC]{\scriptsize{IFIC - Instituto de F\'isica Corpuscular, Edificios Investigaci\'on de Paterna, CSIC - Universitat de Val\`encia, Apdo. de Correos 22085, 46071 Valencia, Spain}}

\address[Colmar]{\scriptsize{GRPHE - Institut universitaire de technologie de Colmar, 34 rue du Grillenbreit BP 50568,\\68008 Colmar, France}}

\address[Roma]{\scriptsize{INFN -Sezione di Roma, P.le Aldo Moro 2, 00185 Roma, Italy}}

\address[Barcelona]{\scriptsize{Technical University of Catalonia, Laboratory of Applied Bioacoustics, Rambla Exposici\'o,\\08800 Vilanova i la Geltr\'u, Barcelona, Spain}}

\address[Genova]{\scriptsize{INFN - Sezione di Genova, Via Dodecaneso 33, 16146 Genova, Italy}}

\address[Erlangen]{\scriptsize{Friedrich-Alexander-Universit\"{a}t Erlangen-N\"{u}rnberg, Erlangen Centre for Astroparticle Physics, Erwin-Rommel-Str. 1, 91058 Erlangen, Germany}}

\address[IRFU/SEDI]{\scriptsize{Direction des Sciences de la Mati\`ere - Institut de recherche sur les lois fondamentales de l'Univers - Service d'Electronique des D\'etecteurs et d'Informatique, CEA Saclay,\\91191 Gif-sur-Yvette Cedex, France}}

\address[UPV]{\scriptsize{Institut d'Investigaci\'o per a la Gesti\'o Integrada de Zones Costaneres (IGIC) - Universitat Polit\`ecnica de Val\`encia. C/  Paranimf 1. , 46730 Gandia, Spain.}}

% ...
\address[IRFU/SPP]{\scriptsize{Direction des Sciences de la Mati\`ere - Institut de recherche sur les lois fondamentales de l'Univers - Service de Physique des Particules, CEA Saclay, 91191 Gif-sur-Yvette Cedex, France}}

% ...
\address[Roma-UNI]{\scriptsize{Dipartimento di Fisica dell'Universit\`a La Sapienza, P.le Aldo Moro 2, 00185 Roma, Italy}}

% ...
\address[IRFU/SAP]{\scriptsize{Direction des Sciences de la Mati\`ere - Institut de recherche sur les lois fondamentales de l'Univers - Service d'Astrophysique, CEA Saclay, 91191 Gif-sur-Yvette Cedex, France}}

% ...
\address[Genova-UNI]{\scriptsize{Dipartimento di Fisica dell'Universit\`a, Via Dodecaneso 33, 16146 Genova, Italy}}

\begin{abstract}
The ANTARES Neutrino Telescope was completed in May 2008 and is the first operational Neutrino Telescope in the Mediterranean Sea. The main purpose of the detector is to perform neutrino astronomy and the apparatus also offers facilities for marine and Earth sciences. This paper describes the design, the construction and the installation of the telescope in the deep sea, offshore from Toulon in France. An illustration of the detector performance is given.
\end{abstract}


\end{frontmatter}

\linenumbers
\setcounter{tocdepth}{5}
\setcounter{secnumdepth}{5}
\tableofcontents
\hyphenation{AN-TA-RES}

\newpage
\section* {Acronyms and abbreviations}
\begin{description}
	\item[ADCP] Acoustic Doppler Current Profiler\vspace*{-.4\baselineskip}
	\item[ARS]	Analogue Ring Sampler\vspace*{-.4\baselineskip}
	\item[AS]	  Acoustic Storey\vspace*{-.4\baselineskip}
	\item[AVC]  Amplitude to Voltage Converter\vspace*{-.4\baselineskip}
	\item[BSS]	Bottom String Socket\vspace*{-.4\baselineskip}
	\item[CTD]	Conductivity Temperature Depth sensor\vspace*{-.4\baselineskip} 
	\item[DAQ]  Data Acquisition\vspace*{-.4\baselineskip}
	\item[DP]		Dynamic Positioning\vspace*{-.4\baselineskip}
	\item[DSP]	Digital Signal Processor\vspace*{-.4\baselineskip}
	\item[DWDM]	Dense Wavelength Division Multiplexer\vspace*{-.4\baselineskip}
	\item[EMC]	(vertical) Electro Mechanical Cable\vspace*{-.4\baselineskip}
	\item[GCN]	Gamma-ray bursts Coordinates Network\vspace*{-.4\baselineskip} 
	\item[GUI]	Graphical User Interface\vspace*{-.4\baselineskip} 
	\item[HFLBL]High Frequency Long Base Line\vspace*{-.4\baselineskip}
	\item[ID]		Inner Diameter\vspace*{-.4\baselineskip}
	\item[IL]		InterLink\vspace*{-.4\baselineskip}
	\item[IL07]	Instrumented Line (deployed in the year 2007)\vspace*{-.4\baselineskip}
	\item[JB]		Junction Box\vspace*{-.4\baselineskip}
	\item[LCM]	Local Control Module\vspace*{-.4\baselineskip}
	\item[LDPE]	Low Density PolyEthylene\vspace*{-.4\baselineskip}
	\item[LFLBL]Low Frequency Long Base Line\vspace*{-.4\baselineskip}
	\item[LPB]  Local Power Box\vspace*{-.4\baselineskip}
	\item[LQS]	Local Quality Supervisor\vspace*{-.4\baselineskip}
	\item[MEOC]	Main Electro Optical Cable\vspace*{-.4\baselineskip}
	\item[MLCM] Master Local Control Modul\vspace*{-.4\baselineskip}
	\item[NWB]	Non-Water-Blocking\vspace*{-.4\baselineskip}
	\item[OD]		Outer Diameter\vspace*{-.4\baselineskip}
	\item[OM]		Optical Module\vspace*{-.4\baselineskip}
	\item[OMF]	Optical Module Frame\vspace*{-.4\baselineskip}
	\item[PBS]	Product Breakdown Structure\vspace*{-.4\baselineskip} 
	\item[PETP]	PolyEthylene TerePhthalate\vspace*{-.4\baselineskip}
	\item[PMT]	Photo Multiplier Tube\vspace*{-.4\baselineskip}
	\item[PU]		PolyUrethane\vspace*{-.4\baselineskip}
	\item[QA/QC]Quality Assurance / Quality Control\vspace*{-.4\baselineskip}
	\item[ROV]	Remote Operated Vehicle\vspace*{-.4\baselineskip}
	\item[SC]   Slow Control\vspace*{-.4\baselineskip}
	\item[SCM]	String Control Module\vspace*{-.4\baselineskip}
	\item[SPE]	Single Photo Electron\vspace*{-.4\baselineskip}
	\item[SPM]	String Power Module\vspace*{-.4\baselineskip}
	\item[SV]		Sound Velocimeter\vspace*{-.4\baselineskip}
	\item[TS]		TimeStamp\vspace*{-.4\baselineskip} 
	\item[TVC]	Time to Voltage Converter\vspace*{-.4\baselineskip} 
	\item[TTS]	Transit Time Spread\vspace*{-.4\baselineskip}
	\item[VNC]	Virtual Network Computing\vspace*{-.4\baselineskip} 
	\item[WB]		Water-Blocking\vspace*{-.4\baselineskip}
	\item[WDM]	Wavelength Division Multiplexer\vspace*{-.4\baselineskip}
	\item[WF]		WaveForm sampling\vspace*{-.4\baselineskip}
\end{description}
\newpage


\section {Introduction}
\label {sec:introduction}
\indent
Neutrino Astronomy is a new and unique method to observe the Universe.  The weakly interacting nature of the neutrino make it a complementary cosmic probe to other messengers such as multi-wavelength light and charged cosmic rays: the neutrino can escape from sources surrounded with dense matter or radiation fields and can travel cosmological distances without being absorbed. This specificity of the neutrino astronomy means that in addition to knowledge on cosmic accelerators seen by other messengers, it may lead to the discovery of objects hitherto unknown. For known high energy sources such as active galactic nuclei, gamma ray bursters, microquasars and supernova remnants, neutrinos will allow to distinguish unambiguously between hadronic and electronic acceleration mechanisms and to localize the acceleration sites more precisely than charged cosmic ray detectors. The ability of neutrinos to exit dense sources means that new compact acceleration sites might be discovered. Furthermore, this feature gives an exclusive signal for indirect searches of dark matter based on the detection of high energy products from the annihilation of dark matter particles which might have been accumulated in the cores of dense objects such as the Sun, Earth and the centre of the Galaxy. Although the search for a diffuse flux of neutrinos from unresolved distant sources is in the research program of neutrino telescopes, the main emphasis of the program is to search for distinct point sources of neutrinos such as the examples mentioned above.
In this matter, the angular resolution of the neutrino telescope is of particular importance: not only to resolve and correlate sources with other instruments using other messengers, but also because it plays an important role in rejecting background.
The flux of neutrinos from interactions of cosmic rays with the atmosphere (``atmospheric neutrinos'') is an irreducible source of background which only differs from the neutrino signal from distant objects in the energy spectrum.
To distinguish a signal from point sources in this background, good angular resolution greatly improves the telescope sensitivity. At a given energy, this angular resolution depends on the optical scattering properties of the medium and on the size of the detector.
\\\indent
The ANTARES detector, located 40~km offshore from Toulon at 2475 m depth\footnote{42$^\circ$48N, 6$^\circ$10E}, was completed on 29 May 2008, making it the largest neutrino telescope in the northern hemisphere and the first to operate in the deep sea. The technological developments made for ANTARES have extensively been built on the experience of the pioneer DUMAND project \cite{bib:dumand_detector} as well as the operational BAIKAL \cite{bib:baikal_detector} detector in Siberia. Some features of the ANTARES design are common with the AMANDA/ICECUBE \cite{bib:amanda_detector} detector at the South Pole. 
\\\indent
The detector infrastructure has 12 mooring lines holding light sensors designed for the measurement of neutrino induced charged particles based on the detection of Cherenkov light emitted in water. The ANTARES telescope extends in a significant way the reach of neutrino astronomy in a complementary region of the Universe to the South Pole experiments, in particular the central region of the local galaxy. Furthermore, due to its location in the deep sea, the infrastructure provides opportunities for innovative measurements in Earth and sea sciences. An essential attribute of the infrastructure is the permanent connection to shore with the capacity for high-bandwidth acquisition of data, providing the opportunity to install sensors for sea parameters giving continuous long-term measurements. Instruments for research in marine and Earth sciences are distributed on the 12 optical lines of the detector and are also located on a 13$^{th}$~line specifically dedicated to the monitoring of the sea environment. 
\\\indent
Another project benefiting from the deep sea infrastructure is an R$\&$D system of hydrophones which investigates the detection of ultra-high energy neutrinos using the sound produced by their interaction in water. This system called AMADEUS (Antares Modules for the Acoustic Detection Under the Sea) is a feasibility study for a prospective future large scale acoustic detector. This technique aimes to detect neutrinos with energies exceeding 100~PeV. The advantage of the acoustic technique is the attenuation length which is about 5~km for the peak spectral density of the generated sound waves around 10~kHz while the attenuation of Cherenkov light in water is about 60~m.
\\\indent
This paper describes the design, construction and operation of the ANTARES Neutrino Telescope with emphasis on the aspects of the infrastructure important for neutrino astronomy. The scope of the present paper is to describe the detector as it was built, the extensive experience obtained in developing this technology will be described in other documents. The marine and Earth sciences aspects of the project are described in other places~\cite{bib:lightbkg, bib:sedimentation, bib:lighttransmission} as is the AMADEUS acoustic detection system~\cite{bib:amadeus}.
\\\indent
Following a summary of the basic concepts of the neutrino detection technique and of the detector architecture, the detector elements are described.
For some aspects of the detector separate papers  have been published and for these the present paper will give a short overview with appropriate references. Those features of the detector which are not described elsewhere are covered in more details. Finally, this paper summarizes the construction and sea deployment of the detector and ends with a description of the detector operation including some performance characteristics.
 
\section {Basic concepts}
\label {sec:basicconcepts}
\subsection{Detection principle}
\label{subsec:detectionprinciple}
\begin{figure}[!ht]
	\centering
		\includegraphics[width=7.5cm]{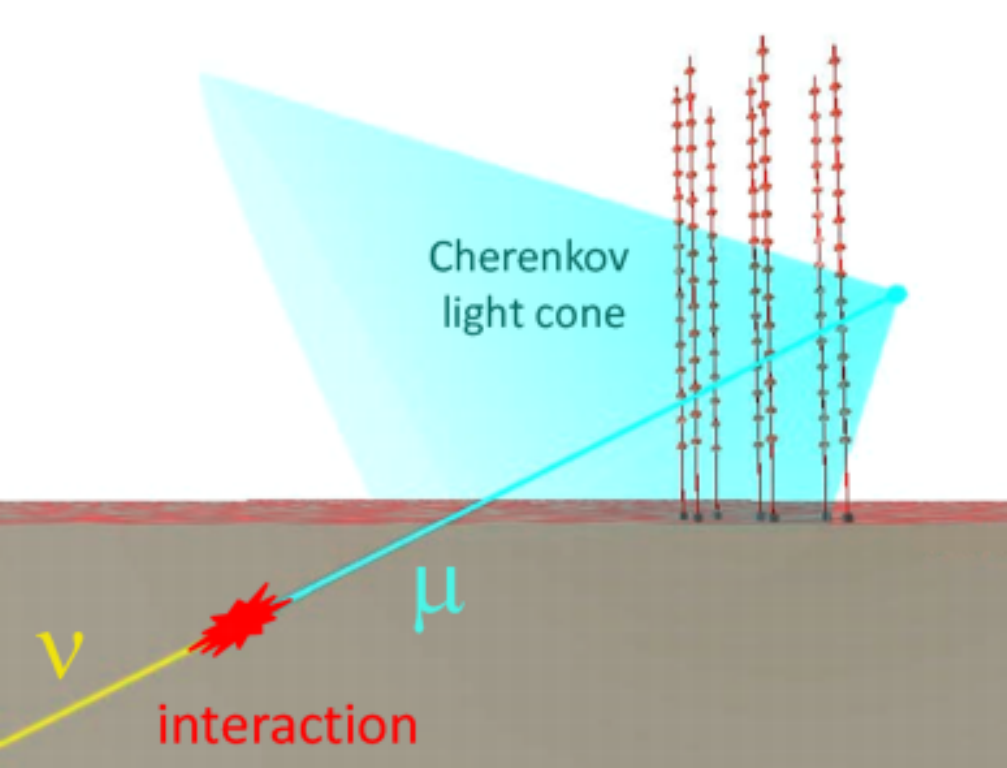}
	\caption{Principle of detection of high energy muon neutrinos in an underwater neutrino telescope. The incoming neutrino interacts with the material around the detector to create a muon. The muon gives Cherenkov light in the sea water which is then detected by a matrix of light sensors. The original spectrum of light emitted from the muon is attenuated in the water such that the dominant wavelength range detected is between 350 and 500 nm.}
	\label{fig:Detection_Principle}
\end{figure}

\indent
The telescope is optimised to detect upward going high energy neutrinos by observing the Cherenkov light produced in sea water from secondary charged leptons which originate in charged current interactions of the neutrinos with the matter around the instrumented volume. Due to the long range of the muon, neutrino interaction vertices tens of kilometres away from the detector can be observed. Other neutrino flavours are also detected, though with lower efficiency and worse angular precision because of the shorter range of the corresponding leptons. In the following the description of the detection principle will concentrate on the muon channel.
\\\indent
To detect the Cherenkov light, the neutrino telescope comprises a matrix of light detectors, in the form of photomultipliers contained in glass spheres, called Optical Modules (OM), positioned on flexible lines anchored to the seabed. The muon track is reconstructed using the measurements of the arrival times of the Cherenkov photons on the OMs of known positions. With the chosen detector dimensions, the ANTARES detector has a low energy threshold of about 20 GeV for well reconstructed muons. The Monte-Carlo simulations 
indicate that the direction of the incoming neutrino, almost collinear with the secondary muon at high energy, can be determined with an accuracy better than 0.3$^\circ$ for energies above 10 TeV. Figure~\ref{fig:Detection_Principle} illustrates the principle of neutrino detection with the undersea telescope.

\subsection{General description of the detector}
\subsubsection{Detector layout}
\label{Detector Layout}
\begin{figure*}[!ht]
	\centering
		\includegraphics[width=0.95\textwidth]{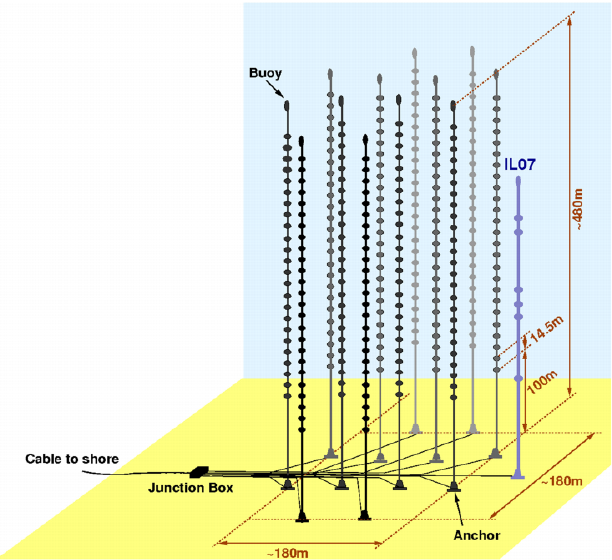}
	\caption{Schematic view of the ANTARES detector.}
	\label{fig:Globalview}
\end{figure*}

\indent
The basic detection element is the optical module housing a photomultiplier tube (PMT). The nodes of the three-dimensional telescope matrix are called storeys. Each storey is the assembly of a mechanical structure, the Optical Module Frame (OMF), which supports three OMs, looking downwards at 45$^\circ$, and a titanium container, the Local Control Module (LCM), housing the offshore electronics and embedded processors. In its nominal configuration, a detector line is formed by a chain of 25 OMFs linked with Electro-Mechanical Cable segments (EMC). The distance is 14.5~m between storeys and 100~m from the seabed to the first storey. The line is anchored on the seabed with the Bottom String Socket (BSS) and is held vertical by a buoy at the top. The full neutrino telescope comprises 12 lines, 11 with the nominal configuration, the twelfth line being equipped with 20 storeys and completed by devices dedicated to acoustic detection (Section \ref{subsec:AcousticDetectionSystem}). Thus, the total number of the OMs installed in the detector is 885. The lines are arranged on the seabed in an octagonal configuration and is illustrated in Figure \ref{fig:Globalview}. It is completed by the Instrumentation Line (IL07) which supports the instruments used to perform environmental measurements.
The data communication and the power distribution to the lines are done via an infrastructure on the seabed which consists of Inter Link cables (IL), the Junction Box (JB) and the Main Electro-Optical Cable (MEOC).

\subsubsection{Detector architecture}
\begin{figure}[!ht]
  \centering
  \includegraphics[width=.95\columnwidth]{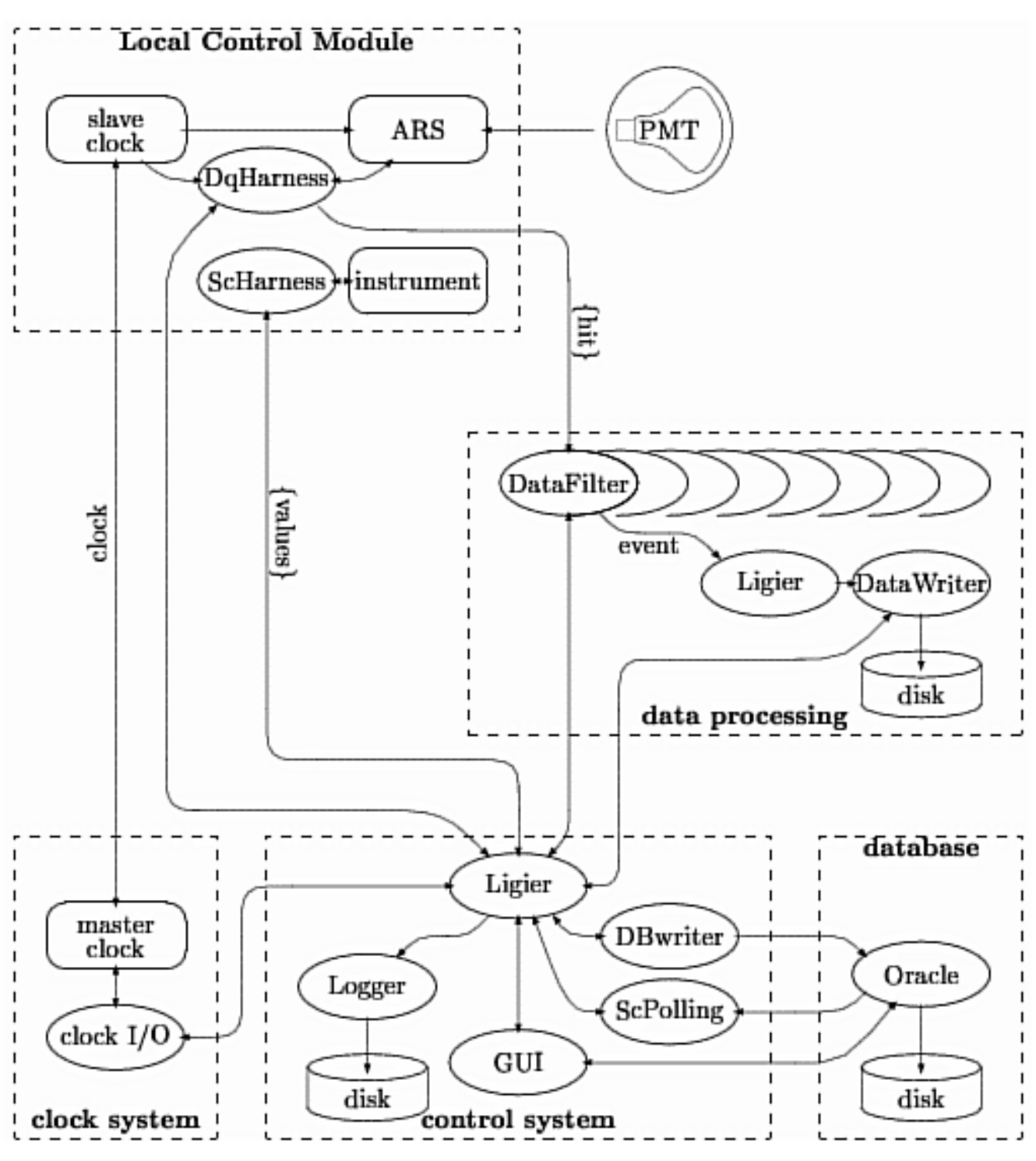}
  \caption{Schematic view of the data acquisition system. The dashed line boxes refer to hardware devices, the ellipses correspond to processes running on those devices. The lines between processes indicate the exchange of information (commands, data, messages, etc.).}
  \label{fig:daq-scheme}
\end{figure}

\indent
The Data Acquisition system (DAQ) is based on the ``all-data-to-shore'' concept \cite{bib:antares-daq-paper}. In this mode, all signals from the PMTs that pass a preset threshold (typically 0.3 Single Photo Electron (SPE)) are digitized in a custom built ASIC chip, the Analogue Ring Sampler (ARS)~\cite{bib:arspaper}, and all digital data are sent to shore where they are processed in real-time by a farm of commodity PCs. The data flow ranges from a couple of Gb~s$^{-1}$ to several tens of Gb~s$^{-1}$, depending on the level of the submarine bioluminescent activity. To cope with this large amount of data, the readout architecture of the detector has a star topology with several levels of multiplexing.
The first level is in the LCM of each storey of the detector, where the data acquisition card containing an FPGA and a microprocessor outputs the digitised data of the three optical modules. The card is also equipped with dedicated memory to allow local data storage and it manages the delayed transmission of data in order to avoid network congestion. The transmission is done through a bi-directional optical fibre to the Master Local Control Module (MLCM), a specific LCM located every fifth storey. It is equipped with an Ethernet switch which gathers the data from the local OMs and from the four connected storeys. Such a group of 5 storeys is called a sector. The switch of each sector is connected via a pair of uni-directional fibres to a Dense Wavelength Division Multiplexing (DWDM) system in an electronics container, the String Control Module (SCM), situated on the BSS at the bottom of each line. The DWDM system is then connected to the junction box on the seabed via the interlink cables. In the junction box the outputs from up to 16 lines are gathered onto the MEOC and sent to the shore station.  In the shore station, the data are demultiplexed and treated by a PC farm where they are filtered and then sent via the commercial fibre optic network to be stored remotely at a computer centre in Lyon\footnote{http://cc.in2p3.fr}. A schematic view of the readout architecture is shown in Figure~\ref{fig:daq-scheme}.

The electrical supply system has a similar architecture to the readout system. The submarine cable supplies up to 4400~VAC, 10~A to a transformer in the junction box. The sixteen independent secondary outputs from the transformer provide up to 500~VAC, 4~A to the lines via the interlink cables. At the base of each line a String Power Module (SPM) power supply distributes up to 400~VDC to each sector. The MLCM and LCMs of the sector are fed in parallel and the power is used by a Local Power Box (LPB) in each storey to provide the various low voltages required by each electronics board.

\subsubsection{Master clock system}
\indent
Precise timing resolution for the recorded PMT signals, of order 1 ns, is required to maintain the angular resolution of the telescope. An essential element to achieve this precision is a 20 MHz master clock system, based onshore, which delivers a common reference time to all the offshore electronics in the LCMs. This system delivers a timestamp, derived from GPS time, via a fibre optic network from the shore station to the junction box and then to each line base and each LCM. The master clock system is self calibrating and periodically measures the time path from shore to the LCM by echoing signals received in the LCM back to the shore station.
\subsubsection{Positioning system}
\label{subsubsec:positioning_system}
\indent
The detector lines connecting the OMs are flexible and are moving continually in the sea current. In order to ensure optimal track reconstruction accuracy, it is necessary to monitor the relative positions of all OMs with accuracy better than 20 cm, equivalent to the ~1 ns precision of the timing measurements. The reconstruction of the muon trajectory and the determination of its energy also require the knowledge of the OM orientation with a precision of a few degrees. In addition, a precise absolute orientation of the whole detector has to be achieved in order to find potential neutrino point-sources in the sky. To attain a suitable precision on the overall positioning accuracy, a constant monitoring with two independent systems is used:
\begin{itemize}
	\item A High Frequency Long Base Line acoustic system (HFLBL) giving the 3D position of hydrophones placed along the line. These positions are obtained by triangulation from emitters anchored in the base of the line plus autonomous transponders on the sea floor.
\vspace*{-0.40\baselineskip}
	\item A set of tiltmeter-compass sensors giving the local tilt angles of each storey with respect to the vertical line (pitch and roll) as well as its orientation with respect to the Earth magnetic north (heading).
\end{itemize}

\subsubsection{Timing calibration systems}
The timing calibration of the detector was performed during the construction and is continually verified and adjusted during operation on a weekly basis. The master clock system measures the time delays between the shore station and the LCMs leaving only the short delays between the electronics in the LCM and the photon arrival at the PMT photocathode as a time offset requiring further calibration. These offsets are first measured after line assembly on shore and then again in the sea after deployment. This {\it in situ} calibration uses a system of external light sources (optical beacons) distributed throughout the detector. There are two types of optical beacons: LED beacons located in four positions on each detector line and laser beacons located on the bottom of two particular lines. In addition, there is an LED inside each optical module which is used to monitor changes in the transit time of the photomultiplier.

\subsection{Detector design considerations}
\indent
The detector location on the seabed at a depth of 2475~m imposes many constraints on the detector design. All components must withstand a hydrostatic pressure between 200 and 256~bar and resist corrosion or degradation in the sea water of 46~mS~cm$^{-1}$ conductivity. The seabed environment has a stable temperature around 13$~^{\circ}$C and little risk of shock or variable mechanical stress. The detector lines sway in the sea current which is typically 10~cm~s$^{-1}$ with variations up to a maximum value of 30~cm~s$^{-1}$. The detector components were designed to take into account possible shocks, vibrations and high temperatures during construction, transport and deployment. All components were chosen with the objective of a minimum detector life time of 10~years.
\\\indent
The materials to be in contact with the sea water were selected according to their known resistance to corrosion: glass, titanium alloys (grade 2 and 5), anode protected carbon steel, polyethylene (LDPE and PETP), polyurethane, aramid and glass-epoxy (syntactic foam and fibre composite). Stainless steel and aluminium alloys were not used due to their reduced corrosion resistance. In addition to this material selection, special attention was paid to prevent any parasitic electrical currents able to induce electrolytic corrosion. Isolating interfaces were used between metals of different nature and the electrical power distribution system was designed to prevent any current leak to the water. 
\\\indent
Avoiding water leaks during operation imposed many constraints on the detector design.  When possible, O-rings in containers, made of Viton\footnote{Viton\circledR, http://www.dupontelastomers.com/products/viton/} or nitrile material were implemented in two seals in a redundant way. The O-ring material hardness, its cross section diameter, the shape and the surface roughness of the groove as well as the characteristics of the matching parts were specified following the recommendations of the manufacturer for the {\it in situ} pressure. Tests under pressure were performed on all the major containers (JB container and glass spheres) and EMC sections. Electronics containers have been tested by sampling. Some tests were performed by the manufacturer of the component (glass spheres and short sections of the EMC ) and others were performed by the collaboration at IFREMER\footnote{IFREMER, www.ifremer.fr}, at the COMEX\footnote{\label{fn:comex}COMEX, www.comex.fr} and Ring-O Valve\footnote{Ring-O Valve SpA, 23823 Colico, Italy.} companies (JB and electronics containers, the rest of the EMC sections). The pressure tests were based on the IFREMER rules
for undersea vessels for a working pressure of 256~bar: a cycle up to 310~bar for 24~h and ten cycles up to 256~bar for 1h with all the pressure changes made at a rate of $\pm$~12~bar per minute. The criterion of success for the acceptance test was the integrity of the tested element, the absence of water inside the containers and the electro-optical continuity of the cable under static pressure conditions.
\\\indent
The maximum static tension along the line is expected to occur during the line deployment in the section below the first storey, which has to sustain the weight in water of the full anchor (BSS + deadweight): $\approx$~3~tons. Dynamic load may reach higher values during the deployment, due to the swell. Since the total mass of the line is 7 tons, an upward acceleration of 1~$g$, for instance, will add a tension of 70 kN in the top part of the line during the descent. In order to minimise the risks of high dynamic loads, the deployment of the lines were required to be performed in quiet sea state ($\leq$~3 on the Beaufort scale, corresponding to waves of $\approx$~60~cm high). However, since the conditions are difficult to predict accurately for the $\approx8$~hours needed for a deployment or a recovery, the general dimensioning rules recommended by IFREMER for deployments in the sea from a surface boat were imposed:
\begin{eqnarray}
Breaking~Load~>~Static~Load \times A 
\label{eq:load_rule}
\end{eqnarray}
\noindent
where $A$ = 1.5 for metal parts (BSS, OMF and buoy equipment) and 4 for organic fibres (the Aramid braid of the vertical EMC). This rule results in a breaking load of more than 7~tons for the OMF and 18~tons for the EMC.

\section{Line structure}
\label {sec:linestructure}
A line is the assembly of an anchor sitting on the seabed, 25 storeys and a top buoy linked by electro-optical mechanical cables. A storey consists of three optical modules, the metal structure that supports them and provides interfaces with the EMCs, the electronics container and additional instrumentation. In order to limit the number of single point failures for a full line, a line is divided in 5 sectors of 5 successive storeys each. The sectors are independent for the power distribution and the data transmission. The distribution of power and routing of clock and acquisition signals toward each sector are performed in electronics containers fixed on the BSS.

\subsection{Optical modules}
\label{sec:OpticalModules}

The optical module, the basic sensor element of the telescope, is the assembly of a pressure resistant glass sphere housing a photomultiplier tube, its base and other components. A detailed description of the ANTARES OM can be found in \cite{bib:om}.
\subsubsection {Photo detector requirements}
The search for a highly sensitive light detector led to the choice of photomultiplier tubes with a photocathode area as large as possible combined with a large angular acceptance. Regarding these criteria, the best candidates are large hemispherical tubes. However, the PMT size is limited by some characteristics which increase with the photocathode area:
\\\indent - the transit time spread (TTS) which has to be small enough to ensure the required time resolution,
\\\indent - the dark count rate which must be negligible compared to photon background rate.
\\\indent In summary, the main requirements for the choice of the ANTARES PMTs are:
\\\indent\indent $\circ\ $photocathode area $>$~500~cm$^2$ 
\\\indent\indent $\circ\ $quantum efficiency $>$~20~$\%$
\\\indent\indent $\circ\ $collection efficiency $>$~80$\%$
\\\indent\indent $\circ\ $TTS $<$~3~ns 
\\\indent\indent $\circ\ $dark count rate $<$~10~kHz (threshold at $1/3$~SPE, including glass sphere)
\\\indent\indent $\circ\ $peak/valley ratio $>$~2
\\\indent\indent $\circ\ $peak width (FWHM)/peak position $<$ 50$\%$
\\\indent\indent $\circ\ $gain of 5$\times$10$^{7}$ reached with HV~$<$~2000~V
\\\indent\indent $\circ\ $pre-pulse rate $<$~1$\%$
\\\indent\indent $\circ\ $after-pulse rate $<$~15$\%$
\subsubsection {Optical module components}
\begin{figure}[!h]
	\centering
	\includegraphics[width=7.5cm]{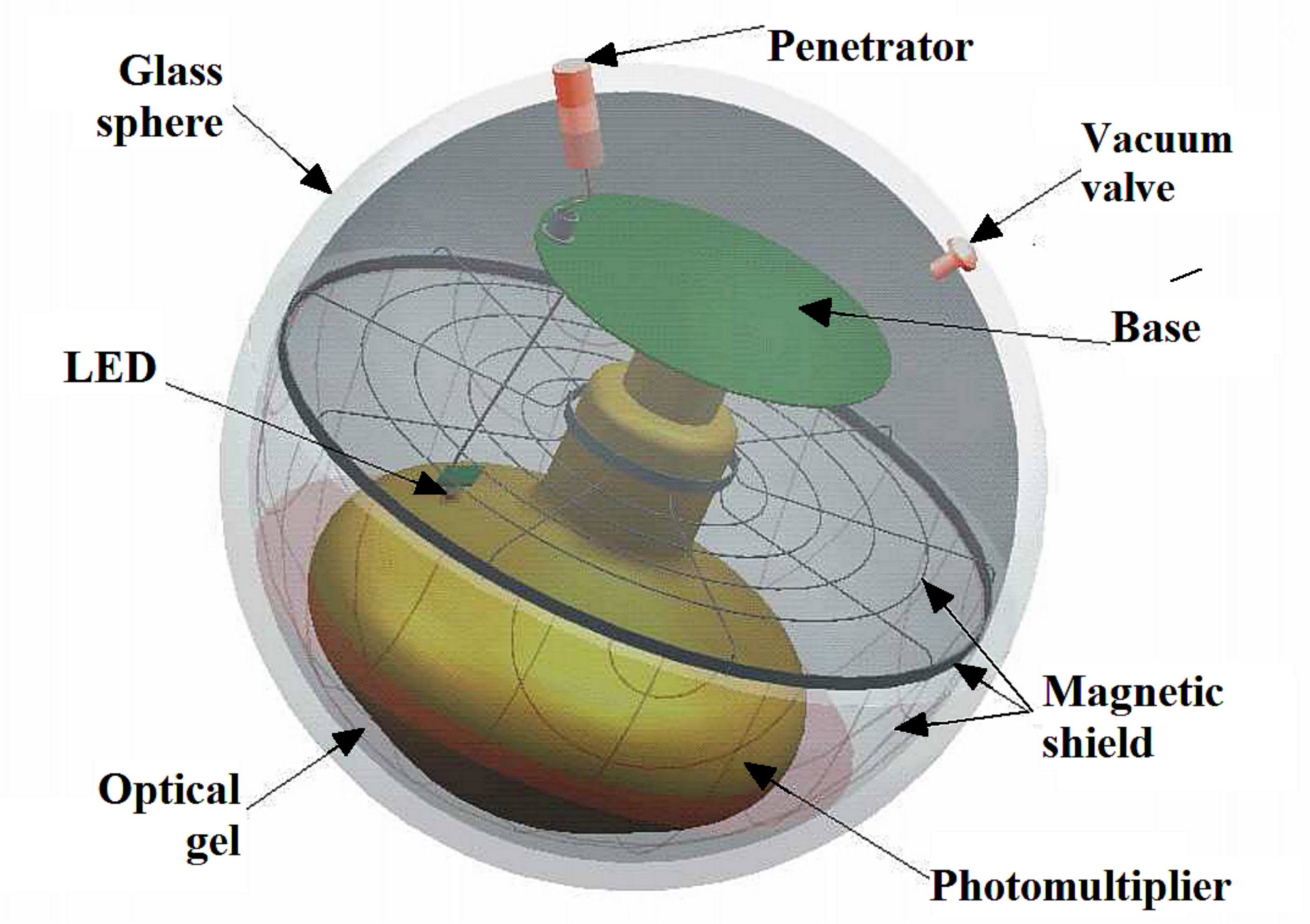}
	\caption{Schematic view of an optical module}
	\label{fig:OMComponents}
\end{figure}

\begin{figure}[!ht]
	\centering
	\includegraphics[width=7.5cm]{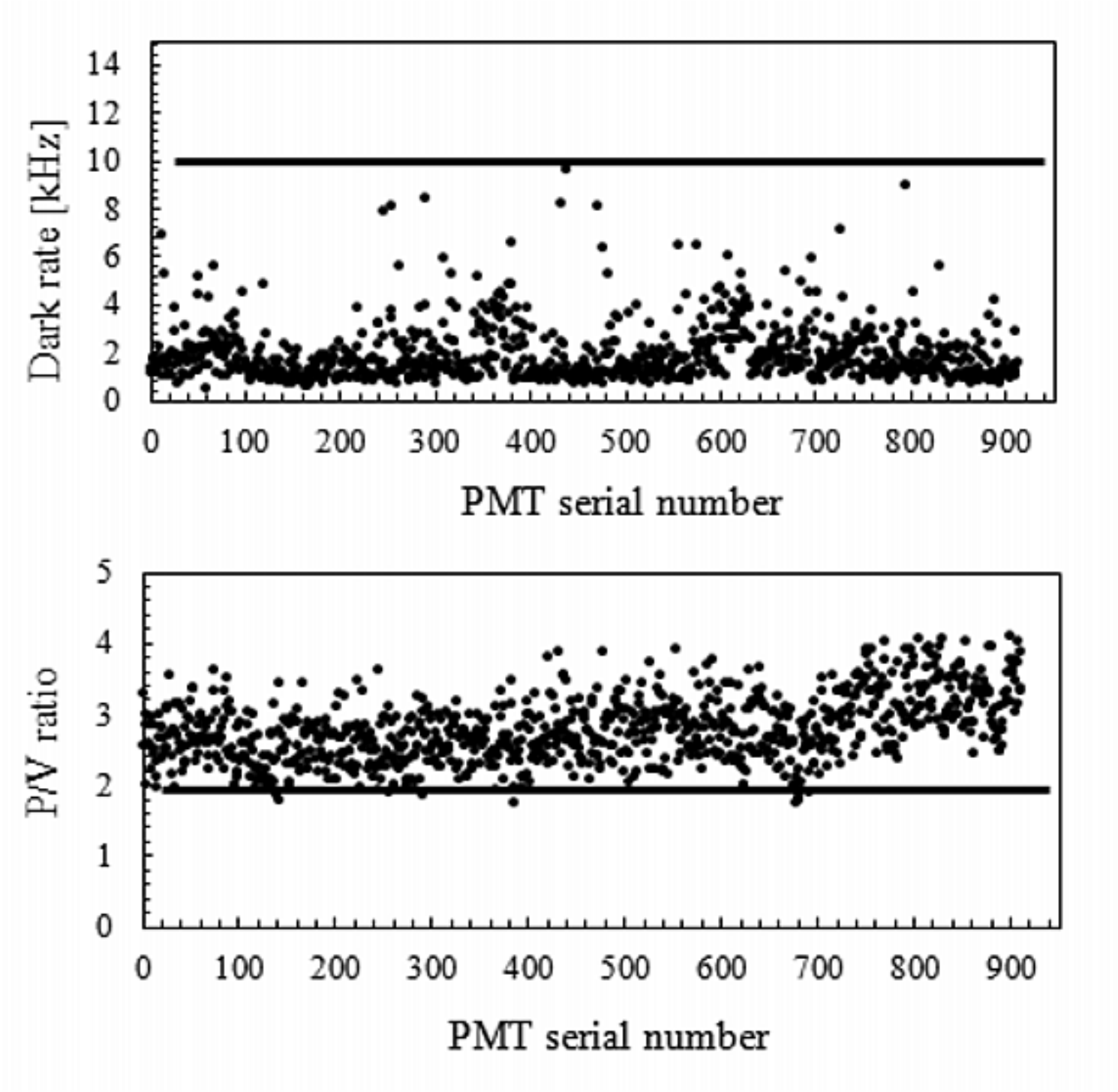}
	\caption{Results of dark count rate (top) and peak/valley ratio (bottom) for the full set of tested PMTs.}
	\label{fig:PMT_prod}
\end{figure}

\begin{figure}[!h]
	\centering
	\includegraphics[width=7.5cm]{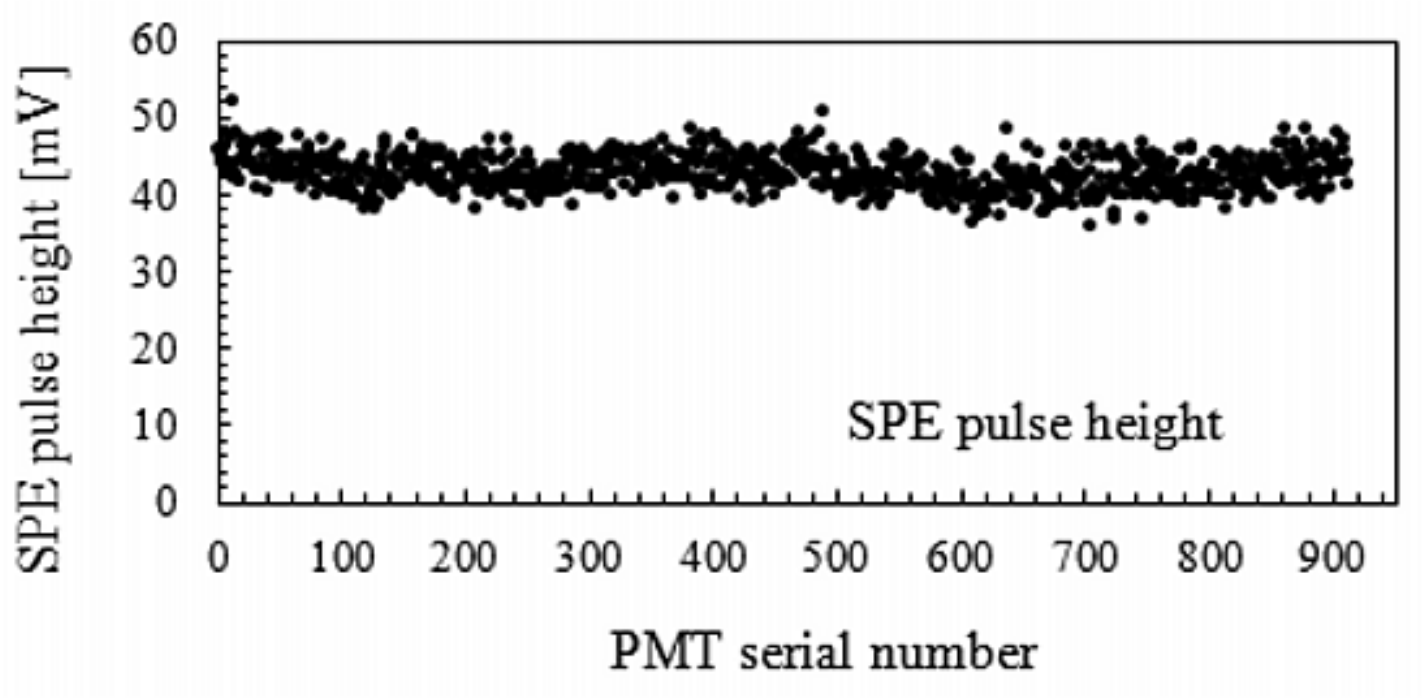}
	\caption{Measured mean pulse height of single photoelectrons for each PMT at nominal gain.}
	\label{fig:SPE_vs_HV_prod}
\end{figure}

Figure~\ref{fig:OMComponents} shows a schematic view of an optical module with its main components. The following sections describe the different components and, when relevant, the assembly process.
\paragraph{Photomultiplier tube}$\ $
\\\indent
In the R\&D phase, an extensive series of tests were performed on several commercially available models of large hemispherical photomultipliers. A summary of this study is presented in \cite{bib:largeom}.
The R7081-20, a 10'' hemispherical tube from Hamamatsu\footnote{Hamamatsu Photonics, Electron tube division, http://www.hamamatsu.com}, was chosen. The full sample of delivered PMTs has been tested with a dedicated test bench in order to calibrate the sensors and to check the compliance with the specifications. The number of rejected tubes was small (17, their peak/valley ratio being too low), these tubes were replaced by the manufacturer. To illustrate the homogeneity of the production, Figure~\ref{fig:PMT_prod} shows the measured values of dark noise rate (top) and of the peak/valley ratio (bottom).
During the testing process, the working point of each PMT, i.e. the high voltage needed to obtain a gain of $5\times10^7~\pm~10~\%$, was determined by measuring the value of the SPE pulse height. The results of these measurements are illustrated in Figure~\ref{fig:SPE_vs_HV_prod}.
\paragraph{Glass sphere}$\ $
\label{subsec:glasssphere}

\indent
The protective envelope of the PMT is a glass sphere of a type routinely used by sea scientists for buoyancy and for instrument housing. These spheres, because of their mechanical resistance to a compressive stress and of their transparency, provide a convenient housing for the photodetectors. Table \ref{tab:DataOnTheOMGlassSphere} summarizes the main characteristics of the Vitrovex$^{\circledR}$ glass spheres\footnote{Nautilus Marine Service GmbH, http://www.nautilus-gmbh.de} used.
\begin{table}[!h]
\renewcommand{\arraystretch}{1.3}
\centering
\footnotesize
	\begin{tabular}[width=0.90\columnwidth]{|l|c|}
	\hline
	Outer diameter & 432 mm (17")\cr
	\hline
	Wall thickness & 15 mm\cr
	\hline
	Type of glass & Borosilicate\cr
	\hline
	Refractive index & 1.47\cr
	\hline
	Light transmission above 350 nm & $>$95\%\cr
	\hline
	Density & 2.23~g~cm$^{-3}$\cr
	\hline
	Pressure of qualification test & 700 bar (70 MPa)\cr 
	\hline
	Diameter shrinking at 250 bar & 1.25 mm (0.3\%)\cr
	\hline
	Absolute internal air pressure & 0.7 bar (70 kPa)\cr
	\hline
	Hole diameters & 20 mm, 5 mm\cr
	(penetrator and vacuum port) & \cr
	\hline
	\end{tabular}
	\caption{Data on the OM glass sphere.}
	\label{tab:DataOnTheOMGlassSphere}
\end{table}
The sphere is provided as two hemispheres: one, referred to as ``back hemisphere'' is painted black on its internal surface and the other, ``front hemisphere'' is transparent.
The front hemisphere houses the PMT and the magnetic shielding held in place by the optical gel.
The back hemisphere has two drilled holes to accommodate the electrical connection via a penetrator and a vacuum port. Around both holes a flat surface is machined on the outside of the sphere for the contact of the single O-ring ensuring water tightness. The back hemisphere is also equipped with a manometer readable from the outside.
The two glass halves have precisely machined flat equatorial surfaces in direct contact (glass/glass) without any gasket or interface.
The risk of implosion and the consequences on the structure were considered since its potential energy is of the order of a megajoule (200~g of TNT) at the depth of the detector. 
Based on tests performed by DUMAND~\cite{bib:dumand_implosion} and further tests performed off Corsica in the year 2000 by the ANTARES Collaboration, it has been concluded that the implosion of a glass sphere at the ANTARES depth would provoke the loss of the two other spheres of the same storey (at centre distances of 770~mm) but not of spheres on adjacent storeys (at a distance of 14.5~m), and would not cut or damage the cable. The rigid storey mechanical frame would be distorted but not destroyed by the implosion.
\paragraph{Optical gel}$\ $

\indent
The optical coupling between the glass sphere and the PMT is achieved with optical gel. The chosen gel is a two-component silicon rubber provided by the Wacker company\footnote{Silgel 612 A/B; Wacker-Chemie AG, http://www.wacker.com}.
The mixture of the components is made in the ratio 100:60. After curing and polymerization, lasting 4 hours at ambient temperature, the optical gel reaches an elastic consistency soft enough to absorb the sphere diameter reduction by the deep sea pressure (1.2~mm) and stiff enough to hold the PMT in position in the sphere. The optical properties of the gel have been measured in the laboratory: the absorption length is 60~cm and the refractive index is 1.404 for wavelengths in the blue domain.

\paragraph{Magnetic shield}$\ $

\indent
At the ANTARES site, the Earth's magnetic field has a magnitude of approximately 46~${\upmu}$T and points downward at 31.5$^{\circ}$ from the vertical. Uncorrected, the effect of this field would be a significant degradation of the TTS, of the collection efficiency and of the charge amplification of the PMT. A magnetic shield is implemented by surrounding the bulb of the PMT with a hemispherical grid made of wires of $\mu$-metal\footnote{Sprint Metal, Ugitech, http://www.ugitech.com} closed by a flat grid on the rear of the bulb. This provides a magnetic shielding for the collection space and for the first stages of the amplification cascade. The efficiency of the screening becomes larger as the size of the mesh is reduced and/or the wire diameter is increased, however the drawback is a shadowing effect on the photocathode. The compromise adopted by the ANTARES Collaboration, a mesh of 68~$\times$~68~mm$^2$ and wire diameter of 1.08~mm, results in a shadowing of less than 4~$\%$ of the photocathode area while reducing the magnetic field by a factor of three. Measurements performed in the laboratory show that this shielding provides a reduction of 0.5~ns on the TTS and a 7~\% increase on the collected charge with respect to a naked, uniformly illuminated PMT.
\paragraph{HV power supply}$\ $

\indent
To limit the power consumption of the HV power supply a high voltage generator based on the Cockroft-Walton~\cite{bib:CroWal} scheme is adopted. The HV generator chosen for the ANTARES detector is derived from the model developed for the AMANDA experiment\footnote{http://icecube.wisc.edu/}, and is manufactured by the iseg company\footnote{PHQ7081-20; iseg Spezialelektronik GmbH, http://www.iseg-hv.de}.
It has two independent high-voltage chains. The first chain produces a constant focusing voltage (800~V) to be applied between photocathode and first dynode. The second chain gives the amplification voltage, which can be adjusted from 400~V to 1600~V by an external DC voltage. The HV generator is powered by a 48~V~DC power supply and has a typical consumption of 300~mW.
\\\indent
The signals of the anode, of the last dynode and of the last-but-two dynode of the PMT are routed to the electronics container together with the PMT ground. A low level voltage image of the actual HV is provided for monitoring purpose.
\paragraph{Internal LED}$\ $

\indent
On the rear part of the bulb of the PMT, a blue LED is glued in such a way to illuminate the pole of the photocathode through the aluminium coating, which acts as a filter of large optical density (optical density~$\approx$~5). This LED is excited by an externally driven pulser circuit and is used to monitor the internal timing of the OM.
\paragraph{Link with the electronics container}$\ $

\indent
The electrical connection of the OM to the electronics container is made with a penetrator\footnote{\label{fn:euroc}EurOc\'eanique S.A., part of MacArtney Underwater Technology, http://www.macartney.com} (Ti socket with polyurethane over moulding). The associated cable contains shielded twisted pairs for the transmission of power, the control of the LED pulser and the setting and monitoring of the DC command voltage of the PMT base. One pair is used to transmit the anode and the last dynode signals. This pseudo differential transmission pair has the advantage of reducing the noise and enhancing the output signal by approximately a factor of two when the subtraction is done at the readout electronics.
The last pair is used to transmit signals from the last-but-two dynode, together with the ground, for the treatment of very high amplitude signals.
\paragraph{Final assembly and tests}$\ $

\indent
The assembly starts with the pouring of the gel into the front hemisphere and a precise sequence of out-gasing is applied in order to avoid the appearance of bubbles during the polymerization phase. Then, the cage and the PMT are positioned by tools which ensure a defined position with respect to marks on the hemisphere. These marks are also used to mount the OM on its support structure, giving each PMT a well-defined and reproducible orientation with respect to the storey mechanical structure.

\begin{figure}[!ht]
	\centering
	\includegraphics[width=7.5cm]{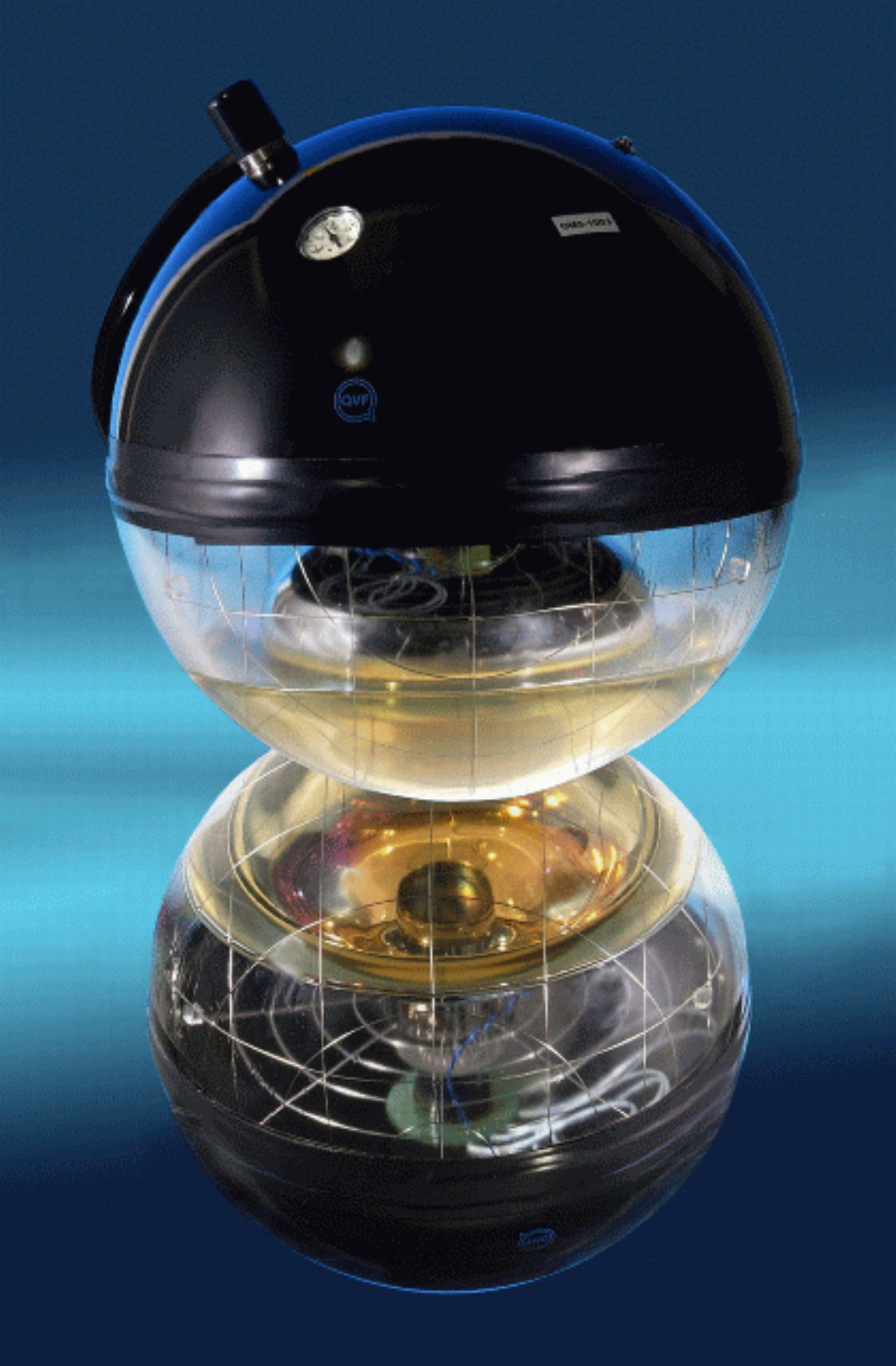}
	\caption{Photograph of an optical module. It is positioned on a mirror to better show the full assembly.}
	\label{fig:OMPicture}
\end{figure}

\indent
After the gluing of the LED, the cabling of the base to the pig-tail of the penetrator and the connection of the PMT, the back hemisphere is placed in contact with the front one. Closure is obtained by establishing an underpressure of $\approx$~300~mbar inside the sphere. The equatorial seam is sealed externally with butyl rubber sealant which is protected by a sealant tape. Figure~\ref{fig:OMPicture} shows an assembled OM.
The same test bench as for the naked PMT is used to test the OM. Dark count rate, gain and LED functionality are checked.

\subsubsection {OM support}
\begin{figure}[!h]
	\centering
		\includegraphics[width=6.5cm]{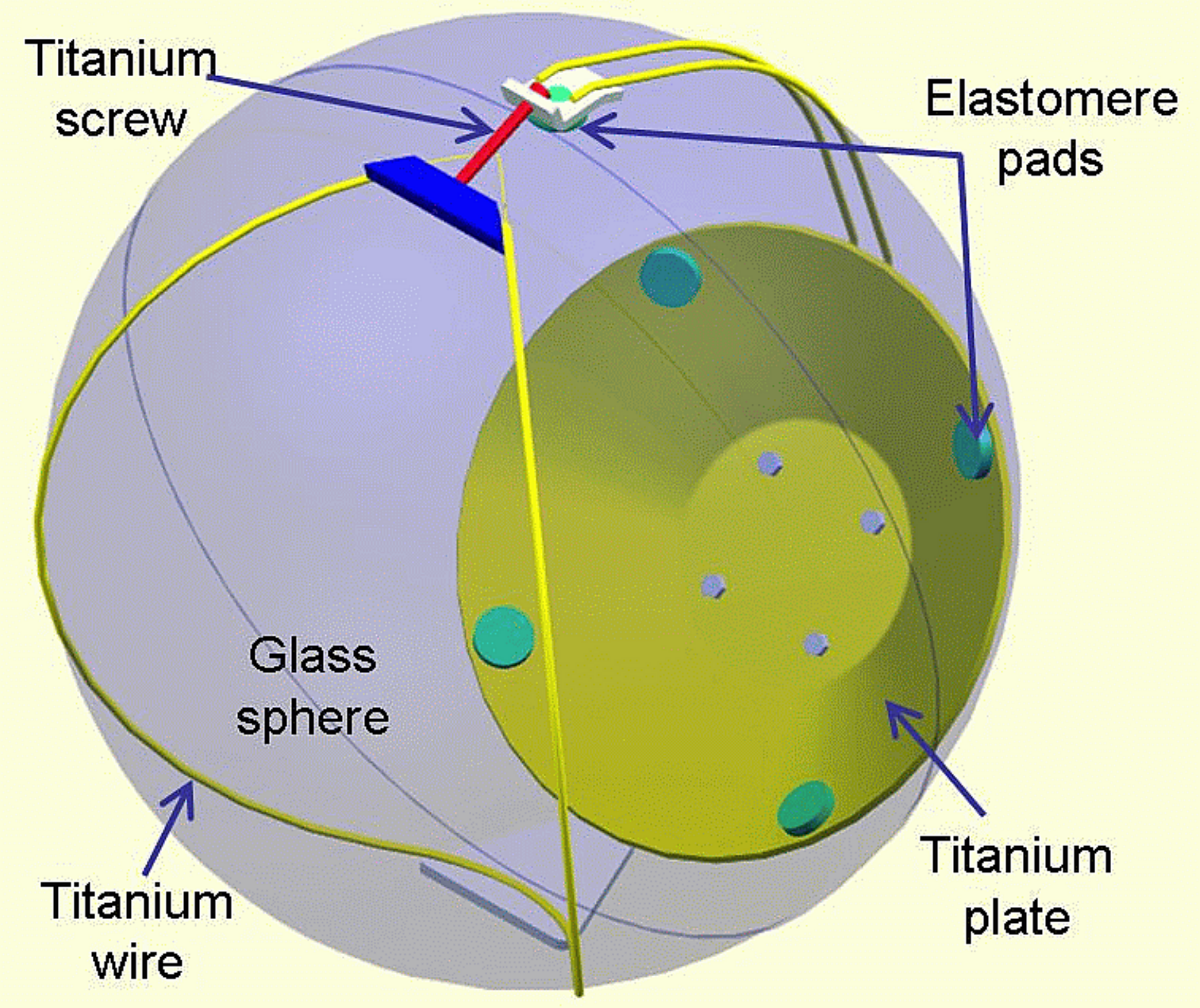}
	\caption{OM support mechanics.}
	\label{fig:SupportOM}
\end{figure}

The OM support is made of a stamped Ti grade 2 conical plate (OD~=~280~mm) on which the OM is pulled by a pair of Ti wires ($\oslash$~=~4~mm) under tension running around the glass sphere (Figure~\ref{fig:SupportOM}).
The wires are designed to follow a great circle of the sphere, which results in their stable equilibrium position on the glass surface. A set of 5 rubber pads are inserted between the metal parts and the OM to protect the glass surface and to keep the assembly under tension in spite of the pressure shrinking. Tests at 250 bar (25 MPa) showed that the support allows the OM to sustain a test torque of 5 Nm without rotating. The titanium plate is also the interface to the optical module frame.
\subsection{Storey}
\subsubsection {Optical module frame}

The role of the optical module frame (OMF) is to hold the three OMs of the storey, the associated LCM container and to connect both mechanical terminations of the EMCs. The OMF and its connections are specified up to a breaking load of 7~tons. Some OMFs also hold optional equipment such as LED beacons~\cite{bib:ledbeacon}, positioning hydrophones and certain oceanographic sensors.
\begin{figure}[!h]
	\centering
		\includegraphics[width=7.4cm]{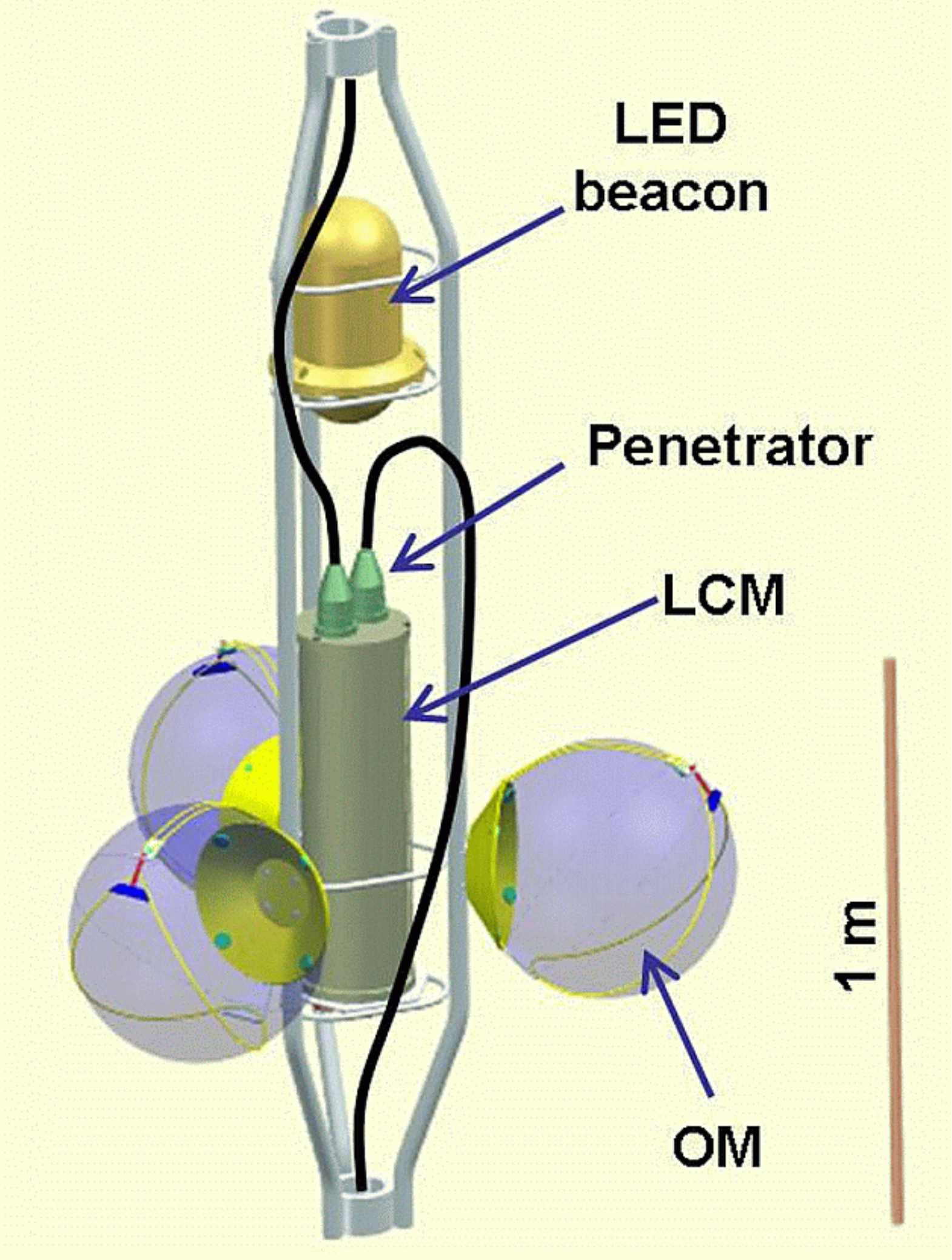}
	\caption{OMF equipped with the 3 OMs, the LCM and an LED beacon. The mechanical parts used for fixing cables toward the upper and the lower storeys are omitted.}
	\label{fig:OMF}
\end{figure}

The OMF (Figure~\ref{fig:OMF}) is a welded vertical structure of Ti (grade 2; chosen for the ease of welding) and of three-fold periodic symmetry around the vertical axis. The main elements are:
\begin{itemize}
	\item at the top and bottom, two rings (ID~=~85~mm) on which are locked the EMC mechanical terminations;
\vspace*{-0.40\baselineskip}
  \item three shaped tubes (OD~=~33.4~mm, thickness~=~3.38~mm) connecting these rings vertically with an overall height of 2.12~m (2~m between EMC mechanical terminations);
\vspace*{-0.40\baselineskip}
  \item four spacers of triangular shape made of 12~mm diameter rod between the three tubes:
\vspace*{-0.40\baselineskip}
  \begin{itemize}
  \item the bottom triangle holds the LCM container on the vertical axis of the OMF;
\vspace*{-0.40\baselineskip}
  \item the next spacer stiffens the structure at the height of the 3 OM fixture plates (80$\times$80$\times$5~mm), welded on each tube at a distance of 195~mm from the axis;
\vspace*{-0.40\baselineskip}
	\item the two top spacers are used to hold the optional LED beacon.
	\end{itemize}
\end{itemize}

All OMFs were validated by applying a traction load of 80 kN, which is in fact higher than the load resulting from the design rule of 7~tons.
\subsubsection {Local control module}
\label{subsec:localcontrolmodule}
\paragraph{Container}$\ $

\indent
The housing of the readout electronics is a Ti grade 5 container made of a hollow cylinder (600~mm long, 179~mm outer diameter and 22~mm wall thickness) closed by two end caps (30~mm thick). The top end cap accommodates the two large penetrators of the EMC linking the storey to its upper and lower adjacent storey. The bottom end cap accommodates three connectors linking the LCM to its three optical modules. In some of the LCMs, a 4$^{th}$ connector is needed for additional equipment.
The fixation of the end caps on the cylinder and of the whole container on the OMF is made with three external threaded rods of 6~mm diameter in Ti grade 2. The thickness of the cylinder and of the end-caps was optimised by Finite Element Method analysis with the goal to stay within the yield strength of the material at an external pressure of 310 bars. The calculations were tested by the collapse under pressure of an Al alloy container of the same configuration. Ti grade 5 was chosen for its yield strength around 900~MPa, compared to that of grade 2 which is around 300~MPa.
\paragraph{Electronics}$\ $

\indent
In order to optimally fill the cylindrical volume offered by the container, a dedicated crate was developed. This crate accepts circular shaped printed circuit boards plugged on a backplane which distributes the signals as well as the DC power supplied by the local power box. The crate was designed to ensure that its mechanical structure acts as a medium that transfers the heat produced by the electronics to the Ti cylinder in contact with the water. After evaluating different metals, the final choice was made for aluminium which can guarantee good performance with light weight and at an affordable price.
Furthermore, boards having high power consumption are equipped with metal cooling bases which are in thermal contact with the crate.
\\\indent
Most of the LCMs contain the same set of electronics cards. However, due to the segmentation of a line in sectors, one in five LCMs, called Master LCM or MLCM, acts as a master for other LCMs of the same sector and houses additional boards. Other differences between individual LCMs are due to electronics necessary for optional equipment on the storey (hydrophone, LED beacon, ...).

A standard LCM contains the following elements:
\begin{itemize}
  \item LPB. Fixed on the crate, the local power box is fed by the 400~V~DC from the bottom of the line, and provides the 48~V for the optical modules and several different low voltages for the electronics boards. An embedded micro-controller allows the monitoring of the voltages, the temperatures and the current consumptions as well as the remote setting of the 48~V for the OMs.
\vspace*{-0.20\baselineskip}
	\item CLOCK. The clock reference signal coming from shore reaches the bottom of the line where it is repeated and sent to each sector. Within a sector, the clock signal is daisy-chained between LCMs. The role of the CLOCK card is to receive the clock signal from the lower LCM, to distribute it on the backplane and to repeat it toward the upper LCM of the sector. It also has the capability to pass commands on the backplane which are coded within the clock signal.
\vspace*{-0.20\baselineskip}
	\item ARS$\_$MB (Figure~\ref{fig:ARS_MB}). The ARS motherboards host the front-end electronics of the OMs (one board per OM). This front-end electronics consists of a 
\begin{figure}[!h]
	\centering
		\includegraphics[width=7.cm]{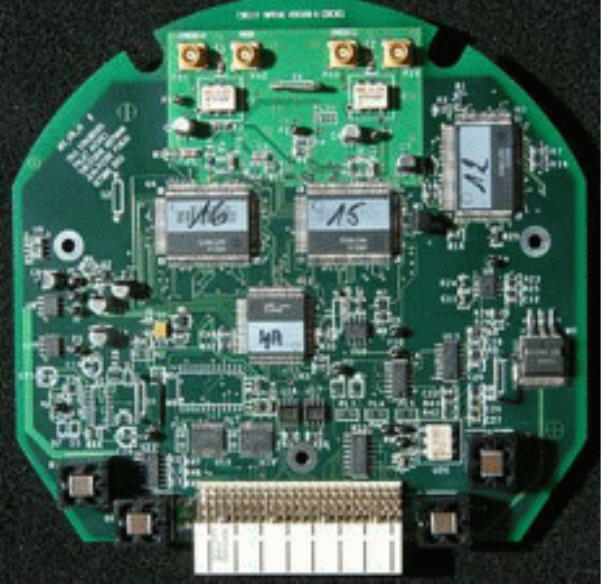}
	\caption{The ARS$\_$MB board with the 2 ARSs (labelled 16 and 15). The 3$^{rd}$ one (top right, labelled 12) is foreseen for trigger purposes.}
	\label{fig:ARS_MB}
\end{figure}
custom-built Analogue Ring Sampler (ARS) chip~\cite{bib:arspaper} which digitizes the charge and the time of the analogue signal coming from the PMTs, provided its amplitude is larger than a given threshold. The level of this threshold is tuneable by slow-control commands. The analogue signal is integrated by an AVC (Amplitude to Voltage Converter) to obtain the charge which is digitized by an ADC. The ARS can also operate like a flash-ADC using analog memories with a sampling tuneable down to sub-nanosecond values. The output consists of a waveform of 128 amplitude samples. The arrival time is determined from the signal of the clock system in the LCM and from a TVC (Time to Voltage Converter) which provides a sub-nanosecond resolution. To minimise the dead time induced by the digitization, each ARS$\_$MB card is equipped with 2 ARSs working in a token ring scheme. For a storey with an optical beacon, a 4$^{th}$ ARS$\_$MB is installed to digitize the signals sent by the internal PMT of the beacon.
\vspace*{-0.20\baselineskip}
	\item DAQ/SC (Figure~\ref{fig:DAQ_board}). The DAQ/Slow-Control card host the local processor and memory. The processor is a Motorola MPC860P which runs the VxWorks real time operating system\footnote{Wind River, http://www.windriver.com} and hosts the software processes~\cite{bib:antares-daq-paper}. These processes are used to handle the data from the ARS chips and from the slow control, respectively. The processor has a fast Ethernet controller (100~Mb~s$^{-1}$) that is optically connected to an Ethernet switch in the MLCM of the corresponding sector. 
\begin{figure}[!h]
	\centering
	\includegraphics[width=7.0cm]{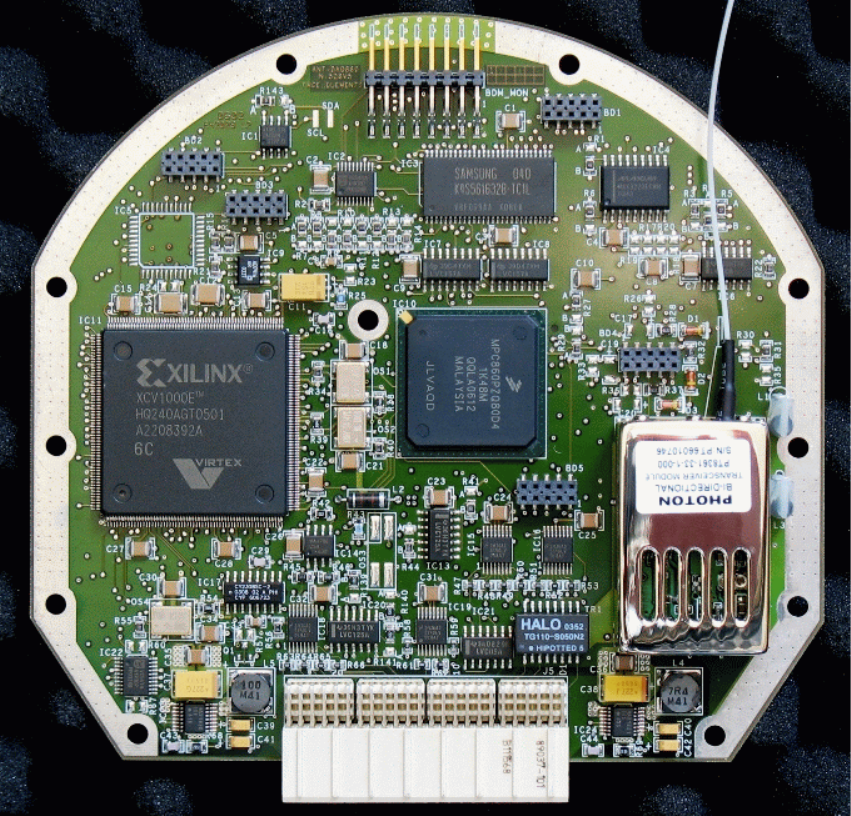}
	\caption{The DAQ/SC board holding the processor (centre), the FPGA (left) and the optical link to the MLCM (right).}
	\label{fig:DAQ_board}
\end{figure}
Three serial ports, two with RS485 links and one with RS232 links, using the MODBUS protocol\footnote{http://www.modbus.org} are used to handle the slow control signals.
The specific hardware for the readout of the ARS chips and data formating is implemented in a high density field programmable gate array\footnote{Virtex-EXCV1000E, http://www.xilinx.com}. The data are temporarily stored in a high capacity memory (64 MB SDRAM) allowing a de-randomisation of the data flow.
\vspace*{-0.20\baselineskip}
	\item COMPASS$\_$MB (Figure~\ref{fig:COMPASS_MB}). The compass motherboard hosts a TCM\footnote{PNI Sensor Corp., http://www.pnicorp.com} sensor which provides heading, pitch and roll of the LCM (i.e.~of the OMF) used for the reconstruction of the line shape and PMT positions. The heading is measured with an accuracy of $1^{\circ}$ over the full cycle and the tilts with an accuracy of $0.2^{\circ}$ over a range of $\pm20^{\circ}$. The same card supports two micro-controllers dedicated to the slow control: they control the measurements of various temperatures and the humidity, and set and monitor the PMT high voltages.
\begin{figure}[!h]
	\centering
	\includegraphics[width=7.0cm]{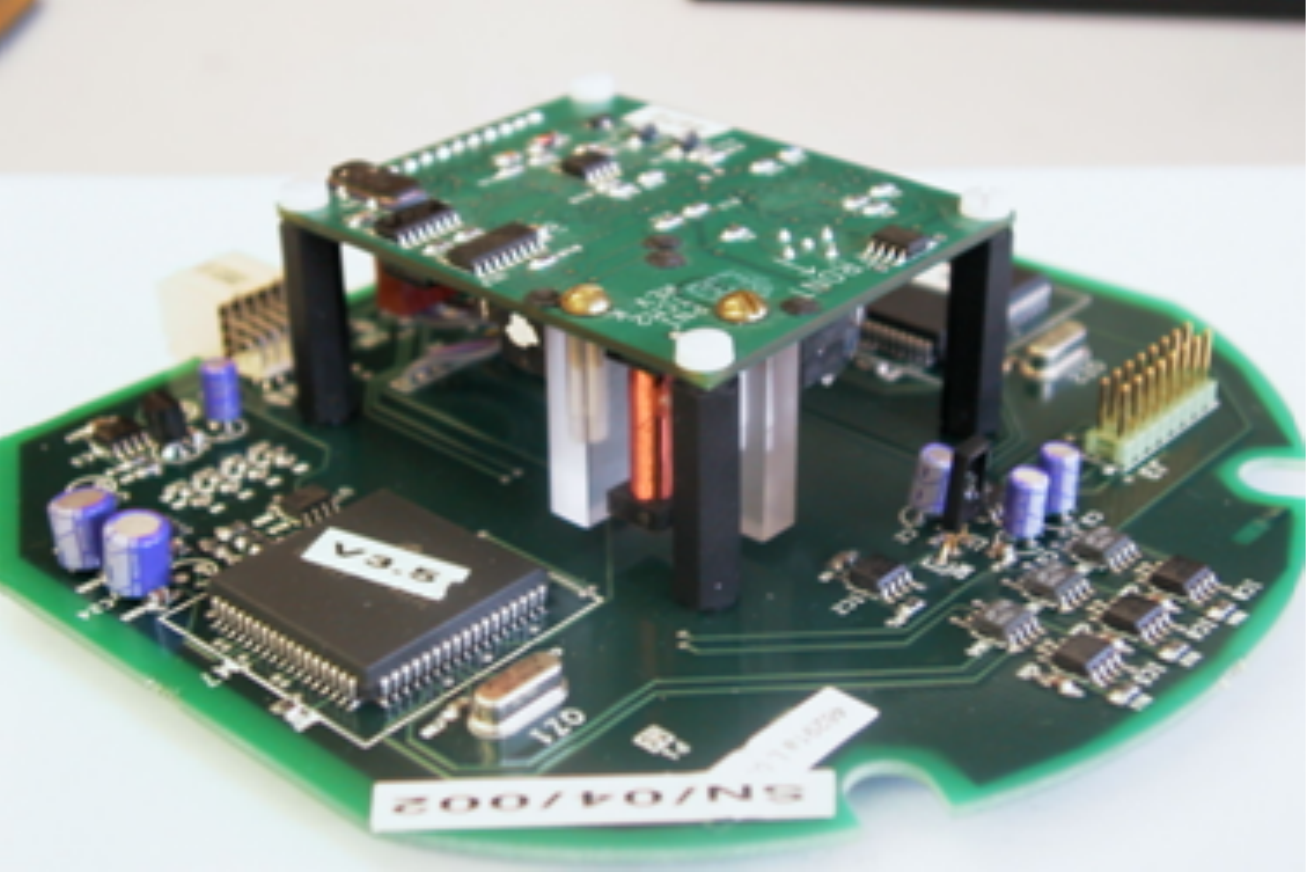}
	\caption{The COMPASS$\_$MB equipped with a TCM2 sensor on a raised daughter card. }
	\label{fig:COMPASS_MB}
\end{figure}
\end{itemize}

For LCMs performing acoustic functions (cf. Section~\ref{subsec:positioningdevices}), there are three additional cards: one housing a pre-amplifier, one a CPU and the third a digital signal processor. These cards are commercial products from ECA\footnote{ECA S.A., http://www.eca.fr}, re-shaped to fit in the crate.
\\\indent
An MLCM holds the following additional cards:
\begin{itemize}
	\item BIDICON. It communicates via bi-directional optical fibres with the four other LCMs of the sector, and performs the electrical$\leftrightarrow$optical conversion of signals transmitted via the backplane to or from the SWITCH card.
\vspace*{-0.20\baselineskip}
  \item SWITCH. An Ethernet switch which consists of a combination of eight 100~Mb~s$^{-1}$ ports and two 1~Gb~s$^{-1}$ ports\footnote{Allayer AL121 and AL1022 respectively.}. One of the 100~Mb~s$^{-1}$ ports is connected to the processor of the MLCM and four to the BIDICON card via the backplane. One of the two Gb~s$^{-1}$ ports is connected to a Dense Wavelength (De)-Multiplexer (DWDM) transceiver. 
\vspace*{-0.20\baselineskip}
  \item DWDM (Figure~\ref{fig:DWDM}). The role of the transceiver is to perform the electrical$\leftrightarrow$optical conversion for the full sector and to communicate with the shore via the SCM located at the bottom of the line. It is electrically connected to the SWITCH card via coaxial cables and optically to the SCM via two uni-directional optical fibres (Rx and Tx) at a connection speed of 1~Gb~s$^{-1}$. For each MLCM (i.e.\ sector) of a line, the laser mounted on the card has a specific frequency chosen in the range from 192.1 to~194.9~THz, the frequency spacing being 400 GHz.
\begin{figure}[!h]
	\centering
		\includegraphics[width=7.0cm]{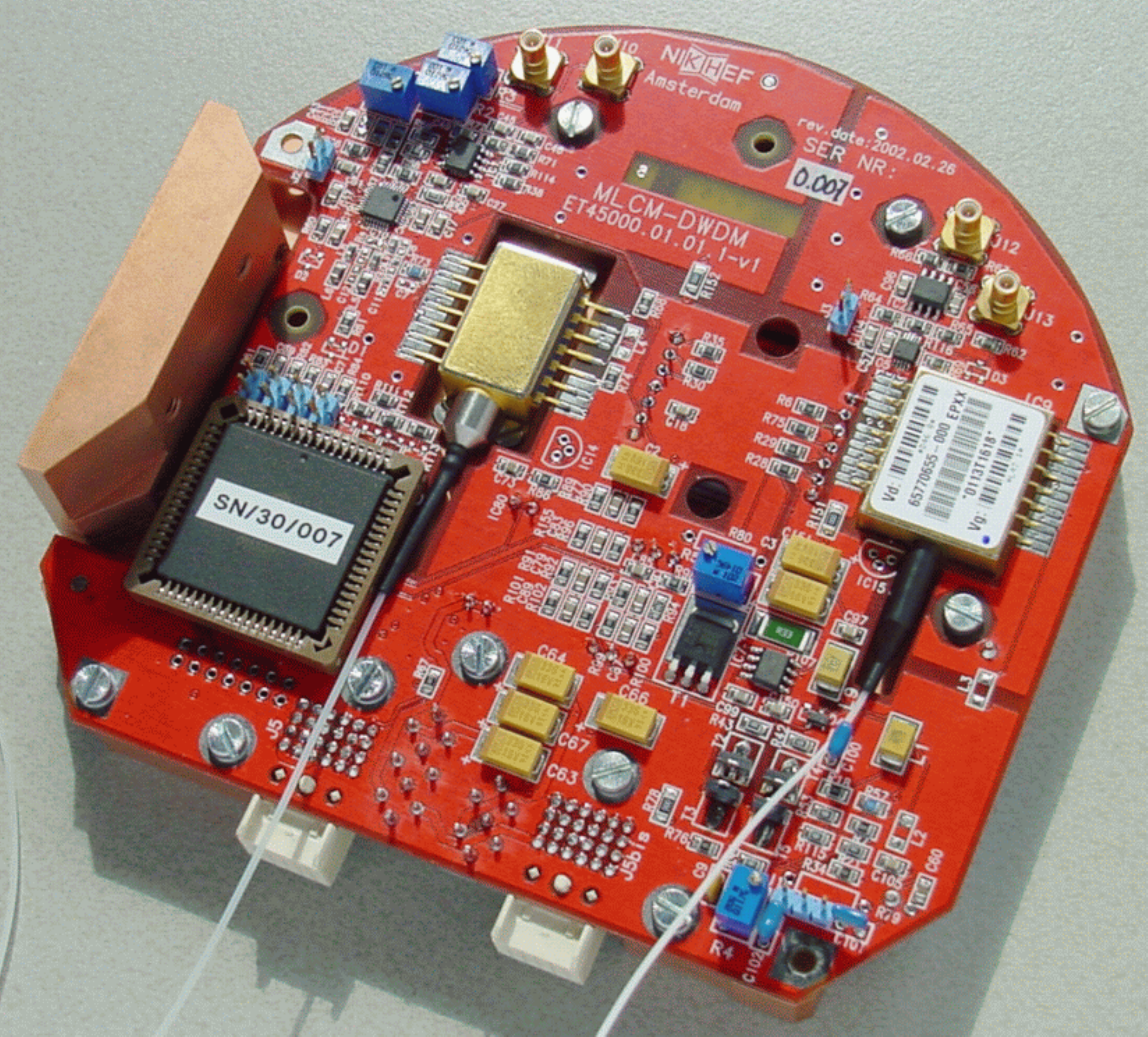}
	\caption{The DWDM board.}
	\label{fig:DWDM}
\end{figure}
\end{itemize}
\begin{figure*}[bht]
	\centering
		\includegraphics[width=.90\textwidth]{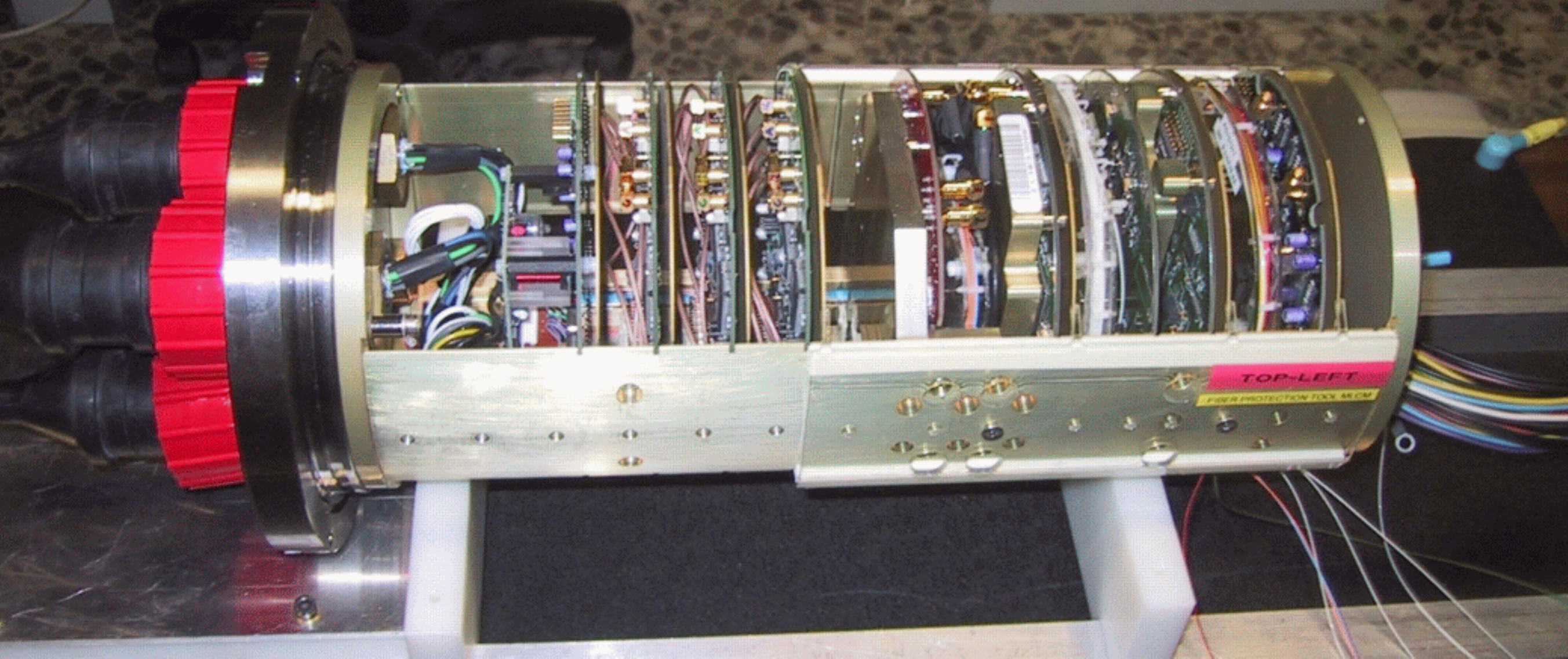}
	\caption{The crate of an MLCM equipped with the electronics boards.}
	\label{fig:MLCM_populated}
\end{figure*}

Figure~\ref{fig:MLCM_populated} shows an MLCM crate equipped with the full set of the electronics cards. A description of the components in the SPM/SCM container will be given later in the BSS Section~\ref{The recoverable part}.

\subsection{Electro-optical mechanical cable (EMC)}
\label{subsec:electroopticalmechanicalcable}

The EMC cable has three roles:
\begin{itemize}
	\item optical data link: 21 single mode optical fibres ($\oslash$=9/125/250 $\upmu$m) run along the cable;
\vspace*{-0.40\baselineskip}
	\item power distribution: 9 electrical conductors (Cu section~=~1~mm$^{2}$ with insulation $\oslash$~=~2.5~mm);
\vspace*{-0.40\baselineskip}
	\item mechanical link: breaking tension above 177 kN.
\end{itemize}

To facilitate the line handling and deployment with its cumulative length of $\approx$~480~m, the minimal allowed radius of curvature of the cable was specified to be less than 300~mm (180~mm for the naked core).

\begin{figure}[!h]
	\centering
		\includegraphics[width=7.5cm]{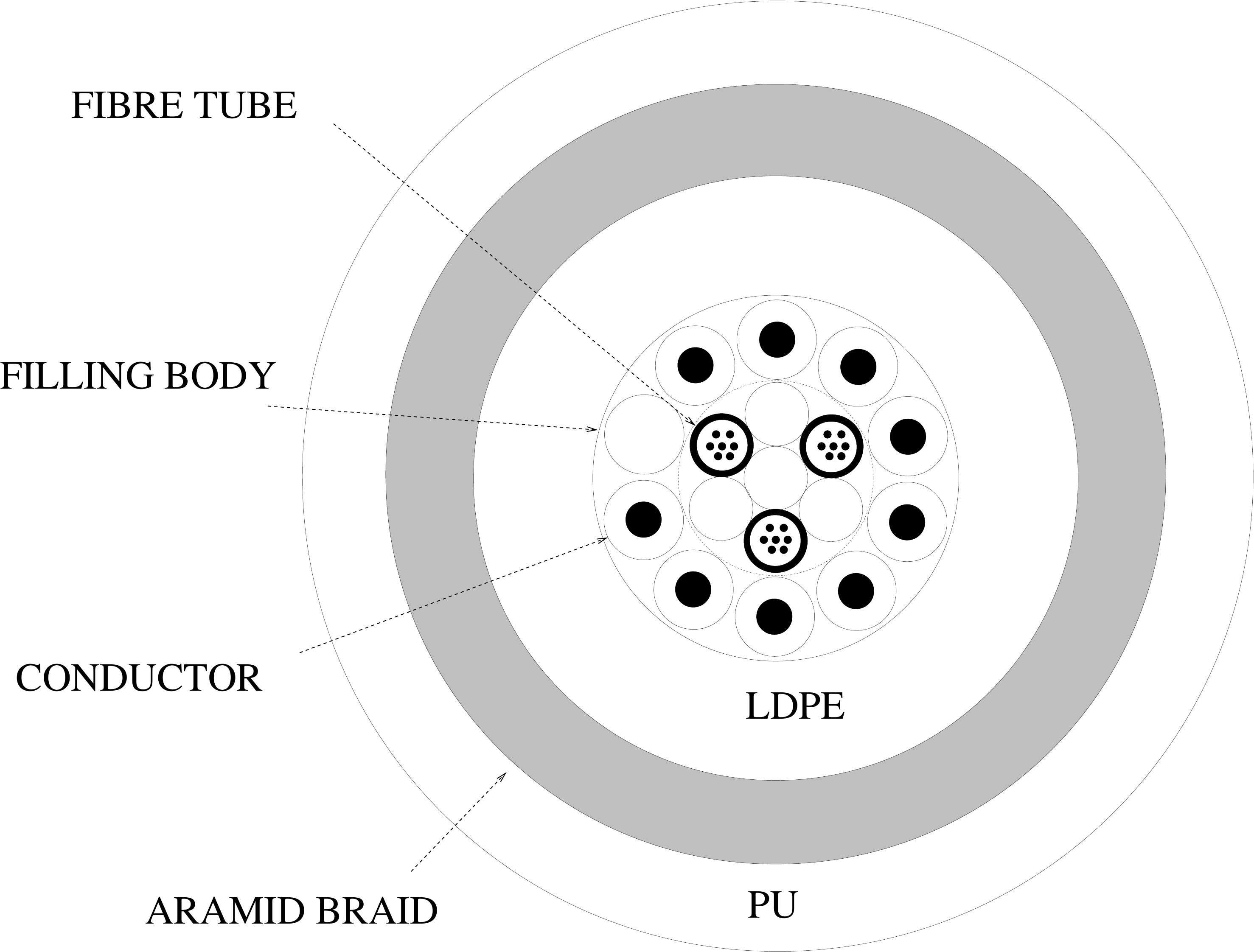}
	\caption{Cross section of the EMC. From centre to outside one can distinguish the layer with 3 tubes, each housing 7 optical fibres, the layer with 9 copper conductors, the LDPE jacket, the aramid braid and the polyurethane sheath. The external diameter is 30~mm.}
	\label{fig:EMCcrossSection}
\end{figure}

\begin{figure}[h!]
	\centering
		\includegraphics[width=6.5cm]{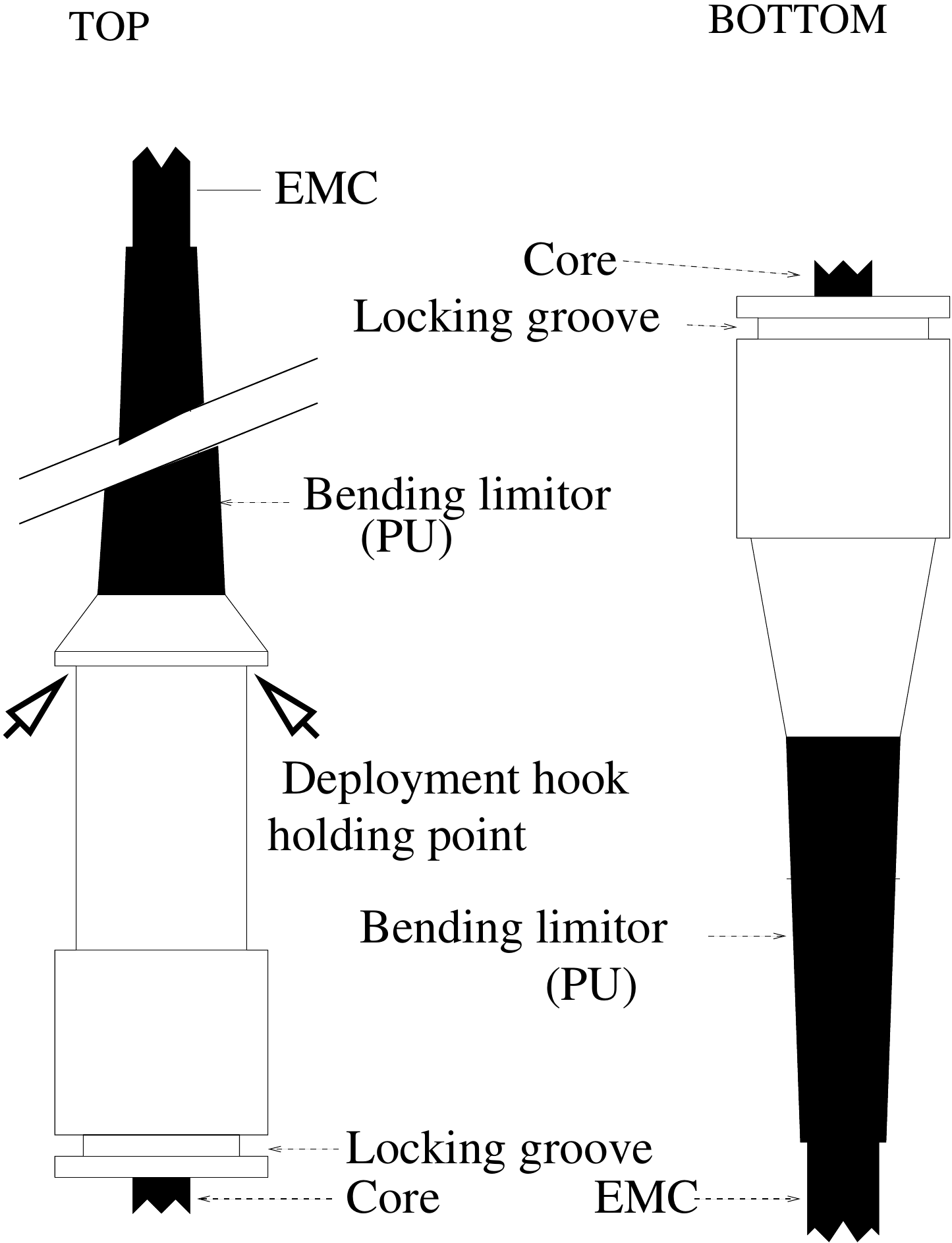}
	\caption{Mechanical termination of an EMC.}
	\label{fig:mechanicalTermination}
\end{figure}

The cable, developed under the responsibility of EurOc\'eanique$^{\ref{fn:euroc}}$ is assembled in successive layers as shown in Figure~\ref{fig:EMCcrossSection}.
The two internal layers are assembled with silicon compound filling the space between the elements. Water can penetrate through the 2 external layers, while the inner polyethylene jacket acts as a water barrier. The polyurethane (PU) sheath is in contact with the water. Its role is to protect the aramid braid and the cable before and during deployment. The two outer layers end inside the mechanical termination where the aramid braid is firmly held by a cone locking-system and the rest of the cable, called the ``core'', continues for a few meters to the LCM penetrators. Each section sustained a static test tension of 50~kN. During this test, the insulation of electrical conductors and the attenuation on the optical fibres are controlled. The cable length between the mechanical terminations is 98~m for the bottom cable section and 12.5~m for the 25 other sections of a line (including the passive section linking the top storey to the buoy), resulting in a pitch between optical modules of 14.5 m. The actual length of each section delivered was measured under a tension of one ton, with an accuracy of $\pm$5~mm, and the results were recorded in a database as input to the line shape reconstruction. Figure~\ref{fig:mechanicalTermination} gives a schematic view of the top and bottom mechanical terminations and their PU bending limitors. 

\begin{figure}[!h]
	\centering
		\includegraphics[width=7.5cm]{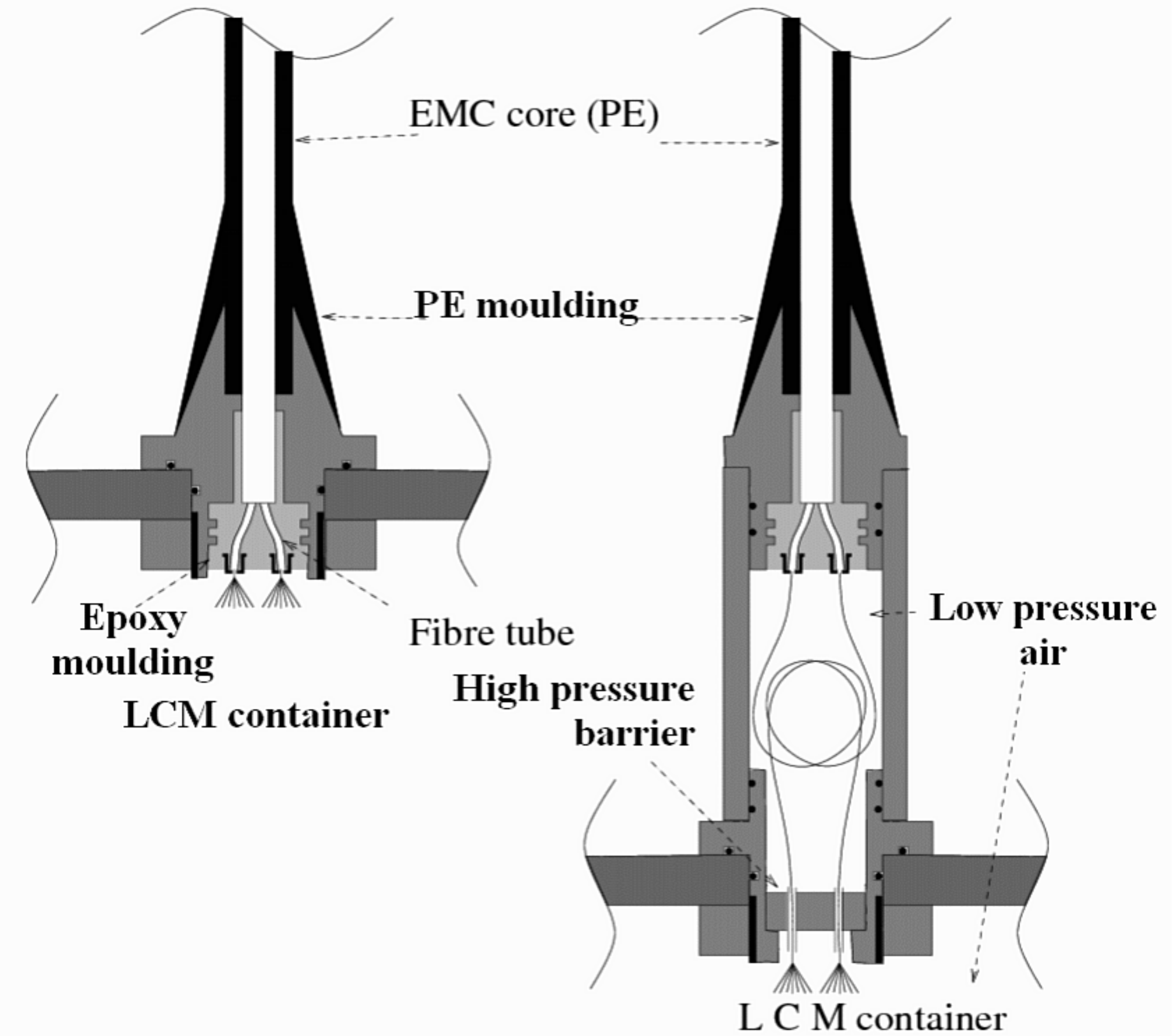}
	\caption{EMC penetrators of the LCM container. Left: non water blocking. Right: water blocking. For clarity, the 3rd fibre tube and the 9 conductors are not shown.}
	\label{fig:penetrator}
\end{figure}

Two different types of LCM penetrators are mounted at the ends of the core: a pair of water-blocking (WB) penetrators for the sections located between sectors and a less expensive pair of non-water-blocking (NWB) penetrators elsewhere. In case of a flooded cable, the WB type stops the propagation of the water along the cable and thus limits the flooded part of the line to one sector (the WB penetrator only stops water propagation from the cable to the container and not in the opposite direction). Figure~\ref{fig:penetrator} gives a schematic view of the NWB penetrator (left) and of the WB penetrator (right). In both cases, the fibre tubes are mechanically blocked in an epoxy moulding, itself blocked in the penetrator body to avoid extrusion when the cable is subject to water pressure.

When subjectted to a uniform horizontal sea current, as present at the ANTARES site, the 3-fold periodic symmetry of the storey induces a torque which is a function of the actual azimuth of the storey. The storey is in stable equilibrium when one of the three OMs is upstream of the current. From measurements performed in a pool, the torque was found to be proportional to the square of the current with a proportionality constant of 9.47~N~s$^{2}$~m$^{-1}$.

Between two adjacent storeys, the EMC acts as a torsion spring tending to keep them at the same relative angle. This torque was measured as a function of the cable tension on a prototype and found to be proportional to the cable torsion angle per unit length and to the tension with a proportionality constant of 1.3~$\times$~10$^{-3}~$m$^{2}~$rad$^{-1}$. In order to specify the minimum torsion strength of the cable, the torsion behaviour of the line was simulated using the above data and for very unfavorable environmental conditions: uniform sea current at 30~cm~s$^{-1}$ slowly increasing in the azimuth angle for several turns. The resulting specification was a maximum torsion angle change per unit length of $\pm~$45$^{\circ}~$m$^{-1}$.

\subsection{Bottom string structure}
\begin{figure}[!h]
	\centering
		\includegraphics[width=7.0cm]{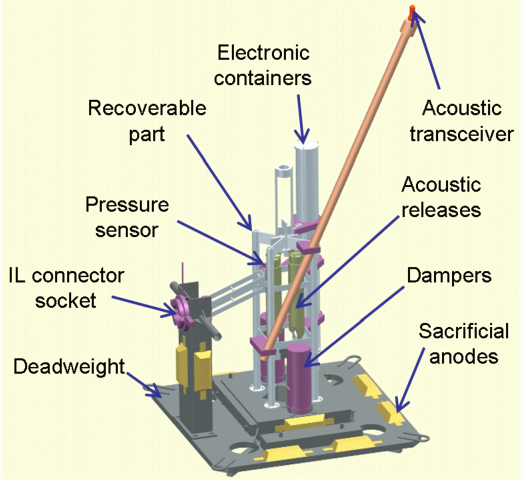}
	\caption{The ANTARES Bottom String Stucture.}
	\label{fig:BSS2}
\end{figure}

The function of the BSS (Figure~\ref{fig:BSS2}) is to anchor the line to the seabed with the capability of a recovery. The BSS is made of two parts: an unrecoverable dead weight laid on the seabed and a recoverable part sitting on top. The two parts are connected by a release system remotely controlled by acoustic signals.

\subsubsection {Dead weight}

The dead weight is a horizontal square plate made of 50 mm thick carbon steel. The line stability requires a dead weight of 1270~kg in water, which means 1.5~tons of steel. Therefore, the dimension of the square side is 1.8~m, resulting in a ground pressure of 4~kPa, while the seabed is believed to sustain safely a pressure up to 5~kPa. Four steel ``wings'' are welded below the square plate to improve the anchoring in the seabed sediment. The total wet surface of steel is 9.5~m$^{2}$. To avoid the galvanic corrosion of this large surface in contact with the sea water, 9 plates of Al-Zn-In alloy, the so-called sacrificial anodes\footnote{BAC Corrosion Control A/S, http://www.bacbera.dk}, are welded on the steel surface with a total mass of 68~kg.

The line is linked to the junction box by the interlink cable (IL), an electro-optical cable laying on the seabed and connected to the line at the level of the BSS by an underwater mateable connector:, developed by ODI company\footnote{\label{fn:odi}Ocean Design Inc. (ODI), http://www.odi.com}.
\\\indent
The remote line release implies an automatic disconnection system for the IL cable: the plug of the IL is fixed to the dead weight while the socket is located on the recoverable part of the BSS, at the end of a pivoting arm in such a way that it is extracted from the plug at the beginning of the ascent. In order to slow down the speed of extraction ($\leq$~5~cm~s$^{-1}$ as recommended by the manufacturer) and to guide the disconnection phase of the ascent, two vertical damping systems are mounted between the two parts of the BSS: a pair of pistons on the dead weight matching a pair of cylinders 
on the recoverable part. These parts are in LDPE and/or PETP, the piston has a diameter of 150~mm and a used height of 670~mm. The damping effect was adjusted in pool tests. Finally, the piston/cylinder gap was set to 0.8~mm and a set of grooves was machined along the pistons to avoid suction effect in water and water inlets were drilled through the cylinder.

\subsubsection {Release system}

The BSS holds two lithium battery powered transponders\footnote{Type RT 861 B2T; IXSEA/Oceano, http://www.ixsea.com} in Ti cylinders which are equipped with release mechanism. The releases are 
mounted on the recoverable part of the BSS in a redundant system which includes a chain, made of Ti and steel, engaged inside a steel part belonging to the dead weight. The chain is pre-tensioned to avoid a gap between the two components of the BSS. The acoustic beacon capability of the transponders is employed in the Low Frequency Long Baseline (LFLBL) positioning and navigation system to monitor the position of the line anchor during its deployment and to determine its geodetic location with a precision of $\approx$~1~m.

\subsubsection {Recoverable part}
\label {The recoverable part}

In addition to the already mentioned transponders, the recoverable part of the BSS holds various equipment:

\begin{itemize}
	\item a 1.8~m long Ti container housing the power module and the control electronics (SPM/SCM) of the line;
\vspace*{-0.40\baselineskip}
	\item a high resolution pressure meter, used for line positioning;
\vspace*{-0.40\baselineskip}
	\item an acoustic transceiver at the top of a 3.6~m long rod of glass-epoxy, used as a reference emitter for the High Frequency Long Base Line (HFLBL) positioning system;
\vspace*{-0.40\baselineskip}
	\item optional sound velocimeter, laser beacon, seismometer depending on the line ~\cite{bib:MILOM};
\vspace*{-0.40\baselineskip}
	\item a weight to keep the line vertical and under enough tension after release, even when the buoy reaches the sea surface.
\end{itemize}

The recoverable part of the BSS is a welded Ti (grade 2) structure sitting on top of a square steel weight (950 mm side length and 160 mm thickness) of 1140~kg total mass (9.6~kN weight in water) and with 2.5~m$^{2}$ wet surface of steel. Depending on the actual equipment of the line, this mass is adjustable by a set of steel plates welded on the weight. The steel parts are anode protected in the same way as the dead weight, with three anodes with a total mass of 30~kg. After two years of immersion of prototypes, the anode consumption rate was measured to be $\approx$~300~g per year and per square meter of wet steel. This value can be extrapolated to an approximate lifetime of the anodes of 40~years.
\\\indent
The SPM/SCM electronics container is the assembly of two cylinders similar to those of the LCMs and is fixed vertically on the structure (Figure~\ref{fig:BSS2}). The SPM part, at the top, contains transformers/rectifiers which deliver the five 400~V~DC supplies needed for the sectors, starting from the 480~V~AC provided by the junction box. An embedded micro-controller operates the remote powering of sectors. Voltages, temperatures and current consumptions are monitored. The micro-controller can also detect anomalies and is programmed to turn off the power in case of over-consumption.
Like a standard LCM container, the SCM cylinder houses a crate equipped with COMPASS, CLOCK and DAQ cards. In order to perform the distribution of the clock signal to the sectors, the SCM crate houses specific boards called:
\begin{itemize}
	\item SCM$\_$WDM (Wavelength Division Multiplexing) (Figure~\ref{fig:WDM}) which receives the 20~MHz clock signal via an optical link from the shore and converts it to an electrical signal which is distributed on the backplane. For redundancy, the clock is transmitted on two fibres from the shore and in case of failure of one fibre, the WDM card automatically switches to the other.
\begin{figure}[!h]
	\centering
		\includegraphics[width=7.cm]{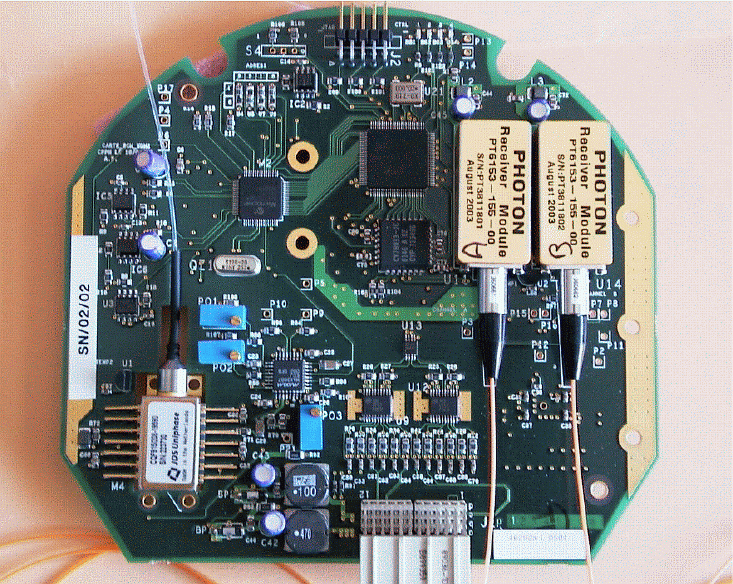}
	\caption{The SCM$\_$WDM board.}
	\label{fig:WDM}
\end{figure}
\vspace*{-0.20\baselineskip}
	\item REP: its role is to operate the reverse conversion. It is equipped with three fibre outputs and two such cards are needed to distribute the clock signals to the five sectors.
\end{itemize}

To communicate with the shore, the SCM is equipped with a DWDM board similar to the MLCM one but working at 100 Mb~s$^{-1}$ and using its own DWDM channel. The pair of fibres of this DWDM is connected, as well as the 5 pairs of fibres coming from the sectors to 6 channels of a 1-to-8 passive optical mux/demux\footnote{Multi-Channel Mux/Demux Module 400~GHz spacing; JDS Uniphase Corp., http://www.jdsu.com} performing the merging/separation of the 6 colours. The BSS being equipped with an acoustic transceiver, the appropriate cards are present in the SCM crate.

\subsection {Top buoy}
The top buoy\footnote{TRELLEBORG CRP, http://www.trelleborg.com/en/offshore} is made of syntactic foam qualified for a depth of 3000~m and with a density around 0.5~g~cm$^{-3}$.
The foam is moulded into an oval shape (horizontal diameter~=~1347~mm, height~=~1530 mm) with a hole along the vertical axis where a Ti rod is inserted. The buoy is held on the rod by a pair of Ti disks. The last EMC section is fixed on the bottom end of this rod and, for the deployment, a releasable transponder is fixed to the top end. The mass of the equipped buoy is 782~kg and its buoyant force 6.7~kN.

\subsection{Mechanical behaviour of a line}
\label{subsec:Mechanical behaviour of a line}
Three rules govern the stability of a line.
\begin{enumerate}
	\item The line must remain firmly anchored on the seabed and must be held close to vertical even in the presence of the strongest sea current considered (30 cm~s$^{-1}$). In all situations, the horizontal displacement of any part of the line must be smaller than the horizontal line spacing (60 m) to avoid any possible contact between two lines. The fact that, in a uniform sea current, two adjacent lines will lean in the same direction gives an extra safety factor.
  \item The tension in the release chain while the line is on the seabed, must be above 4 kN in order to overcome any possible blocking of the release systems and in order to reach the surface in one hour or less.
\vspace*{-0.40\baselineskip}
  \item During the recovery, while the buoy is floating on the surface, the EMC tension must be above 2 kN everywhere along the line to allow a safe operation of an automatic hook system which must slide down along the EMC.
\end{enumerate}

The buoyancy is provided by the top buoy but also by the storeys, since each OM has a buoyant force of 0.22~kN. A sector of five storeys with their cables has a buoyant force of 1.42~kN. The weight is mainly provided by the BSS recoverable part and, until the release, its dead weight. The global recovery force is the small difference between two large quantities: the weight in air of the line (without the dead weight) which amounts to 5.5 tons and the weight of the sea water displaced by the line (6~tons). To limit the uncertainty on this force to 10\%, a measurement of the mass and of the volume of all the line components within 0.5\% is required. Table \ref{tab:tensionAtTheBottomOfTheSpecifiedCable} summarises the resulting tension along the line for three static periods of the life of a line without sea current: 
\begin{itemize}
	\item during the deployment, held by the deep sea cable of the surface boat (maximum stress conditions);
\vspace*{-0.50\baselineskip}
	\item in operation on the seabed (rule 2 above applies);
\vspace*{-0.50\baselineskip}
	\item during the recovery, while the buoy floats at the surface (rule 3 applies).
\end{itemize}

These data are based on a detailed list of measurements and calculations which take the acceleration of gravity of 9.805~m~s$^{-2}$ (computed for the site location), the specific mass of 1033~kg~m$^{-3}$ for deep sea water, 998~kg~m$^{-3}$ for fresh water (in which some components were weighed) and 0.9\% for the volume shrinking of the three OMs.

\begin{table}[!h]
\renewcommand{\arraystretch}{1.2}
	\centering
	\footnotesize
		\begin{tabular}{|l|c|c|c|}
		\hline
		 & Deployment & Seabed & Sea surface\cr
		\hline
		Deployment cable & 7.7 & & \cr
		\hline
		26$^{th}$ EMC section & 14.3 & 6.6 & 2.0\cr
		\hline
		21$^{th}$ EMC section & 15.7 & 8.0 & 3.4\cr
		\hline
		16$^{th}$ EMC section & 17.3 & 9.5 & 4.8\cr
		\hline
		11$^{th}$ EMC section & 18.6 & 10.8 & 6.3\cr
		\hline
		6$^{th}$ EMC section & 20.0 & 12.4 & 7.6\cr
		\hline
		1$^{st}$ EMC section & 21.4 & 13.6 & 8.9\cr
		\hline
		Release chain & 12.5 & 4.7 & \cr
		\hline
		\end{tabular}
	\caption{Tension (kN) at the bottom of the specified cable or chain, for 3 periods in the life of the line and at 8 positions along the line.}
	\label{tab:tensionAtTheBottomOfTheSpecifiedCable}
\end{table}

\begin{figure}[!h]
	\centering
		\includegraphics[width=7.cm]{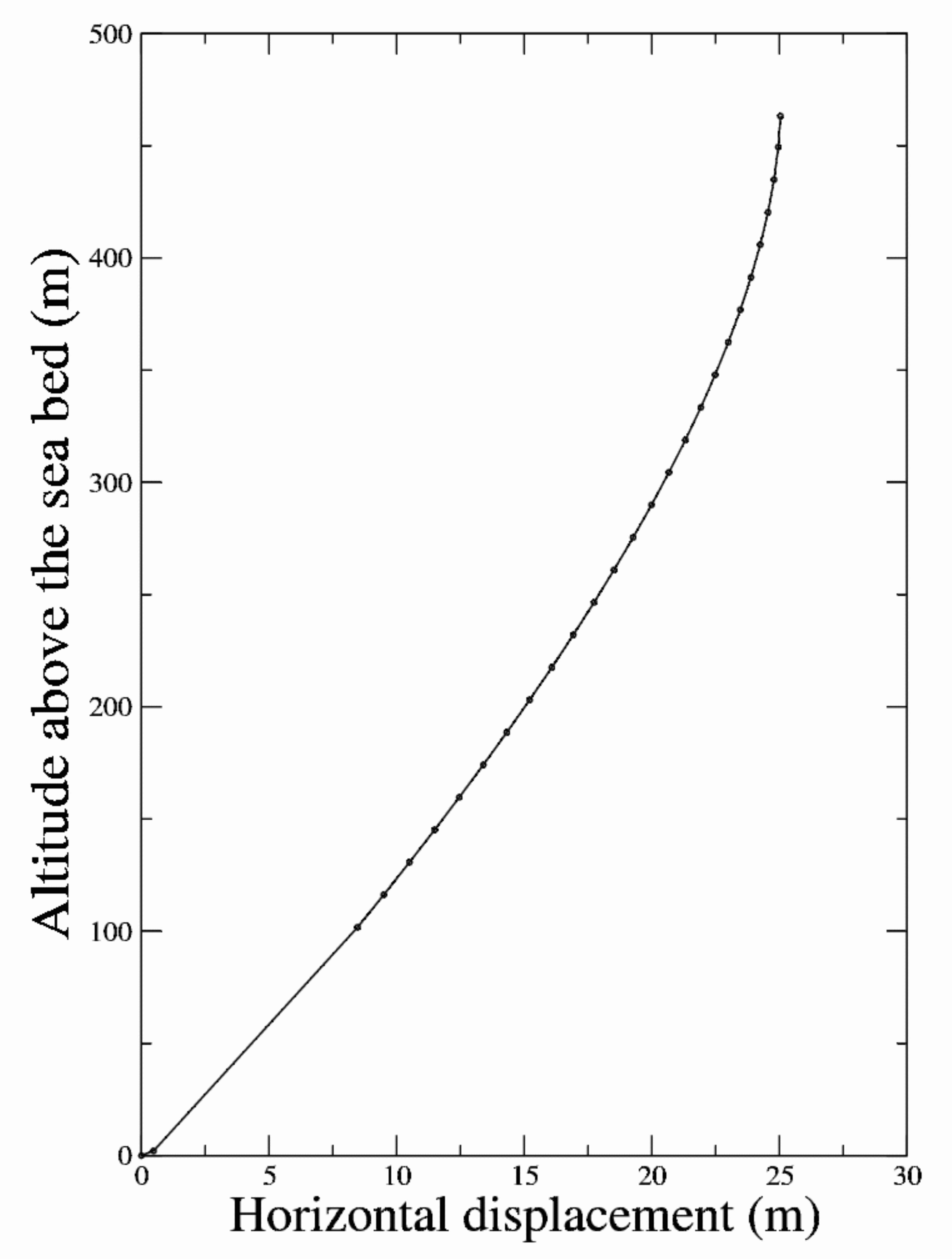}
	\caption{Line shape for a sea current velocity of 25 cm~s$^{-1}$ velocity. The horizontal scale is enhanced to better illustrate the line shape.}
	\label{fig:lineShape}
\end{figure}

Figure \ref{fig:lineShape} shows the results of a calculation based on the buoyant and drag forces on horizontal displacements for a current of 25 cm~s$^{-1}$.
Even in this unfavorable condition, the maximum displacement of the line compared to the vertical is $\lesssim$~25~m. The displacement scales with the square of the current velocity. Such a current increases the cable tension by 0.6\% in the worst case, at the bottom of the line, scaling with the fourth power of the velocity. 

\subsection{Timing calibration devices}
\label{subsec:calibrationdevices}
For the timing calibration of the apparatus, pulsed light sources are used. They are of two types: LED beacons and laser beacons \cite{bib:ledbeacon}. They are distributed in specific locations throughout the detector.

\subsubsection{LED beacons}
\begin{figure}[h!]
	\centering
		\includegraphics[width=6.5cm]{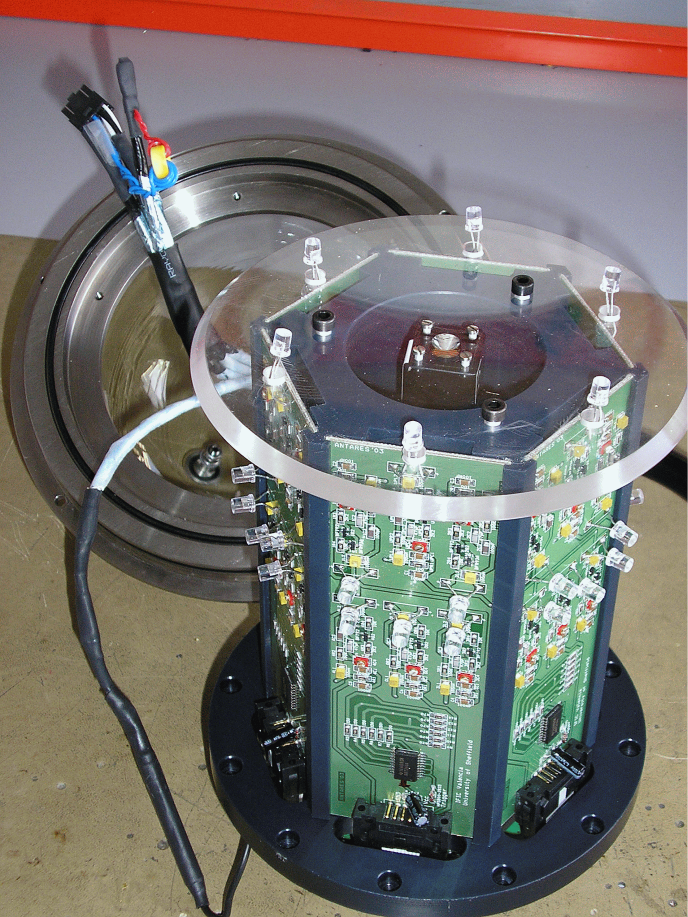}
	\caption{The electronic boards and light sources of an LED beacon.}
	\label{fig:LEDBeacon}
\end{figure}

An LED beacon is a point-like light source which can be triggered remotely. The electronics and individual light sources (LEDs) are enclosed in a glass container (same manufacturer as the OM sphere). This container is a cylinder completed by two hemispheres and joined by titanium flanges. The overall dimensions are 210~mm for the outer diameter and 443~mm for the full length. They are positioned at the top of the OMF (Figure~\ref{fig:OMF}) on storeys 2, 9, 15 and 21 (numbering from the bottom storey) of each line.
The pulsed light source is composed of 36 blue\footnote{HLMP-CB15; Agilent Technologies Inc., http://www.agilent.com} LEDs in groups of six on 6 printed circuit boards (Figure~\ref{fig:LEDBeacon}). These boards are assembled into a hexagon configuration and contain the pulser circuits and components to allow an individual tuning of the timing for each LED. The geometrical arrangement of the LEDs is such that the emitted light is almost isotropic in azimuth. The number of boards as well as the number of LEDs flashing can be varied.
A pencil PMT\footnote{H6780-03; Hamamatsu, http://www.hamamatsu.com} sits in the centre of the hexagon and is exposed to the emitted light in order to provide the precise pulse time.
\subsubsection{Laser beacons}
\begin{figure}[h!]
	\centering
		\includegraphics[width=7.5cm]{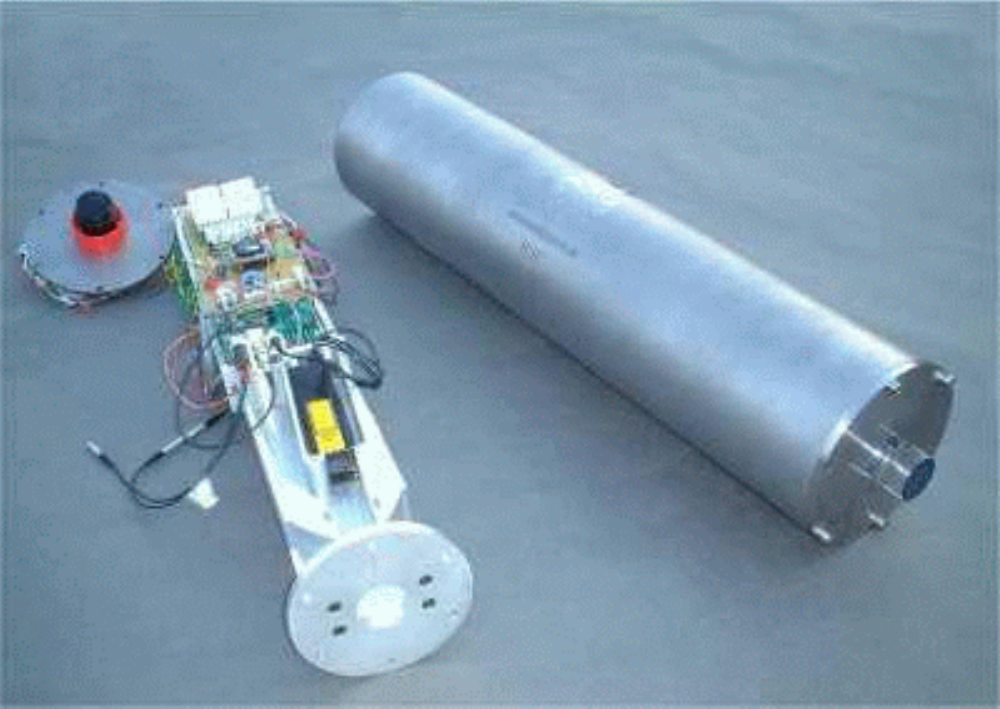}
	\caption{Components of the laser beacon.}
	\label{fig:LaserBeacon}
\end{figure}

Due to their positions, LED beacons are not efficient for the timing calibration of the lowest storeys of the lines and between lines. Hence, they are complemented by light sources sitting on the BSS. However, because of the larger distances, the required light intensity demands the use of a laser.
This laser\footnote{NG-10120-120; Nanolase, presently part of JDS Uniphase Corp., http://www.jdsu.com} is housed in a cylindrical titanium container 705~mm in length and 170~mm in diameter (Figure~\ref{fig:LaserBeacon}). Inside the container, an aluminium inner frame holds the laser and its associated electronics. The laser beam points upwards and leaves the container through a flat disk diffuser coupled by bonding to a quartz cylinder (n = 1.54). This output window configuration is needed in order to minimize transmission losses due to underwater sedimentation and biofouling which affect mainly horizontal surfaces.
The actual time of laser emission is obtained from a fast photodiode integrated into the laser head. Two lines located in a central position in the detector are equipped with laser beacons.

\subsection{Positioning devices}
\label{subsec:positioningdevices}
Each ANTARES detector line is equipped with an acoustic transceiver (RxTx module) fixed on its anchor and receiving hydrophones (Rx modules) fixed on storeys 1, 8, 14, 19 and 25. There are five Rx modules per line, one is placed on the bottom storey and one on the top storey. The others are distributed in order to obtain a larger density of hydrophones in the top third part of the line, where the maximum curvature of the line shape is expected. The RxTx module is composed of a transducer (emitting and receiving hydrophone) placed at the top of a pole on the line anchor and six electronic boards (preamplification, CPU, two DSPs, power, emission) integrated in the SCM. It emits the acoustic signals in emission mode and acts as an Rx module in reception mode. The Rx module is composed of a receiving hydrophone placed on the storey and three electronic boards (preamplification, DSP, CPU) included in the LCM. 
Since the position measurements are based on the travel time of acoustic signals, the knowledge of the sound velocity {\it in situ} is mandatory: sound velocimeters are distributed on some lines. Data from these devices are used to reconstruct by triangulation the positions of the hydrophones. In order to obtain the optical modules positions, the following complementary information is used: 
\begin{itemize}
	\item orientation of the OMFs provided by the COMPASS$\_$MB sitting in each LCM;
\vspace*{-0.50\baselineskip}
	\item a model for the line shape, see Section~\ref{subsec:operation_positioning}.
\end{itemize}

\subsection{Instrumentation line}
\begin{figure}[!h]
	\centering
		\includegraphics[width=7.5cm]{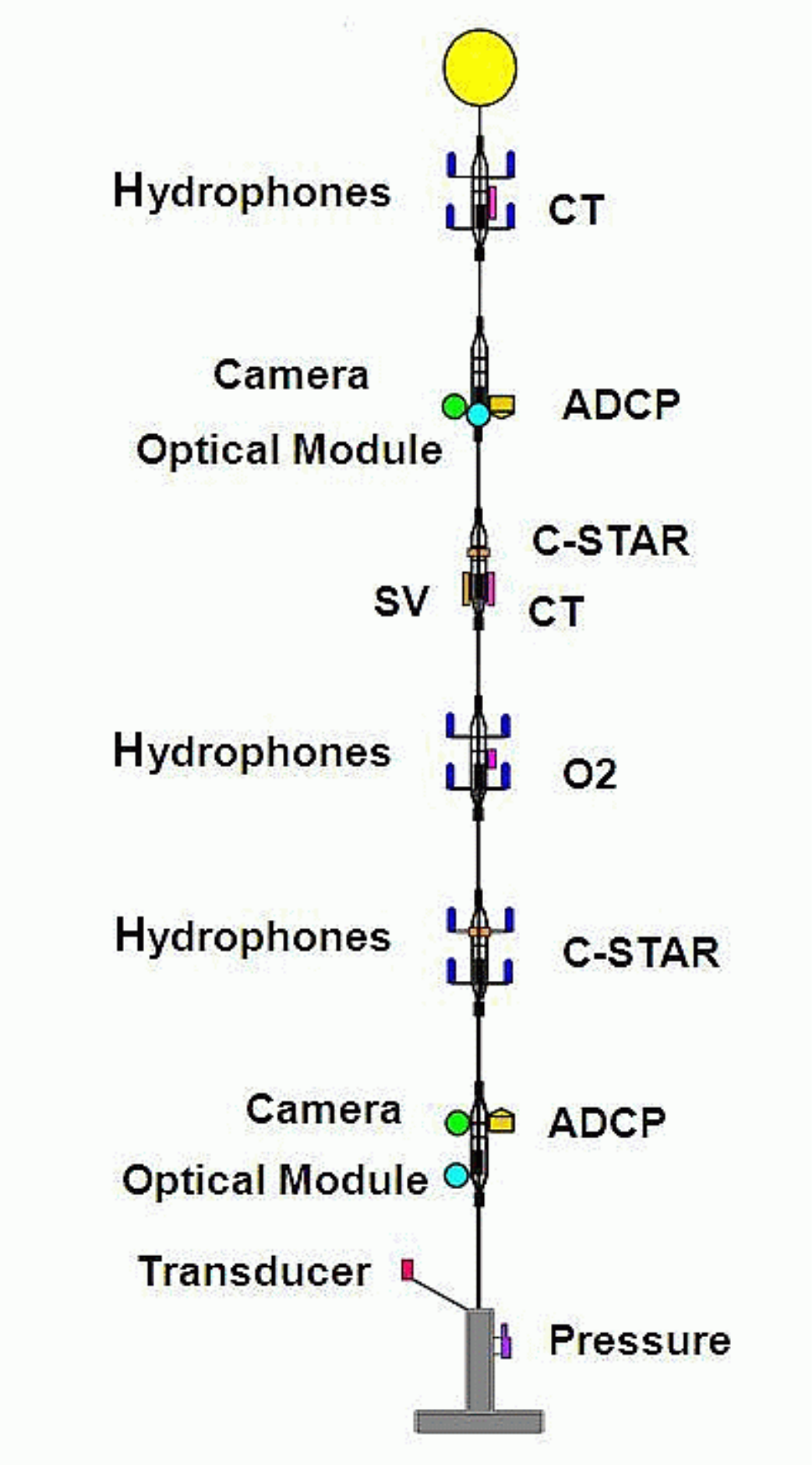}
	\caption{The instrumentation line IL07.  Elements are indicated schematically; not drawn to scale. }
	\label{fig:IL07}
\end{figure}

\indent
The instrumentation line has evolved in time from the ``MILOM'' line \cite{bib:MILOM}, which was operational from March 2005 to June 2007 to the ``IL07'' line which has been operational since December 2007.
\\\indent
This IL07 instrumentation line has six storeys of which three house elements of the acoustic detection system, which will be described in Section~\ref{subsec:AcousticDetectionSystem}. The storeys of the line house various oceanographic instruments: Acoustic Doppler Current Profilers (ADCP) to monitor the intensity and direction of the underwater flow; a sound velocimeter to record the local value of the sound velocity; probes to measure the conductivity and temperature (CT) of the sea water; transmissiometers to monitor the light attenuation of the water (C-STAR); a dissolved oxygen sensor (O2) widely used by physical oceanographers to characterize mixing and ventilation of water masses and two cameras continuously connected in order to record images of bioluminescent organisms. A schematic layout of the instruments on the line is shown in Figure~\ref{fig:IL07} and details of the instruments are given in Table~\ref{tab:IL07_Devices}. 

\begin{table*}[!ht]
\renewcommand{\arraystretch}{1.3}
	\centering
	\scriptsize
		\begin{tabular}{|m{.6cm}|m{.8cm}|m{2.0cm}|m{1.9cm}|m{1.7cm}|m{2.9cm}|}
\hline	
Storey & Height above seabed & Device type & Manufacturer & Model & Measured parameters\cr
\hline
\multirow{2}{*}{6} & \multirow{2}{*}{305~m} & 6 hydrophones & HTI & HTI-90-U & sound level, transients\cr
\cline{3-6}
& & CTD & SEABIRD & SBE37-SMP & conductivity, temperature\cr		
\hline
\multirow{3}{*}{5} & \multirow{3}{*}{290~m} & Optical module & ANTARES & custom & light level\cr
\cline{3-6}
& & ADCP & TeledyneRD & Workhorse & sea current velocity\cr
\cline{3-6}
& & Camera & AXIS & AXIS221 & images\cr
\hline
\multirow{3}{*}{4} & \multirow{3}{*}{210~m} & Transmissometer & WETLabs & C-Star & water transparency\cr
\cline{3-6}
& & SV & GENISEA/ECA & QUUX-3A(A) & sound velocity\cr
\cline{3-6}
& & O$_2$ probe & AANDERAA & Optode 3830 & oxygen level\cr 
\hline\	
\multirow{2}{*}{3} & \multirow{2}{*}{195~m} & 6 hydrophones & Erlangen & custom & sound level, transients\cr
\cline{3-6}
& & CTD & SEABIRD & SBE37-SMP & conductivity, temperature\cr		
\hline
\multirow{2}{*}{2} & \multirow{2}{*}{180~m} & 6 hydrophones & HTI & HTI-90-U & sound level, transients\cr
\cline{3-6}
& & Transmissometer & WETLabs & C-Star & water transparency\cr
\hline
\multirow{3}{*}{1} & \multirow{3}{*}{100~m} & Optical module & ANTARES & custom & light level\cr
\cline{3-6}
& & ADCP & TeledyneRD & Workhorse & sea current velocity\cr
\cline{3-6}
& & Camera & AXIS & AXIS221 & images\cr
\hline
\multirow{2}{*}{BSS} & \multirow{2}{*}{0~m} & pressure sensor & GENISEA/ECA & & pressure\cr
\cline{3-6}
& & Transponder & IXSEA & RT661B2T & acoustic positioning\cr
\hline
			
		\end{tabular}
	\caption{List of the instruments on the line IL07.}
\label{tab:IL07_Devices}
\end{table*}

\subsection{Acoustic detection system AMADEUS}
\label{subsec:AcousticDetectionSystem}
\indent
The acoustic neutrino detection is integrated into ANTARES in the form of Acoustic Storeys (AS) which are modified standard storeys with the PMTs replaced by acoustic sensors with custom-designed electronics for signal processing. AMADEUS consists of six ASs, three of them located on the instrumentation line and three on Line 12. Each AS comprises six acoustic sensors that are arranged at distances of roughly 1~m from each other. The ASs on the instrumentation line IL07 are located at 180~m, 195~m, and 305~m above the seabed, respectively. Line 12 is anchored at a horizontal distance of about 240~m from the IL07, with the ASs positioned at heights of 380~m, 395~m, and 410~m. With this setup, the maximum distance between two ASs is 340~m.

\section{Detector infrastructure}
\label {sec:detectorinfrastucture}
\indent
The infrastructure required to power and control the offshore detector includes the onshore buildings to house the electronics for monitoring and data acquisition, the main electro-optical cable providing the electrical power and the data link between the detector and the shore, the junction box and the interlink cables to distribute the power and the optical fibres to the 13 lines.
\subsection{Interlink cable}
\label{subsec:InterlinkCable}
\begin{figure}[h!]
	\centering
		\includegraphics[width=7.5cm]{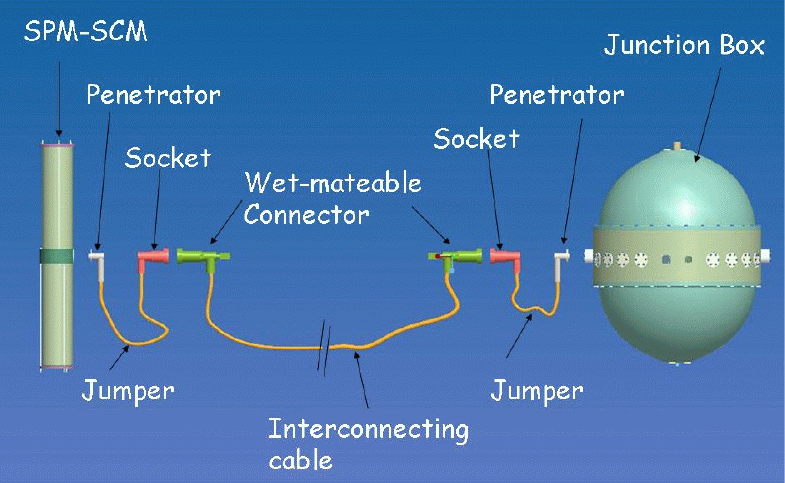}
	\caption{A schematic view of the complete link between the SPM/SCM container of a line and the JB container.}
	\label{fig:linkjbspm}
\end{figure}

The connection between the JB and each line is provided by the interlink cables. These cables, produced by the ODI company$^{\ref{fn:odi}}$, contain four monomode optical fibres and two electrical conductors. A complete link between the JB and a line is composed of three parts: two short cables at each end (jumpers) and the IL itself as shown in Figure \ref{fig:linkjbspm}. Each jumper is terminated at one end by a penetrator equipped with a water blocking system and at the other end by a socket fixed on a strong mechanical structure. Whereas the mounting of the jumpers is performed on shore, the completion of the connection is realised by a Remotely Operated underwater Vehicle (ROV) which lays the IL on the seabed, plugs it on the JB side then on the line side. After each step of the connection operation, both the electrical and optical contacts are checked from shore to be within specifications.

\indent
In order to compensate for failures experienced in some of the 16 outputs of the JB, interlink cables of a special design are used in the seabed infrastructure. Each of these special cables connects two separate lines of the detector with one single JB output. Due to their particular shape, these cables are denoted as ``Y'' links. In this configuration, the cable coming from the JB is split and then linked to the two lines with the same system that is used for the other lines. The two connected lines share the power provided by the JB output. Their DWDM systems are tuned on two different frequency domains. The splitting of the electrical conductors and the splicing of the optical fibres are performed in a titanium container located at the end of the common path, at 10~m distance from the JB. 

\subsection{Junction box}
The JB is a pressure resistant titanium container mounted to the JB Frame (JBF). The JB and JBF provide the following facilities:
\begin{itemize}
	\item connection of the MEOC and of the sea return power electrode;
\vspace*{-0.40\baselineskip}
	\item power transformer housing;
\vspace*{-0.40\baselineskip}
	\item line over-current protection system;
\vspace*{-0.40\baselineskip}
	\item remote diagnostic system;
\vspace*{-0.40\baselineskip}
	\item 16 electro-optical sockets to plug the interlink cables.
\end{itemize}

\subsubsection {Junction box mechanical layout}

\begin{figure}[!h]
	\centering
		\includegraphics[width=7.4cm]{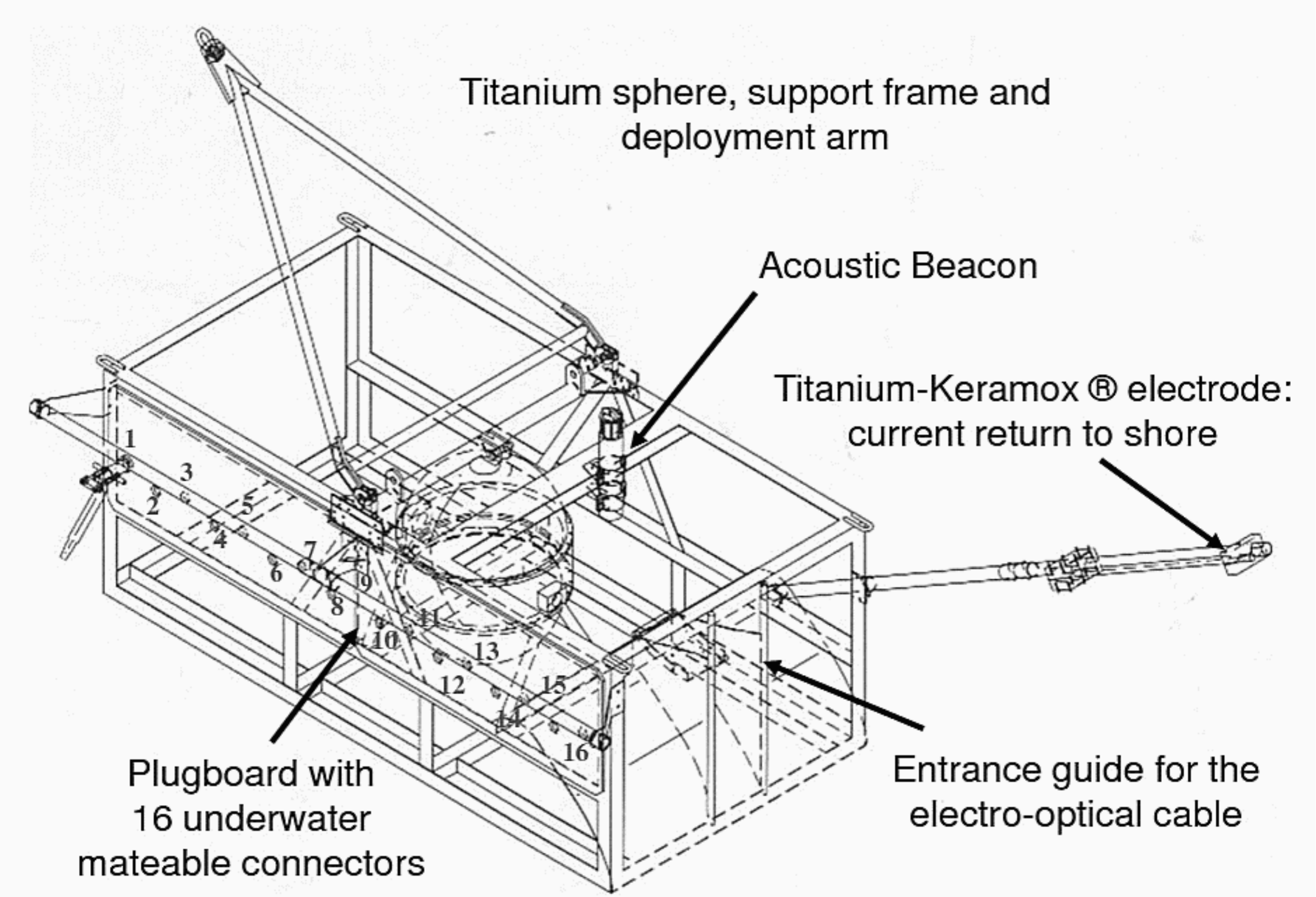}
	\caption{The junction box container and its support frame.}
	\label{fig:JBGeneralView}
\end{figure}

The junction box structure, illustrated in Figure~\ref{fig:JBGeneralView}, is based on a 1~m diameter titanium pressure sphere,

whose hemispheres are separated by a central titanium cylinder (``belt'') through which all power and data connections pass to the exterior. The junction box internal pressure is 1 bar, the external water pressure is $\approx$~250~bars. Each hemisphere is sealed to the belt with two concentric O-rings. The lower hemisphere contains a transformer immersed in oil\footnote{Nynas 10GBN napthalene based transformer oil, meeting ASTM spec D3487 type 1; Nynas AB, http://www.nynas.com}, the upper hemisphere contains the power system slow control electronics. Following component installation, the junction box sphere was qualified in a 24 hour pressure test at 310 bar (20~\% overpressure) in a 2.5~m diameter caisson$^{\ref{fn:comex}}$.

\begin{figure}[!h]
	\centering
		\includegraphics[width=7.5cm]{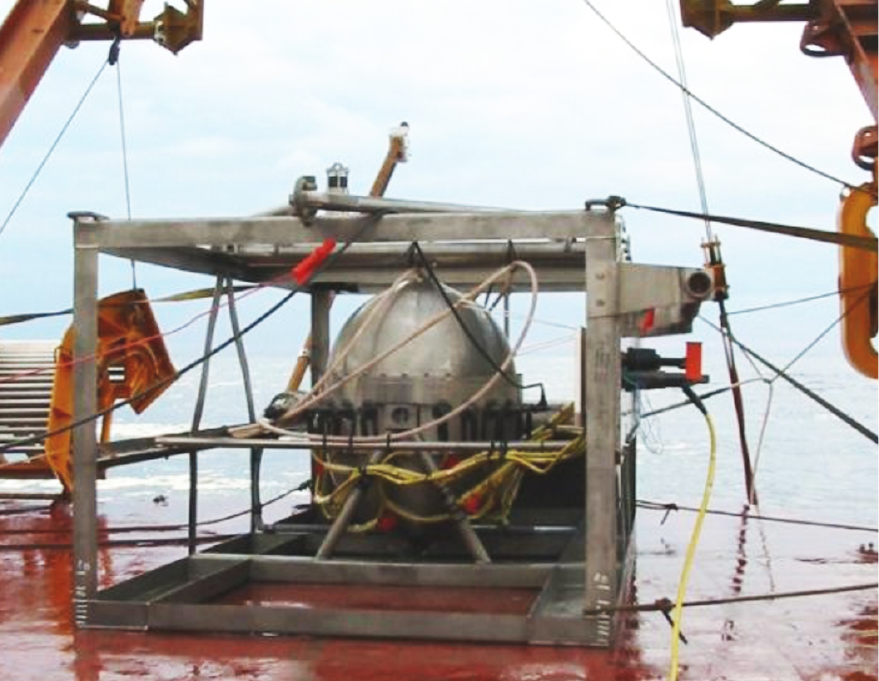}
	\caption{Junction box on the deck of the deployment ship.}
	\label{fig:JBonDeck}
\end{figure}

\indent
 The sphere is supported within a rectangular titanium transit frame. The cage incorporates an acoustic transponder to allow triangulation of the junction box position during deployment, an electrode for the return of the current to shore, and an entry guide to protect the undersea cable from scuffing during the deployment procedure and bending too sharply. In addition, it is equipped with a plug board with 16 deep sea wet mateable electro-optical sockets for the interlink cable connections. Figure~\ref{fig:JBonDeck} shows the junction box on the deck of the deployment ship. The sphere, the cable penetrations
through the belt and the plug board of wet-mateable connectors are visible.

\subsubsection {Junction box cabling}
\begin{figure}[!h]
	\centering
		\includegraphics[width=7.5cm]{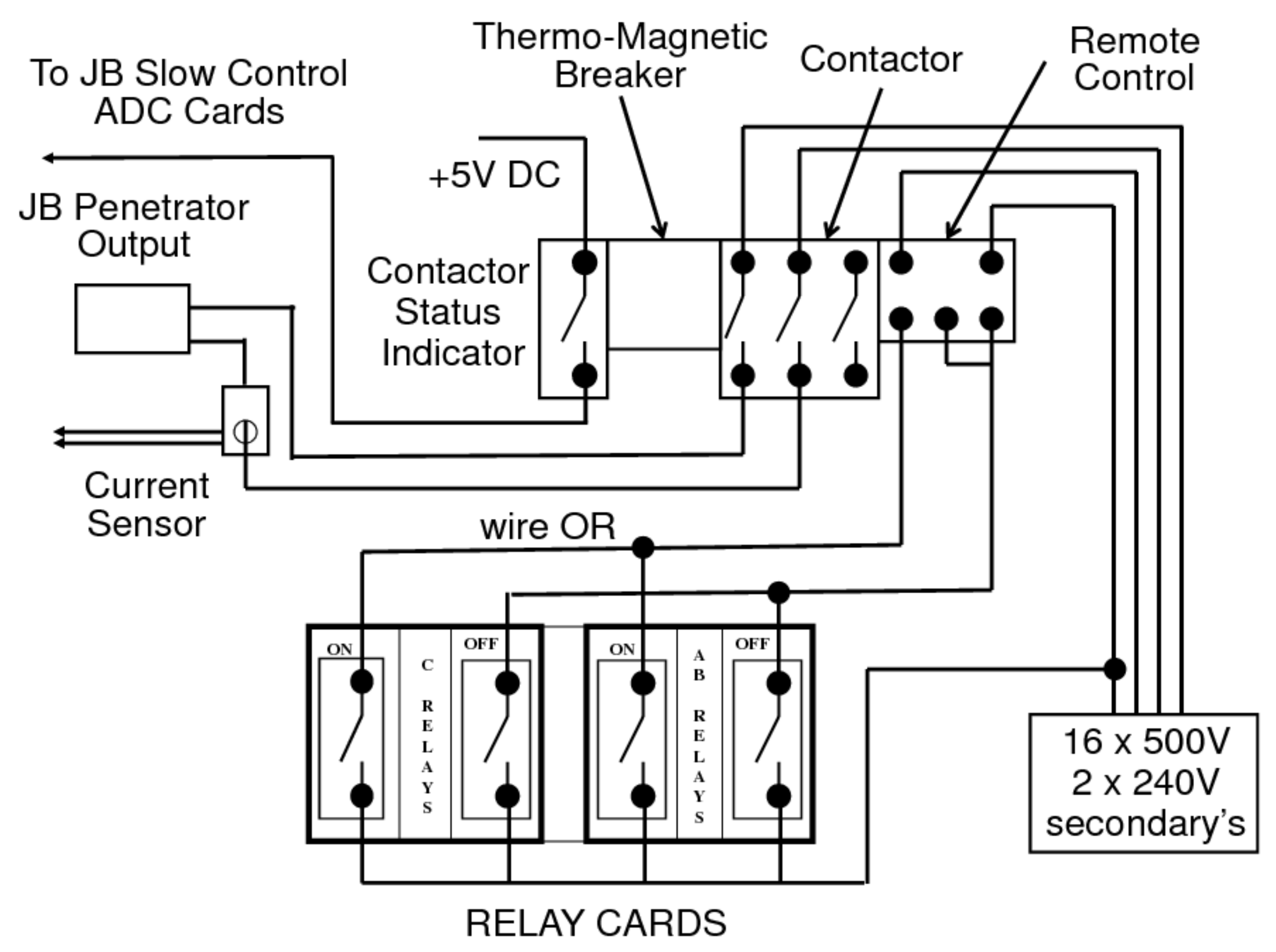}
	\caption{Junction box output breaker management.}
	\label{fig:JBBreaker}
\end{figure}

The junction box is equipped with 16 outputs for connection of the detection and instrumentation lines. The typical power drawn for a detection line is around 1~kW. The junction box outputs are galvanically separated through a transformer with 16 individual secondaries rated at 500~V. Two additional windings rated at 240~V power the junction box internal slow control systems. Each output is protected by a thermo-magnetic breaker\footnote{PKZ2/ZN6 with RE-PZK2 remote control block; Moeller, http://www.moeller.fr}  set to a 5~A threshold. In addition, each breaker can be rearmed or opened by remote control (Figure~\ref{fig:JBBreaker}) should the leakage current monitored by an inductive current sensor\footnote{``MACC plus'' Zero flux current transformer, 10 A full scale, 100~$\mu$A resolution; Hitec BV, http://www.hitecups.com} exceed a safe threshold. Each output has four optical fibres: one pair used for data up- and down-links and the other for duplicated distribution of the central clock pulse train. In addition, each output contains a pair of electrical conductors providing AC power in the range 435-480~V. The conductors and optical fibres of unmated output connectors are protected from sea water exposure by a shutter which opens only during the final phase of cable insertion.
Breakers corresponding to unused outputs are kept in the closed (powered) position, both to minimize the number of output breaker operations and to allow early detection of water infiltration past the shutter of a connector, which would be manifested as an increase in leakage current monitored on the corresponding current sensor. Breaker manipulation is possible with any of three independent control channels through the wired-OR powering of intermediate 240~V relays.

\subsubsection{Junction box slow control electronics}

\begin{figure}[!h]
	\centering
		\includegraphics[width=7.5cm]{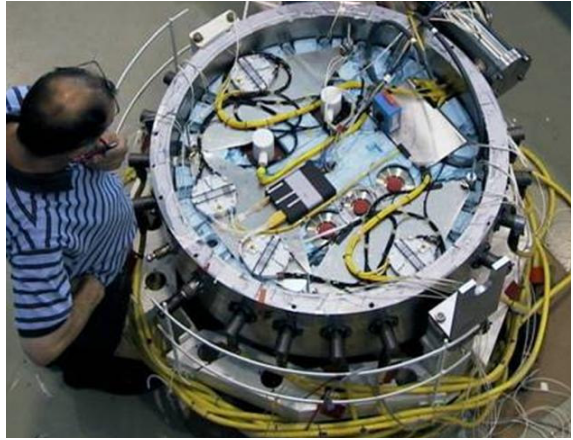}
	\caption{View inside the open JB: heat-spreader disk, transformer connections and primary circuit current sensor, passive fibre optic splitters and cassettes protecting fibre fusion splices.}
	\label{fig:JBHeatDisk}
\end{figure}

Output breaker manipulation and measurement of currents, temperatures and humidity are the main activities of the triply redundant junction box slow control system.
The system uses eight of the 48 fibres in the undersea cable and is based on three control cards built in two different technologies. 
\\\indent
In the first of these, two identical cards communicate with the shore station through 160~Mb~s$^{-1}$ links, using a transmitter/receiver chip set\footnote{HDMP-1022/1024; Agilent Technologies Inc., http://www.agilent.com} with the Photon Techno PT5543-13-3-SC laser emitter and PT6143-155-SC receiver operating at 1550~nm. Associated firmware is embedded in FPGAs\footnote{7256S; Altera Corp., http://www.altera.com}.
Each card can simultaneously stream 16-bit digitized data from 48 internal temperature and humidity sensors, and 24-bit data at 2.6~kHz sampling from a group of 4 ``MACC plus'' inductive current sensors. This latter data is passed onshore to a DSP\footnote{TMS320C5510 200~MHz; Texas Instruments, http://www.ti.com} and used for sinusoid reconstruction and RMS current calculation.
A third card, designed for ultra-low power operation and powered by lithium batteries\footnote{Eight  SAFT LSH20 Lithium elements of 3.6~V, 13~Ah each.}, is based on a microcontroller\footnote{MSP430F149; Texas Instruments, http://www.ti.com} with 60~kbyte flash and 2048~byte RAM memory equipped with eight 12-bit ADC entries and 45 digital I/O ports. This card communicates,even in case of JB power failure, at very low speed (1200 Baud) using an NDL7701 laser uplink operating at 1550$\:$nm and an LPD80 pin diode receiver. This channel has a power consumption of 5$\:\upmu$A in sleep mode and 60$\:\upmu$A when active. When the uplink is transmitting, the maximum power consumption of the card is 100$\:$mA for short periods.
\\\indent
A hermetic stainless steel diaphragm separates the lower transformer compartment from the hemisphere containing the slow control system and fibre optic routing devices. The electronics is mounted on an aluminium heat sink disk (Figure~\ref{fig:JBHeatDisk}) making thermal contact with the titanium belt of the junction box. 
The diaphragm and heat spreader disk sandwich a thermal insulation blanket of silica aerogel\footnote{Spaceloft$^{\circledR}$; Aspen Aerogels Inc., http://www.aerogel.com} which also serves as a getter for water vapour.

\subsubsection{Fibre optic signal distribution in the junction box}

Each junction box output connector contains four optical fibres with the following functions:
\begin{itemize}
	\item DAQ Rx$_n$ (data downlink from shore; n=1$\rightarrow$16);
\vspace*{-0.40\baselineskip}
	\item DAQ Tx$_n$ (data uplink to shore; n=1$\rightarrow$16);
\vspace*{-0.40\baselineskip}
	\item Clock channel A;
\vspace*{-0.40\baselineskip}
	\item	Clock channel B.
\end{itemize}

DAQ Tx and Rx are specific to each line, and are accommodated using 32 fibres in the undersea cable, which are point-to-point spliced in the junction box hub to their respective fibres in the 16 output connectors. 
\\\indent
The central clock signal, vital for time referencing  of photomultiplier data to subnanosecond precision, is transmitted with a 4-fold redundancy. Pulse trains from two independent, identical clock transmitters at the shore station are split for broadcast on four undersea fibres. In the junction box they are routed via dual-input 16-way passive splitters so that clock pulse trains from either or both transmitters are available on every output connector. 
All internal fibre connections are made by fusion splicing, resulting in an optical loss of around 0.01~dB per joint. The laser diode intensity in the shore based clock system is sufficient to maintain an optical power margin of 12 dB over the attenuation in the undersea cable and passive splitters.

\subsection{Main electro-optical cable}
The main electro-optical cable provides the electrical power link and the optical data link between the shore station and the detector. The selected cable, a standard telecommunications type, satisfies the electrical and optical transmission specifications as well as the environmental and mechanical criteria such as temperature tolerance, bending radius and mechanical strength.
\subsubsection{Cable}
\begin{figure}[!h]
	\centering
		\includegraphics[width=7.5cm]{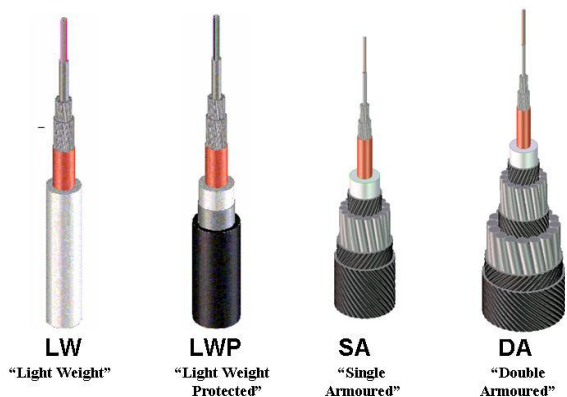}
	\caption{Sections of the undersea cable.}
	\label{fig:MEOCSections}
\end{figure}

Prior to deployment of the MEOC, surveys by a ROV have been carried out to select the best possible offshore site for the apparatus in terms of flatness of the sea bottom and the absence of obstacles. The MEOC  has been deployed from the site to the shore by a specialized cable-laying ship and crew under the responsibility of Alcatel. The cable was tested for optical and power transmission prior to the deployment operation.
Figure~\ref{fig:MEOCSections} shows the structure of the different cable sections used at different depths and Table~\ref{tab:MEOCChar} gives the main characteristics of the undersea cable\footnote{Alcatel URC3 Type 4 (unrepeatered); Alcatel-Lucent, http://www.alcatel-lucent.com}, which contains 48 monomode optical fibres\footnote{Type G24B DE 1302XB (BBO) WB B1, $\oslash$ = 125 $\mu$m.} in a stainless steel tube surrounded by a ``pressure vault'' of two windings of steel armour wires. A tubular copper power conductor surrounds the vault and delivers current up to a maximum of 10~A to the junction box. A standard undersea cable configuration with a single conductor (normally used for series powering of repeaters in long-haul cables) was chosen to minimize the cable cost and weight. The use of the sea for current return reduces ohmic losses by a factor 4 compared with a cable sharing equivalent cross-section between supply and return conductors.

\begin{table}[!h]
\renewcommand{\arraystretch}{1.3}
\centering
	\begin{tabular}{|l|l|}
	\hline
	Electrical resistance & 1~$\Omega~$km$^{-1}$\cr
	\hline
	Fibre attenuation	& 0.18~dB$~$km$^{-1}$\cr
	\hline
	Fibre chromatic dispersion & 21~ps$~$nm$^{-1}~$km$^{-1}$\cr
	\hline
	\end{tabular}
\caption{Characteristics of the undersea cable.}
\label{tab:MEOCChar}
\end{table}

\begin{figure}[!h]
	\centering
		\includegraphics[width=7.5cm]{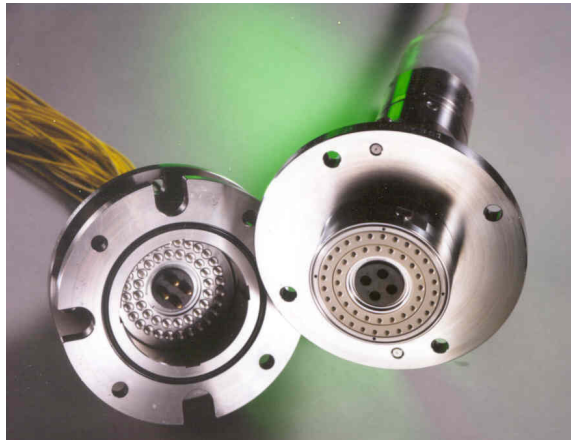}
	\caption{Electro-optical plug and receptacle for connection of the undersea cable to the junction box.}
	\label{fig:MEOCTerm}
\end{figure}

On the deep, smooth seabed the cable exterior terminates in a 21~mm diameter polyethylene sheath (``Light-Weight'' configuration). An additional polypropylene jacket (LWP) protects the cable in the zone of shelving seabed. In shallower water with risk of damage from fishing or boat anchors, the cable has an additional single layer of armour wires (SA) and a coating of tarred polyurethane yarn. The final short section in very shallow water has an additional armour layer (DA). This sequence  with cable sections lengths and water depths is summarized in Table~\ref{tab:MEOCLayout}.
\begin{table}[!h]
\renewcommand{\arraystretch}{1.2}
\centering
	\begin{tabular}{|c|l|c|}
	\hline
	\textbf{LW} & Length & 17.4 km\cr
	'Light-Weight' & Depth & $>$ 2300 m\cr
	\hline
	\textbf{LWP} & Length & 10.2 km\cr
	'LW-Protected' & Depth & 422 to 2300 m\cr
	\hline
	\textbf{SA} & Length & 12.1 km\cr 
	'Single Armoured' & Depth & 27 to 422 m\cr 
	\hline
	\textbf{DA} & Length & 1.6 km\cr
	'Double Armoured' & Depth & $<$ 27 m\cr
	\hline
	\end{tabular}
\caption{Layout of the undersea cable.}
\label{tab:MEOCLayout}
\end{table}
On the sea side, the cable terminates in a titanium shell dry-mated electro-optical connector\footnote{SeaCon Europe Ltd, http://www.seaconbrantner.com} (Figure~\ref{fig:MEOCTerm}) mating with a receptacle in the junction box belt. The overall weight of the deployed cable is 88~tons for a total length of 41.3~km.

The central electrical conductor connects with the HV pole of the primary winding of the junction box transformer. The LV pole returns the current through an external sea electrode\footnote{Titanium with Keramox$^{\circledR}$ coating, of length 1.6 m and diameter 40 mm; Magneto BV, http://www.magneto.nl} to the power hut, which has receiving electrodes buried in the nearby ground.

\subsubsection{Power supply to the junction box}
\begin{figure}[!h]
	\centering
		\includegraphics[width=7.4cm]{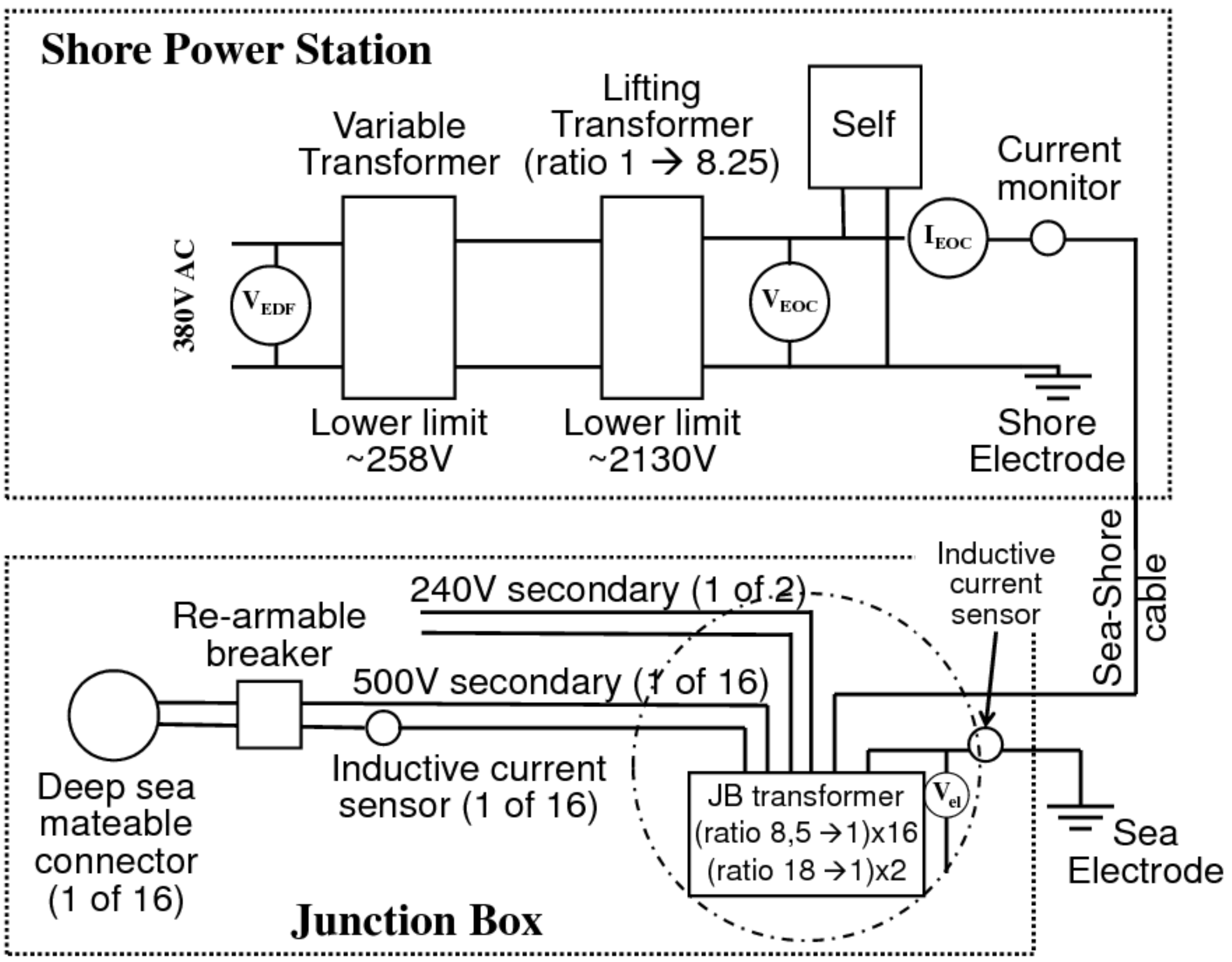}
	\caption{Power distribution system from the shore to the underwater junction box.}
	\label{fig:JBPowerDistrb}
\end{figure}

Figure~\ref{fig:JBPowerDistrb} illustrates the power system of the detector up to the outputs of the junction box.
The detector shore power supply draws 400~V 50~Hz AC power from the electricity grid. The power supply located in a building (the Power Hut) near the cable landing beach raises the voltage to the range 3700-4100~V for passage through the undersea cable to the junction box. The voltage at the cable input is adjusted using a motor driven variable transformer, depending on the load requirement (i.e. the number of lines to be powered). The 50~Hz AC power system was chosen as the best compromise for power delivery over the 42~km transmission length. Although direct and indirect losses are increased relative to a DC system having the same voltage and current limits, the AC option was preferred since it allows for more reliable variation of the cable entry voltage, using passive (transformer) elements. It also has a greater simplicity and reliability at the seabed distribution node through the use of a transformer with multiple windings in the junction box. The 9 $\upmu$F cable capacitance needs to be compensated; this is largely achieved through the use of a 1.4~H self inductance at the shore end of the cable. The self inductance strongly reduces the reactive component produced by the cable capacitance. The dissipation in the cable is therefore mostly resistive and corresponds to about 10\% of the 36~kW (9.6~A at 3800~V) leaving the shore station.

\subsection{Shore facilities}
The onshore infrastructure consists of two separate buildings, the Shore Station housing control and data management infrastructure and providing space for onsite personnel, and the Power Hut devoted to power distribution requirements. 
The shore station is situated at La Seyne-sur-Mer.

The building has three rooms dedicated to the operation of the ANTARES experiment: a computer room, a control room, and a service room. The computer room hosts the racks for the clock crate, the 13 onshore DWDM crates (counterparts of the DWDM boards of the 13 lines) and the PC farm for data filtering and storage. The control room contains various computers for apparatus control and status monitoring.

The Power Hut is located near to the MEOC landing point at Les Sablettes, and is connected to the latter with a fibre optical link 1.5 km long. The Power Hut is connected to the 60~kVA, 400~V three-phase electrical distribution from \'Electricit\'e de France. The building has been adapted to house the transformer 400/4000~V, the MEOC to shore link rack as well as  the return current electrodes. 

\section{Construction}
\label {sec:construction}
\subsection{Generalities}
The construction of the apparatus started in 2001 with the installation of the long distance electro-optical cable. In late 2002 the underwater junction box was installed; the deep sea end of the cable was recovered for this purpose, the junction box dry connected to it, tested on the deck of the ship and finally deployed.
During the following years several prototype lines were installed and operated {\it in situ}, allowing the validation and optimization of their design, as well as the evaluation of possible long-term effects. 
The first detection line was installed in early 2006 and the last two lines of the apparatus were put into operation in May 2008. A European-wide effort was mobilized for the construction of the detector. While all optical modules were assembled in just one laboratory, the production of different mechanical parts and electronics boards was performed by a number of laboratories in different countries. A large effort was subsequently devoted to the assembly of the electronics modules and of the complete lines. Three sites fabricated electronics modules in order to feed two laboratories which worked independently on the assembly of new lines. When running at full speed, two lines were produced every three months. 

Since the same integration tasks had to be performed at different sites, special care was devoted to the development of dedicated tools, the definition of detailed procedures and the distribution of the expertise among the different teams, under a unified Quality Control approach. A coherent scenario of tests to be performed at the different integration levels made it possible to identify faulty components in order to avoid delays in the integration of the lines. Logistics was an important issue; the Cellule Logistique of IN2P3\footnote{ULISSE, http://ulisse.in2p3.fr} was used for managing the transportations. 

\subsection{Quality assurance and quality control}
The organization of the Quality Assurance/Quality Control activities was based on the methodologies defined by the rules of ISO~9001:2000\footnote{http://www.iso.org/iso/catalogue$\_$detail?csnumber=21823}, with an effort to guarantee a high level of adaptability and flexibility. A special attention was given to defining rules for proper management of the documentation, with different levels of approval established for the most critical documents, such as the integration and test procedures. All documents were stored in a centralised repository, accessible through a password-protected website. Problems and changes in the organization were traced through appropriate documents: a Non-Conformity Report was the tool used to report the problems found at all levels in the construction of the apparatus. Improvements of the organization were sometimes defined after the treatment of such reports, and implemented in response to a Design Change Request.

A key document for the construction of the apparatus was the Risk Analysis, whose output of served as the basis, first for the implementation of detailed prototyping campaigns and later for the definition of the test criteria to be adopted during the construction of the apparatus. A general Quality Plan was then put in place, defining the main guidelines for Quality management in the Collaboration, and all laboratories participating in the construction of the apparatus were required to define their local Quality Plans, to be applied under the control of Local Quality Supervisors. A Quality Plan was also required from the external providers at the time of placing the orders. A program of audit activities was also set up for all laboratories in order to continually improve the system.

A central database was used for collecting traceability information and, when applicable, calibration data of all products, which were individually identified by a bar-code label built according to a well defined Product Breakdown Structure of the apparatus and a serial number. Detailed information of which products were integrated in which parts of the apparatus was also stored in this DB, so that all necessary information for the configuration of the apparatus at the time of operation was immediately available.

\subsection{Assembly}
\subsubsection{Control module integration}
Construction of the electronics modules required a very high level of reliability since the failure of one module could lead to the loss of functionality of a whole sector of a line. 

The integration of the electronics modules was a delicate task because of the design of the mechanical crate, the fact that electronics boards were densely packed inside it and the need to have careful handling of optical fibres at all times. Detailed procedures were therefore defined and dedicated tools developed. A full functionality test was performed on all integrated modules in order to find and cure all possible problems. Calibration of the front-end electronics was also performed during these tests.

\subsubsection{Line integration}
Line integration took place at two different sites. The sharing of expertise, the usage of the same tools and procedures and the respect of quality rules ensured that the level of quality was the same in the two laboratories. This was confirmed by a cross calibration between the two sites. The lines were integrated from the bottom to the top. Optical splices were used on all optical fibre connections for maximum reliability. Once a sector was completed, a calibration in a dedicated dark room (or in dark boxes) for the optical modules was performed. Simultaneously, the integration of a new sector started. The purpose of the tests in the dark room was to verify the full functionality of the sector, as well as to provide an initial time and charge calibration for all optical modules in the final configuration of the line. A calibration of the tiltmeters in each storey was also performed. Then, the storeys were arranged on a line transportation pallet. The optical modules were temporarily connected to their storeys for the tests, but were then removed and transported separately from the line for maximum safety.

\subsubsection{Deployment preparation}
The final steps of integration took place in a dedicated hangar at the port of La Seyne-sur-Mer. Here, a final functionality test of the lines was performed. Then, in preparation for deployment, the storeys were arranged on wheeled carts, equipped with the optical modules and the instrumentation and moved onto a deployment pallet. The top buoy and the bottom deadweight were finally added. An integrated line was arranged on a single pallet which was then installed on the deck of the ship for the deployment. 

\subsection{Line deployments and connections}
The vessel Castor of the Foselev Marine Company was used for installation of all ANTARES lines. 
       
The deployment of a line proceeds as follows: once the ship reaches the site, the first package to be launched under the boat frame winch is the heavy BSS. Then, the storeys are put into the water one by one until the top buoy of the line. Two 5-ton winches are used on the deployment ship, each equipped with a specially designed remote release hook which made it possible to avoid the use of divers during the deployment. Once the top buoy is in water it is connected to the deep sea cable winch through a hook equipped with an acoustic release.
The transponders mounted on the BSS are localized while paying out cable until the seabed is reached. The ship then adjusts its position using its Dynamic Positioning (DP) capabilities in order to place the BSS on the target location.
This procedure allows the positioning of the lines within a few metres from their target points.
 
A team of 12 people from the Collaboration is needed for a line deployment in addition to the 4 deck crew. The typical duration of activities on site, including DP station tests, acoustic position tests, launch and deployment of the line and cable recovery is about 8 hours.

As explained previously, the connection between the junction box and the lines is made with electro-optical cables of suitable length (ranging from 120 to 350 m), equipped with a wet-mateable connector at each end. These interlink cables are prepared on turrets which are deposited on the seabed, either being deployed with the deep sea cable winch or in free falling mode. An underwater vehicle is then used for the subsequent actions: it moves the turret close to the JB and connects one end of the cable to a free output of the JB. Once a good connection is established at the level of the junction box, the underwater vehicle moves the turret towards the base of the line to be connected, while routing the cable on the seabed. Finally, the connection to the BSS is performed. Each operation is monitored from the shore station where tests are made in order to test electrical and optical continuity.

All connection operations were performed by means of the ROV Victor of IFREMER, except the connection of the second line of the apparatus which was performed with the manned submersible Nautile of IFREMER.

The weather conditions permitting the safe operation of an underwater vehicle depend on the support vessel used. Wind limits of 20 knots\footnote{1 knot = 1.852~km~h$^{-1}$.} were found when Victor was operated onboard Castor, while larger vessels allowed operation with winds up to around 35 kts. The seabed conditions have also to be acceptable, since operation of the ROV becomes difficult when the sea current exceeds 10 cm~s$^{-1}$.

The ROV was also used during the detector construction for other tasks, such as:
\begin{itemize}
\item inspection and test of the outputs of the junction box;\vspace*{-0.40\baselineskip}
\item survey of optical modules;\vspace*{-0.40\baselineskip}
\item deployment of the seismograph;\vspace*{-0.40\baselineskip}
\item measurement of the electrical current and visual survey of the MEOC;\vspace*{-0.40\baselineskip}
\item survey of the acoustic transponders installed around the apparatus;\vspace*{-0.40\baselineskip}
\item change of the interlink cables.
\end{itemize}

\subsection{Maintenance}
A simplified scheme of the construction organization is still operating today for maintenance of the apparatus. The possibility of recovering lines is foreseen in case of severe functionality problems while no routine maintenance of the offshore apparatus is scheduled. A recovery operation is performed as follows: once the ship is on site, the hook holding the BSS to its deadweight is opened by means of a release command issued acoustically from onboard. Once the deadweight is released, the line comes up to the surface freely in about 40 minutes. In order to perform this operation safely, the sea current conditions must be suitable with deep sea currents not exceeding 5 cm~s$^{-1}$ in order to prevent the released line from colliding with the other lines of the detector. When the top buoy of the line reaches the surface, it is dragged to the ship. Then, the recovery of the rest of the line takes place in a way similar to a reversed deployment procedure. 

A set of spare components for all different products is available, so that generally the lines could be repaired without delays for new productions of elements. The different laboratories remain, however, in charge of the products they have originally provided, in case new productions must be launched, so components would be provided with the same quality level as during the construction. One laboratory is still active for assembly of new electronics modules while another laboratory is still active for line dismounting and re-integration. A line recovery gives also the opportunity to inspect all parts of the lines for any effects induced by the long-term operation at large depth.

\section{Operation}
\label {sec:operation}

\subsection{Apparatus control}
Control of the apparatus is performed from the shore station in Institute Michel Pacha, which is manned during the day for this purpose, although full control can also be performed remotely from all institutes participating in the experiment by means of a VNC (Virtual Network Computing) application\footnote{http://www.realvnc.com}. All information for apparatus control is stored in the central database of the experiment, located together with the resources for mass storage of data at the IN2P3 Centre de Calcul in Lyon\footnote{http://cc.in2p3.fr}. The Oracle database is also regularly updated with the slow control information from the data acquisition system so as to maintain a detailed record of the performance of each element of the apparatus. 

The operator controls the data acquisition operations by means of two main programs, both provided with a Graphical User Interface (GUI), one for monitoring and control of the power delivery system, and the second one for control of the data acquisition. The former program is capable of delivering the commands set by the operator to the onshore power system facility or directly to the junction box, as appropriate. It can also be used to retrieve monitoring data from all sensors of the junction box. The sensor data are converted into engineering units for on-screen presentation and are written at regular intervals into the database. The environmental conditions inside the power distribution hut are also monitored and recorded regularly. Alarm thresholds are set for each sensor in the database so as to define different levels, and priorities, of alarms: low priority alarms alert the operator with on-screen messages while higher priority alarms can also generate SMS text messages, and depending on the criticality of the sensor, may trigger a power shutdown after a predetermined delay.

In total, the data acquisition control system involves about 750 processes (300 offshore processes for data acquisition, 300 offshore processes for slow control, and about 120 processes running on the onshore computers for data processing and filtering, monitoring and user interface). These processes implement the same state machine diagram \cite{bib:antares-daq-paper}. Transitions between different states are decided by the operator and handled by the main control GUI. All relevant configuration information is extracted from the database. A message logging system keeps track of all operations in a designated file (that is archived regularly); warnings generated by any process are captured, recorded in the same file and shown on the computer screen for operator alert. In order to archive data efficiently, the main control GUI updates the run number regularly. The database system is also used to keep track of the history of the detector integration and the data taking. A number of monitoring programs have been developed to monitor different data coming from the apparatus so that the operator can have a detailed view of the working conditions in the apparatus at a glance; this is very important for an undersea apparatus, since depending on the optical background conditions, the user has to choose the best data taking configuration in order to maximise the data quality. Monitored quantities include environmental conditions inside the electronics modules (temperature, humidity), position information retrieved by the compasses and tiltmeters inside the electronics modules, PMT hit rates, the measured sea current direction and speed. A fraction of the data is reconstructed online and reconstructed events are also displayed.  

\subsection{Data acquisition}
The operations onshore are optimized so as to maximize the time devoted to data taking. The data collected offshore are temporarily stored in high capacity buffers on the LCMs which allow a de-randomisation of the data flow. The data are packed offshore as  arrays of hits of predefined time frame duration of about 100~ms. Depending on the hit rate of the PMTs, the size of these data packets amounts to  60--200~kB. The data are then sent to shore in such a way that the data collected for the full detector for the same time frame are sent to a single data filter process in the onshore data processing system. The data flow to the different data filter processes is staggered to avoid network congestion.

The onshore data processing system consists of about 50 PCs running the GNU/Linux operating system. To make optimal use of the multi-core technology, four data filter processes run on each PC. The physics events are filtered from the data by the data filter process using a fast algorithm, as described in the next subsection. For one processor, the typical time needed to process 100~ms of raw data amounts to 500~ms. The available time allows the application of concurrent software triggers to the same data. On average, the data flow is reduced by a factor of about 10,000. The filtered data are written to disk in ROOT\footnote{http://root.cern.ch} format by a central data writing process and copied every night to the computer centre in Lyon. The count rate information of every PMT is stored together with the physics data. The sampling frequency of these rate measurements is about 10~Hz.

The data from the readout of the various instruments are transferred as an array of parameter values and stored in the database via a single process. The readout of the various deep sea instruments is scheduled via read requests that are sent from shore by a designated process. 
The frequency of these read requests is defined in the database. A general purpose data server based on the tagged data concept is used to route messages and data~\cite{bib:controlhost}. For instance, there is one such server to route the physics events to the data writer which is also used for online monitoring. 

\subsection{Trigger}

The data filter algorithm applied onshore is based on different trigger criteria, including a general purpose muon trigger, a directional trigger, muon triggers based on local coincidences, a minimum bias trigger for monitoring the data quality, and dedicated triggers for multi-messenger investigations. 

\begin{figure}[!h]
	\centering
		\includegraphics[width=7.4cm]{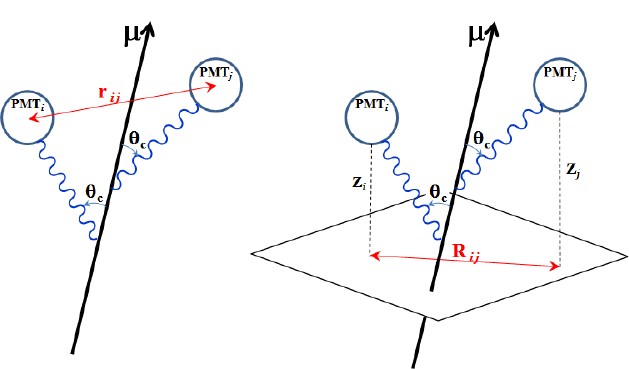}
	\caption{Definitions of the symbols used in equation 2 (left) and in equation 3 (right).}
	\label{fig:symboldefinitions}
\end{figure}

The general purpose (``standard'') muon trigger makes use of the general causality relation:
\begin{eqnarray}
   \left| t_i - t_j \right| \le r_{ij} \times \frac{n}{c}
   \label{eq:3D}
\end{eqnarray}
\noindent
where $t_i$ ($t_j$) refers to the time of hit $i$ ($j$), $r_{ij}$ to the distance between PMTs $i$ and $j$, $c$ is the speed of light and $n$ the index of refraction of the sea water (Figure~\ref{fig:symboldefinitions},~left).
The direction of the muon, and hence of the neutrino, being not used, this trigger is sensitive to muons covering the full sky. To limit the rate of accidental correlations (i.e. to increase the purity of the event samples), the hits have to be preselected. This preselection provides the L1 signals, i.e. either coincidences in a time window of 20~ns between two neighbouring PMTs in the same storey or the occurrence of large pulses (number of photoelectrons typically greater than 3 in a single PMT). Then, the trigger criteria consist either in a set of at least 5 L1 hits that are causally related or in a local cluster of neighbouring L1 hits. The efficiency and the purity of this trigger have been determined with a simulation of the detector response to muons and accounting for the observed background~\cite{bib:antares-daq-paper}. The efficiency grows fast above 10 detected photons and reaches $\approx 1$ at 40 detected photons. The typical threshold for the neutrino energy is a few hundred GeV. The purity is of the order of 90\%, the remaining impurity being mainly due to (low-energy) muons which in combination with the random background produce a trigger; only a small fraction of the events ($\ll1\%$) is found to be due to accidental correlations. The observed trigger rate is thus dominated by the background of atmospheric muons and amounts to \mbox{5--10~Hz} (depending on the trigger conditions). The standard trigger can operate with hit rates in each PMT up to about 250~kHz.

In addition to the standard trigger, a directional trigger has been implemented to maximize the detection efficiency of tracks coming from predefined directions. Currently, this trigger is used to look for events coming from the Galactic centre. This trigger makes use of the following direction specific causality relation:
\begin{eqnarray}
\lefteqn{(z_i - z_j) - R_{ij}\tan{\theta_C}} \nonumber \\ 
& \le c(t_i - t_j) \le (z_i - z_j) + R_{ij}\tan{\theta_C}
   \label{eq:1D}
\end{eqnarray}
\noindent
where $z_i$ refers to the position of hit $i$ along the neutrino direction, $R_{ij}$ refers to distance between the positions of hits $i$ and $j$ in the plane perpendicular to the neutrino direction and $\theta_C$ is the Cherenkov angle in water (Figure~\ref{fig:symboldefinitions},~right). Compared to equation \ref{eq:3D}, this condition is more stringent because the 2-dimensional distance $R_{ij}$ is always smaller than the corresponding 3-dimensional distance. Furthermore, this distance corresponds to the distance travelled by the photon (and not by the muon). Hence, it can be limited to several absorption lengths without loss of detection efficiency. This restriction reduces the combinatorics significantly. As a consequence, all hits can be considered for the directional trigger and not only the preselected hits used for the standard trigger, without compromising the purity of the physics events. In a field of view of about 10~degrees around the selected direction and for neutrino energy below 1~TeV, the detection efficiency with the directional trigger is 2 times higher than that obtained with the standard trigger.

Additional trigger schemes have been implemented to allow multi-messenger searches. The onshore data processing system is linked to the Gamma-ray bursts Coordinates Network (GCN)\footnote{http://gcn.gsfc.nasa.gov/}. There are about 1 to 2 GCN alerts distributed per day and half of them correspond to a real gamma-ray burst. For each alert, all raw data are saved to disk during a preset period (presently 2 minutes). The buffering of the data in the data filter processors is used to save the data up to about one minute before the actual alert. Furthermore, ANTARES is capable of distributing proper event alerts to external detectors. A collaboration with the TAROT \cite{bib:tarot} optical telescope has been recently established in this respect. The direction of interesting neutrino triggers (two neutrinos within 3 degrees within a time window of 15 minutes or a single event of very high energy) are sent to the TAROT telescope in Chile in order that a series of optical follow-up images can be taken. Such procedures are well-suited to maximize the sensitivity for transient sources such as gamma-ray bursters or flaring sources. 

\subsection{Calibration}

\subsubsection{Position determination}
\label{subsec:operation_positioning}
Accurate position information for each OM is needed for good event reconstruction (cf.~Section~\ref{subsubsec:positioning_system}). The shape of each line is reconstructed by performing a fit based on all the available measurements: positions coming from the acoustic positioning system, headings provided by the compasses and tilt angles provided by the tiltmeters. These measurements are performed every two minutes. The relative positions of the OMs are then deduced from the reconstructed line shape and from the known geometry of the storeys: a hydrophone is mounted on a storey offset from the centre of the storey. The acoustic positioning system described below allows to determine the position of the hydrophone. Combining the hydrophone position with the tilt and heading information of the same storey one obtains the position and the orientation of that storey.
Five storeys of a line are equipped with hydrophones. From the position and orientation of these five storeys and from the tilt and heading measured in the other storeys, the shape of the line can be reconstructed and the position of every OM can be determined.

The reconstruction of the line shape is based on a model which predicts the mechanical behaviour of the line under the influence of the sea water flow taking into account the weight and drag coefficients of all elements of the line. The zenith angle $\Theta$ in one point of the line can be computed from the vertical forces $F_z$ (buoyancy minus weight) and the horizontal drag forces $F_{\bot}$~=~$\rho$~$C_WAv^2$~/~2, where $\rho$ is the water density, $A$ is the cross-section area of the element considered, $v$ is the sea current velocity, and $C_W$ is the drag coefficient. The drag coefficient was determined by a hydro-dynamical study of the storey in the IFREMER pool facility. Since $\tan(\Theta)= dr/dz$, the radial displacement $r$ as a function of the vertical coordinate $z$ can be obtained by integration along the line, resulting in the expression:  
\begin{eqnarray}
r(z) = av^2z - bv^2\ln[1-cz],
\label{eq:rz}
\end{eqnarray}
where $a$, $b$ and $c$ are known constants, and the horizontal components of the sea current velocity $v^2=v_x^2+v_y^2$ are treated as free fitting parameters. The values of sea current velocity inferred from the reconstructed shapes of the different lines can be compared among themselves and to the measurements provided by the ADCP installed on the instrumentation line in order to have a test of the accuracy of the reconstruction procedure. 

The measurements with the acoustic positioning system are performed as follows. 
Acoustic sinusoidal wave packets of short duration (typically 2 ms) are broadcasted from the emitters at the bottom of each line and detected by the hydrophones installed on all lines. Various fixed frequencies between 40 and 60 kHz are used in turn to differentiate the sound emissions and to avoid possible interference due to successive emissions of acoustic waves with the same frequency. Detection of the acoustic signal by the hydrophones is done by comparison of the amplitude of the numerically filtered signal to a preset threshold. The gain of the preamplification as well as the detection threshold are set for each receiver depending on the emission cycle, the emission frequency and the attenuation due to the distances travelled.
In this way the travel time between the emitter and the receiver can be determined independently. Knowing the sound velocity profile, the distance between one emitter and one receiver is deduced from the travel time measurement. Positions of all hydrophones and transducers are then computed from the measured distances using the triangulation principle and a least-mean-square minimization procedure.

Prior to positioning measurements, configuration messages are sent from the shore station to all acoustic modules. The configuration defines whether modules will act as receivers or, in the case of the devices at the bottom of the lines, as emitters for a given measurement cycle. In addition the frequency and the detection gain are set. The emission of acoustic signals is triggered by a synchronization signal sent by the master clock system. The timestamp of each detected signal is obtained by starting a counter in the acoustic module with the synchronization signal, and by stopping the counter when the signal is detected. The accuracy of this counter is 100 ns.

Autonomous transponders installed around the detector are used in the measurements in order to enlarge the triangulation basis. These transponders are autonomous emitter-receiver beacons fixed on pyramidal structures anchored on the seabed and powered by batteries. Each transponder responds at one unique frequency whilst the interrogation occurs at a common communication frequency. The transponders can thus be activated and de-activated by a transceiver using an acoustic modem dialogue. 

The acoustic travel times have to be corrected for the delays of the signal due to emission and detection delays including the frequency-matching numerical filter. Such delays depend on the ratio between the detection threshold and the measured signal amplitude. The global delay has been measured and found to be in the range from 140 to 180~$\mu$s. They can be modelled according to a third-order polynomial. This polynomial correction is then applied to the detection timestamp. The accuracy of the acoustic travel time measurement is primarily determined by the jitter of the detected signal. This jitter has been measured to be less than 4~$\mu$s corresponding to a distance of 6~mm for a sound velocity of 1500~m~s$^{-1}$, even in the presence of a 30~dB white-noise background.

For the determination of distances from the measurement of the acoustic travel time, the knowledge of the sound velocity within the detector is needed. The detector is equipped with several Sound Velocimeters (SV) placed at different locations along the detector lines, in order to determine the sound velocity and its variations. In sea water, the sound velocity depends on thermodynamic parameters such as the conductivity, the temperature and the pressure, which depend on the depth. Sound velocity can be inferred by combined measurements of these quantities performed with a CTD detector (conductivity, temperature and depth probes), according to the Chen \& Millero model \cite{bib:chen_millero, bib:unesco}. One SV-CTD has been also installed in the apparatus in order to have independent sound velocity determination, and also to get an estimation of the salinity and temperature gradients within the detector.

\begin{figure}[!h]
	\centering
		\includegraphics[width=7.4cm]{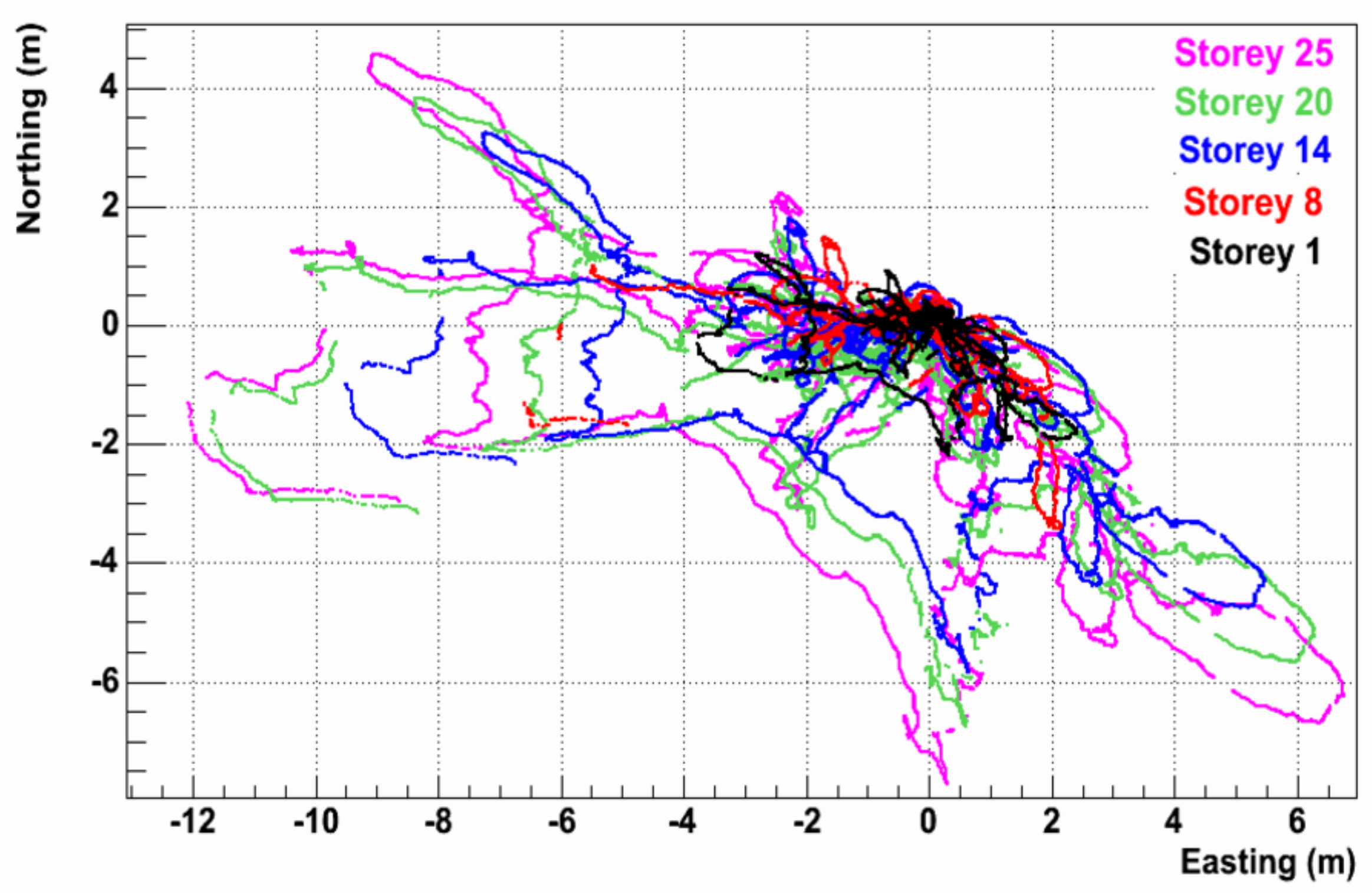}
	\caption{Displacements in the horizontal plane of the five storeys equipped with positioning hydrophones of a line as determined by the positioning system.}
	\label{fig:positions}
\end{figure}

The behaviour of the positioning system using the first ANTARES data is described in \cite{bib:pkeller, bib:mardid}. As an example, Figure~\ref{fig:positions} shows the x-y displacement in the horizontal plane of the five hydrophones at different heights along a line as a function of time for a period of 6 months (from July to December 2007). A detailed analysis of the system performance indicates that the resolution is better than the 20~cm specification at which value it does not degrade the angular resolution.

\subsubsection{Timing calibration}
\begin{figure}[!h]
  \centering
    \includegraphics [width=7.4cm]{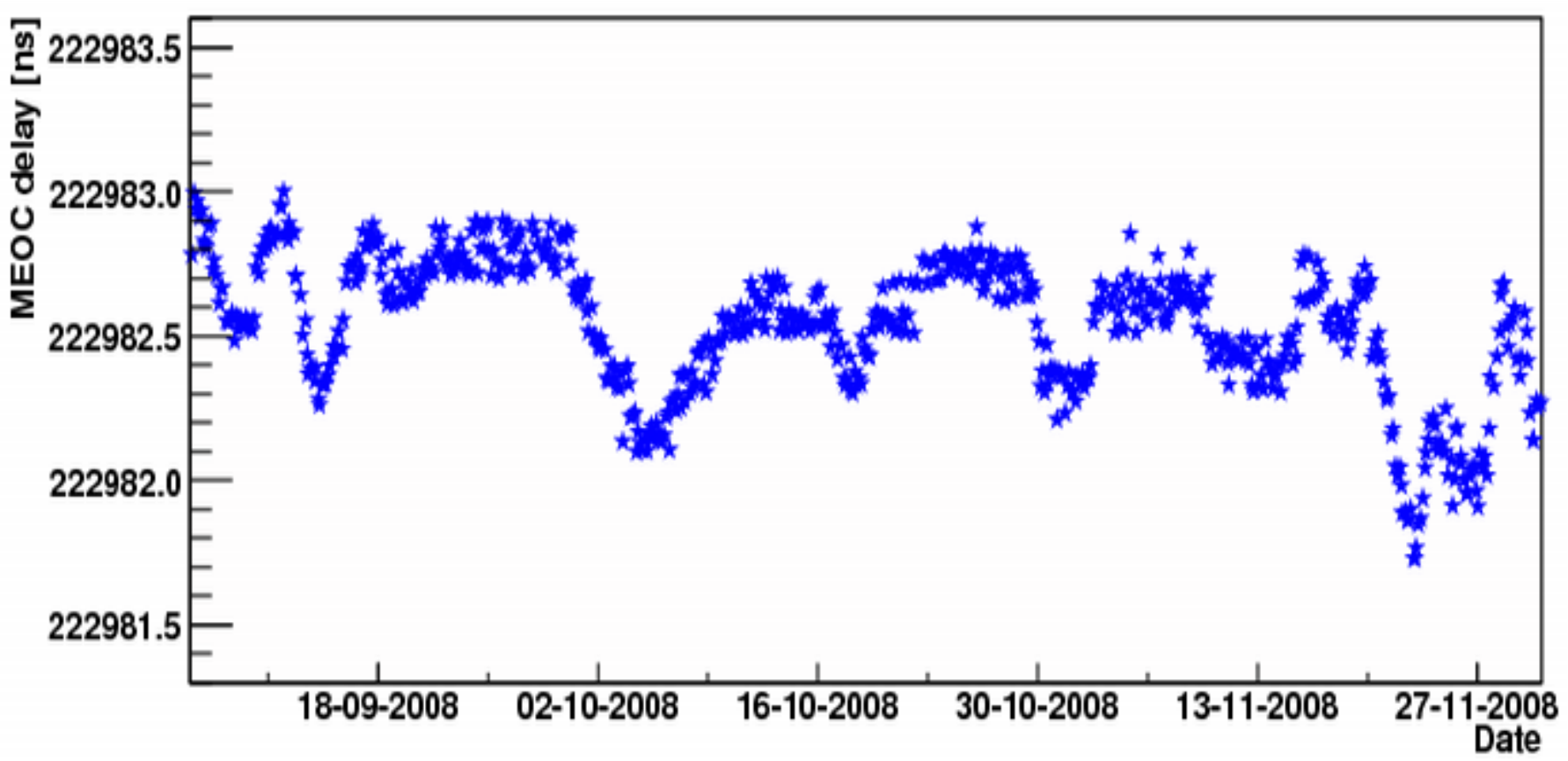}
    \caption{Measurement of the round trip time for clock signals between shore and one of the electronics module of the apparatus.}
    \label{clock}
\end{figure}

The timing calibration \cite{bib:timing} can be divided in two parts. On one hand, the master clock which provides a common synchronization signal to the whole apparatus can be used to measure the time path from shore to each electronics module. This information is useful to check the overall stability of the system and to measure the {\it in situ} time delays after the connection of a detector line. Figure~\ref{clock} shows the round trip time measured for one electronics module. The stability and accuracy of the measurements are at the sub-nanosecond level, as required. 

The time offsets for each specific channel are then calibrated {\it in situ} by means of the optical beacons installed on the lines and the LED pulsers mounted inside each Optical Module. The list of optical beacons include LED beacons, distributed at different levels along each line, and two laser beacons located at the bottom of two central lines. These devices are operated in a similar way. While the laser beacons are mainly used for cross-check of the timing calibration of the OMs of different lines, the LED beacons remain the main tool for {\it in situ} timing calibration. These beacons are flashed in turn for short time periods in order to illuminate the surrounding optical modules. From the comparison between the measured and the expected time of the hits, taking into account the propagation time of the light, one can infer the time offset for each OM. 

\begin{figure}[!h]
  \centering
    \includegraphics [width=7.5cm]{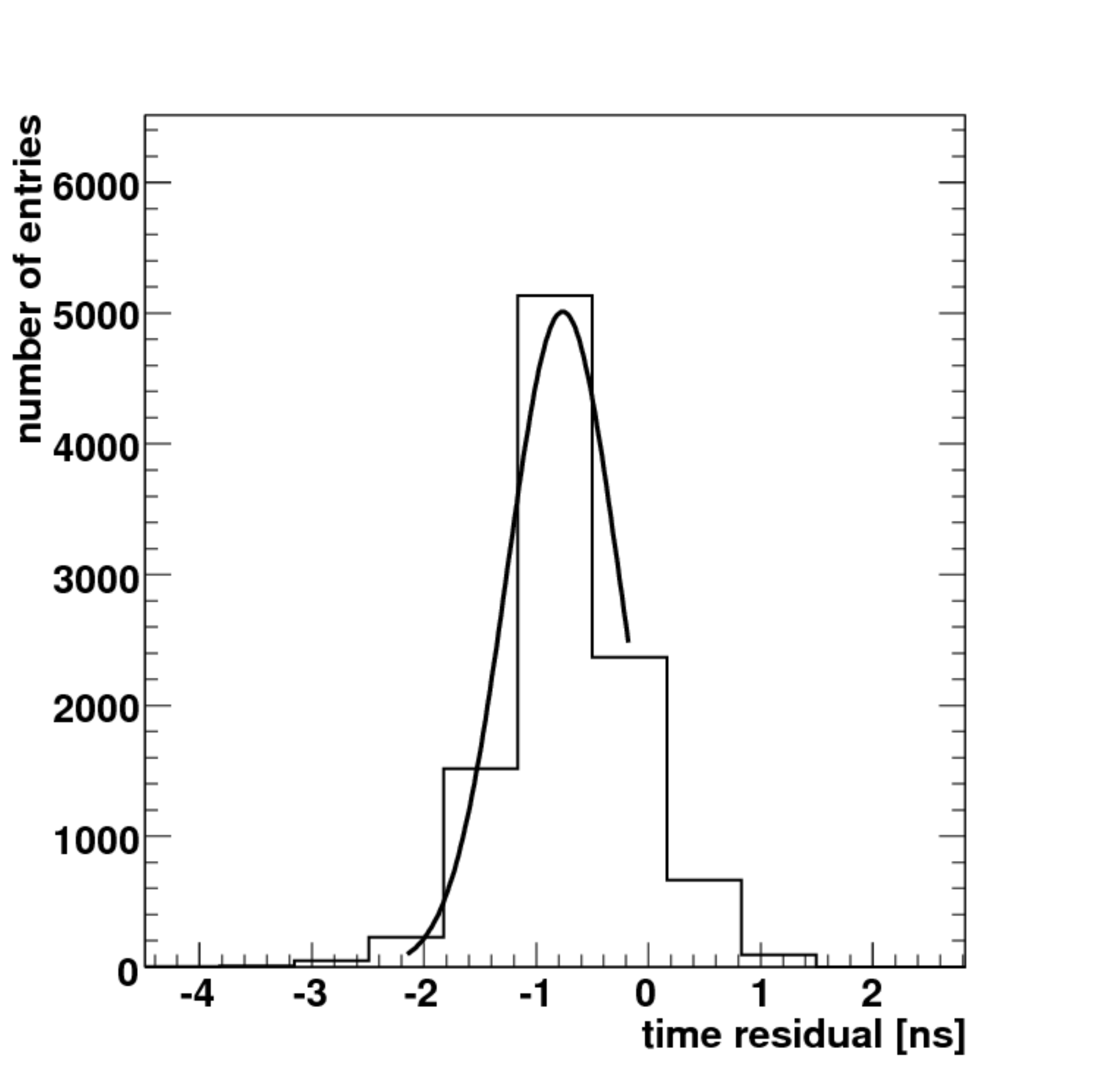}
    \caption{Time residual distribution of the signals in an OM located two storeys above a flashing LED Beacon. The curve is a Gaussian fit with a sigma of 0.5~ns.}
   \label{OMLOBResiduals}
\end{figure}

\begin{figure}[!h]
  \centering
    \includegraphics [width=7.5cm]{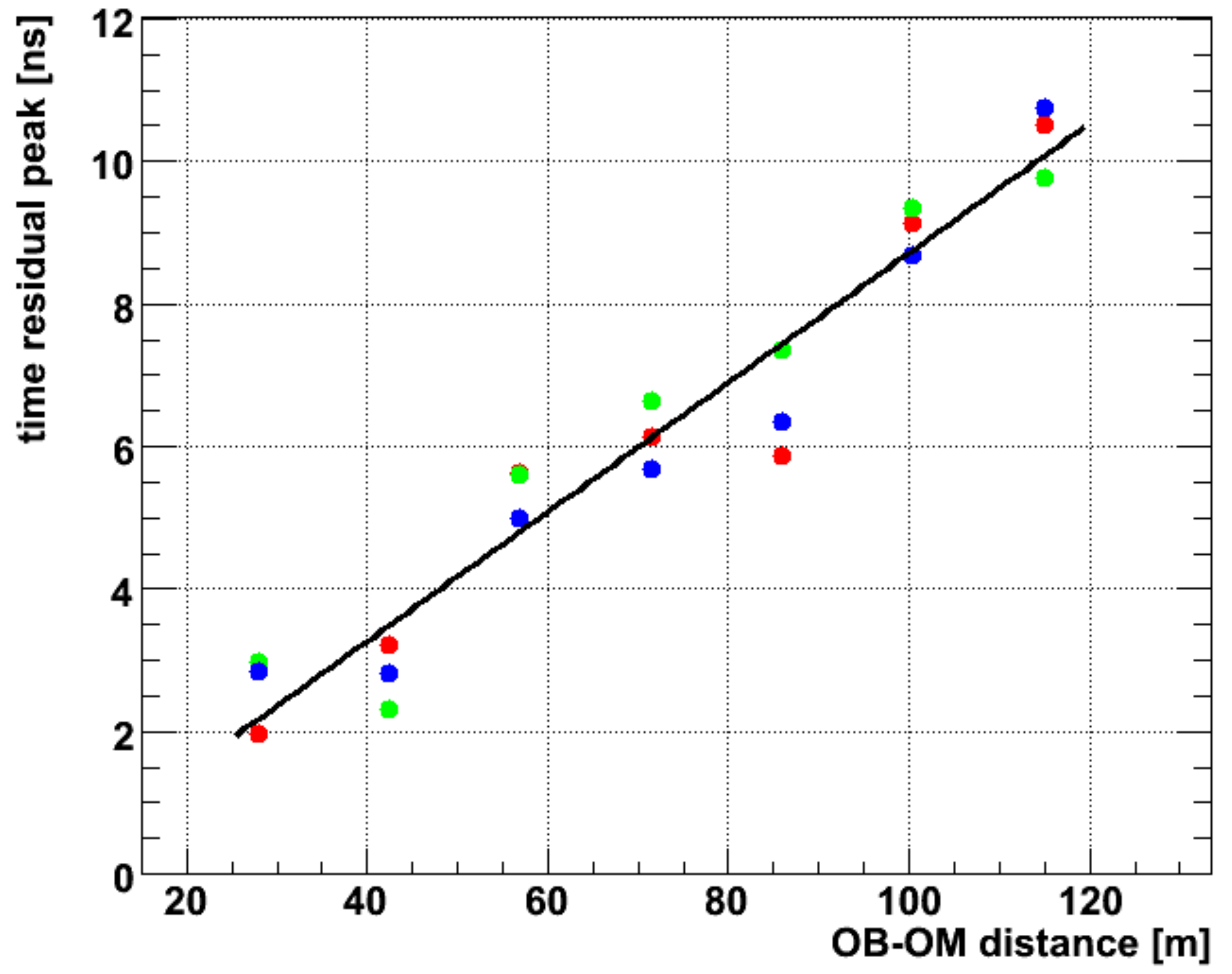}
    \caption{Time residual peak position as a function of the distance between a flashing LED beacon and the OMs along seven storeys above. The three points at each distance correspond to the three OMs in each storey. The additional delay with distance is due to the early photon effect.}
   \label{fig:lineLOB}
\end{figure}

Figure~\ref{OMLOBResiduals} shows the time residual distribution for one particular OM obtained from one calibration run. The tail on the right part of the distribution can be attributed to light scattering.
The position of the leading edge can be determined with a Gaussian fit to the left part of the distribution, which is less affected by scattering. The distribution of the leading edge as a function of the distance (or, equivalently, the storey number) shows a linear trend, which is ascribed to the ``early-photon effect''. This effect is due to the duration of the light pulse (FWHM~$\approx~4$~ns) and the intensity of the detected light. The closer the OM, the more light it receives and therefore the sooner the PMT signal passes the preset threshold of the ARS, an effect which is further emphasized by time walk. A straight line fit is then applied to the data and deviations from this fit are understood as the corrections to be made on the time offsets obtained by the calibration onshore. An example is given in Figure~\ref{fig:lineLOB}.
In most cases these corrections are small, and only for $\approx$~15\% of cases they are larger than 1~ns. This method provides an average improvement of $\approx$~0.5~ns over the timing calibration performed onshore. 
 
The time offset variations of each optical module, due to variations in the transit time of the photomultiplier for instance, can be monitored by operating the LED pulser placed in each optical module. These data show a good stability of the time delays when the HV of the PMT and the settings of the ARS are not changed.

\begin{figure}[!h]
  \centering
    \includegraphics [width=7.4cm]{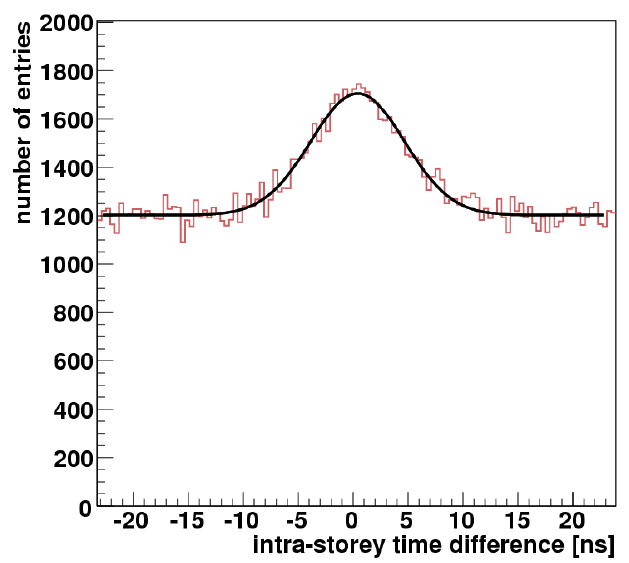}
    \caption{Time difference between signals measured by a pair of PMTs in storey 1 of line 1. The peak is due to single $^{40}$K decays , whereas the flat background is due to accidental coincidences of $^{40}$K decays and bioluminescence. The solid line is a fit of the sum of a Gaussian distribution and a flat background to the data.}
    \label{fig:K40}
\end{figure}

\begin{figure}[!h]
  \centering
    \includegraphics [width=7.5cm]{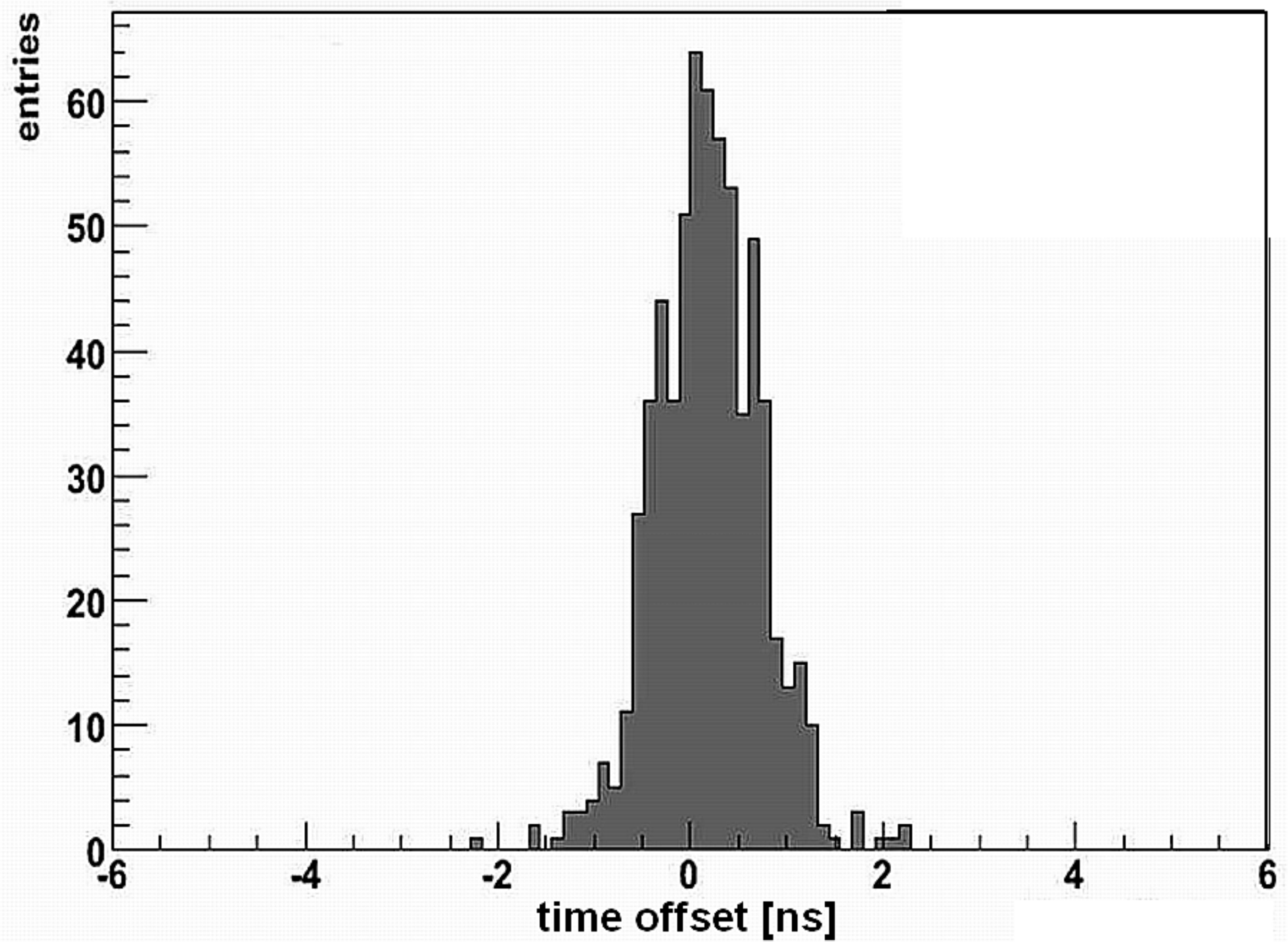}
    \caption{Differences between the time offsets inferred from the calibrations with the LED beacons and independently determined by the $^{40}$K coincidence method for all photomultiplier tubes.}
    \label{time_offset_accuracy}
\end{figure}

An additional check of the timing calibration accuracy may come from the detection of coincidences of PMT signals induced by Potassium-40 decays ($^{40}$K). This radioactive isotope is naturally present in the sea water. From its decay, electrons with a kinetic energy up to 1.3~MeV are produced. This energy exceeds the Cherenkov threshold for electrons in water (0.26~MeV), and is sufficient to produce up to 150 Cherenkov photons. If the decay occurs in the vicinity
of a detector storey, a coincident signal may be recorded by a pair of PMTs. In Figure~\ref{fig:K40} the distribution of the measured time difference between hits in neighbouring PMTs of one storey is shown. The peak around 0 ns is mainly due to single $^{40}$K decays producing coincident signals. The fit to the data is the sum of a Gaussian distribution and a flat background. The full width at half maximum of the Gaussian function is about 9~ns. This width is mainly due to the spatial distribution of the $^{40}$K decays around the storey. The positions of the peaks of the time distributions for different pairs of PMTs in the same storey are used to cross-check the time offsets computed with the timing calibration. This is illustrated in Figure~\ref{time_offset_accuracy}  which shows a comparison of the time offsets calculated from the optical beacon calibration and those extracted from the analysis of $^{40}$K coincidences.The rms of the distribution is about 0.6~ns. 

The coincidences induced by $^{40}$K decays provide also a powerful tool for monitoring the relative efficiencies of the individual OM, with an accuracy of about 5\%.

\subsubsection{Amplitude calibration}
Amplitude calibration of each channel is routinely performed. During special runs, the output signal of the PMT is digitized at random times. This allows for a measurement of the corresponding pedestal value of the AVC channel. The single photoelectron peak is studied with minimum bias events. The optical activity due the $^{40}$K decays and bioluminescent bacteria produces primarily single photons at the photocathode level. The knowledge of the position of the single photoelectron peak and of the pedestal is used to determine the charge conversion over the full dynamical range of the ADC. The charge measurements, performed inside the ARS, appear to be influenced by the time measurements in the TVC channel (the inverse effect does not apply). This cross-talk effect is corrected on an event-by-event basis. The maximal size of this correction amounts to 0.2 photoelectrons. The effect is thought to be due to a coupling between the capacitors inside the pipeline in the chips. This correction is inferred with {\it in situ} measurements of the AVC value versus the TVC value. 

\begin{figure}[!h]
  \centering
    \includegraphics [width=7.5cm]{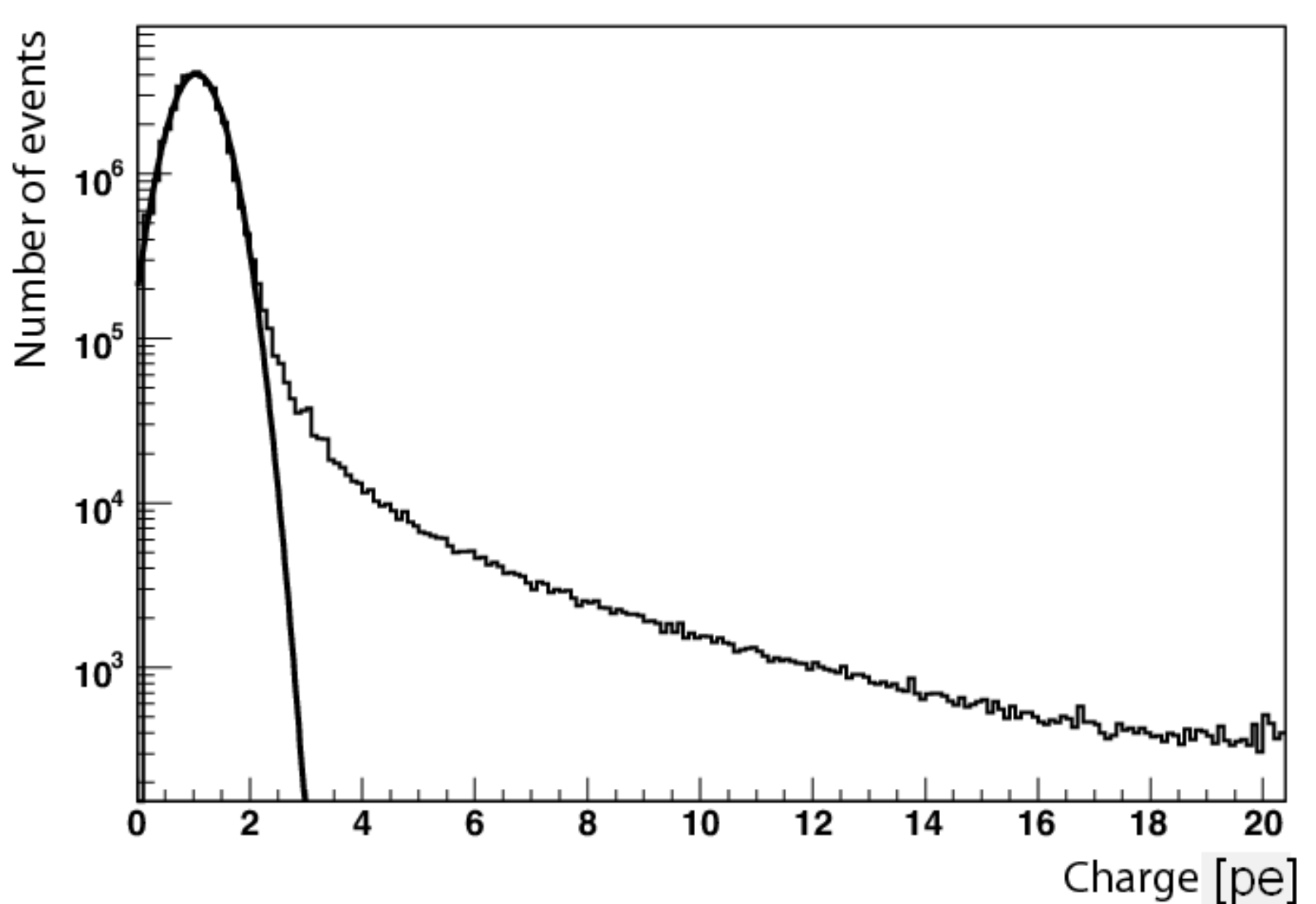}
    \caption{Calibrated charge distribution combining all PMTs in the detector.}
    \label{fig:pulseheight}
  \end{figure}

Once the cross-talk correction is made, the charge calibration is applied to reconstruct the amplitude of the individual signals detected from the optical modules. As shown in Figure~\ref{fig:pulseheight}, this distribution is peaked at one photoelectron as expected from $^{40}$K decay and bioluminescence.

The front-end chip has also the capability to perform full waveform sampling (WF) of the PMT signal in addition to the charge measurement of the PMT pulse and its arrival time. This functionality is primarily meant for recording double pulses or large amplitude signals. However, it is also exploited during the calibration in order to cross-check the computation of the charge by the integrator circuit of the front-end chip and for determining the shape of the SPE signals in order to correct for the so-called walk effect (i.e., the dependence of the threshold crossing time on the signal amplitude). 


\subsection{Performance of the apparatus}

\begin{figure*}[!htb]
	\centering
		\includegraphics[width=0.90\textwidth]{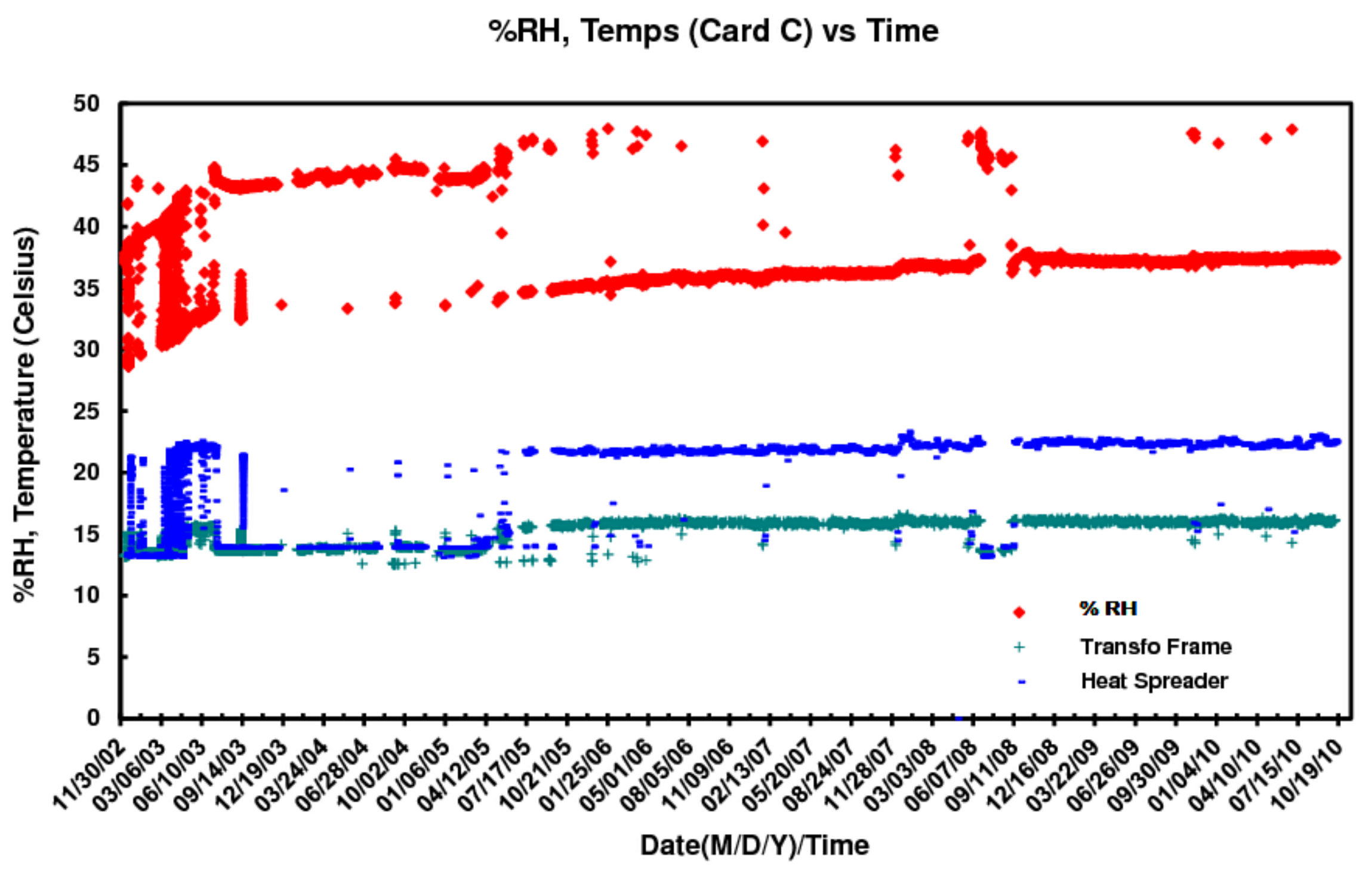}
	\caption{Long-term monitor of relative humidity (RH) (top curve) and temperature (bottom curves) in the underwater junction box over a period of eight years. From 2002 to 2005, during the prototyping phase, the detector was powered off a large part of the time while from 2006, with operational detector lines, the power was mostly on. The two temperature curves come from probes located at different positions in the junction box.}
	\label{fig:JBResults}
\end{figure*}

There are a number of criteria which can be considered for assessing the performance of a complex apparatus like ANTARES. The first criterion concerns stability of the operating conditions. The junction box has the longest operation history since it was installed in December 2002. Temperature and relative humidity inside the junction box have been continuously logged during this period by a battery-powered monitor system. A sample of such measurements, taken during the first years of operation, is shown in Figure~\ref{fig:JBResults}. The relative humidity is seen to plateau at 50\% when the detector is not powered and to drop during periods when the junction box is warm with the transformer powered for the operation of prototype detection lines. 

Another important parameter for assessing the apparatus performance is the fraction of time dedicated to data taking. This should be as high as possible in order to maximize the statistics of events collected and to allow for a maximum probability of detecting transient phenomena. Since the start of the operation of the detector in March 2006, the data taking live time has been better than 90~\%, the larger fractions of dead time being due to construction/maintenance activities ($\approx$~4\%) and calibrations ($\approx$~3\%). The trigger rate, which is dominated by cosmic ray muons, is at the level of a few tens of Hz. Neutrino events are recorded at a rate of about four per day.

Detection efficiency and angular resolution are the parameters which mainly determine the apparatus sensitivity  to neutrino sources; neutrino energy resolution is also significant, as it helps to discriminate between neutrinos of astrophysical origin and those created in cosmic ray showers in the atmosphere. These three parameters have been studied using a detailed  Monte Carlo simulation of the detector response to muons and neutrinos. The performance of the apparatus has been reported in references \cite{bib:paper_L01, bib:paper_muonflux, bib:paper_5lines}.

Angular resolution, which is a key element for separating a point source neutrino signal from the atmospheric neutrino background, depends on the timing resolution, the accuracy of the OM positioning system and the water scattering properties. 
Detection efficiency is affected by different factors, the most significant being the light transmission parameters in water and the OM detection efficiency; the latter in turn depends on several factors such as light transmission losses, photocathode quantum efficiency, electron collection efficiency and the threshold setting.

\begin{figure*}[!htb]
	\centering
		\includegraphics[width=0.9\linewidth]{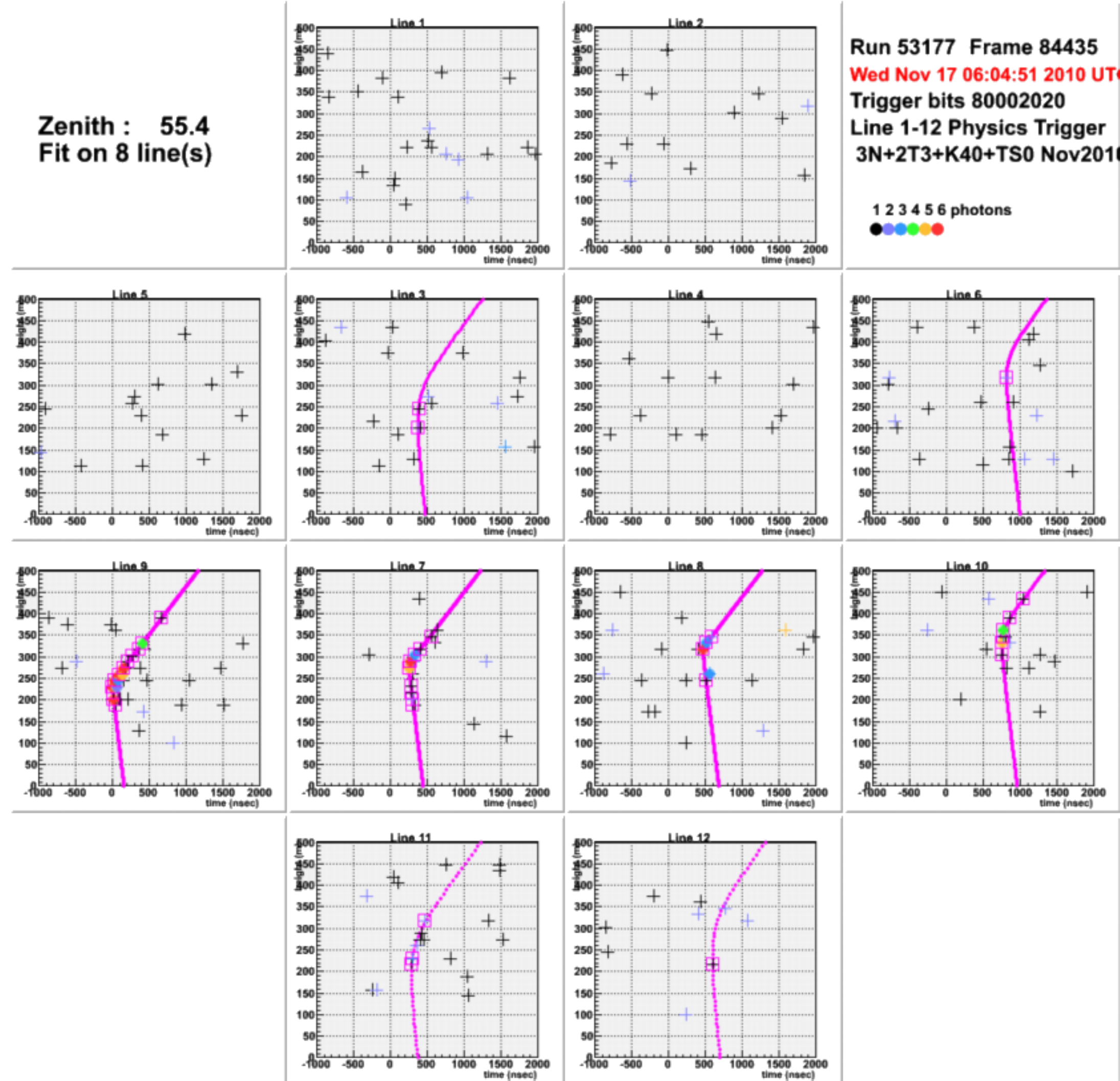}
	\caption{A graphical representation of a neutrino-induced event. For each detector line, a panel shows the vertical position (y-axis) and the arrival time of the hits (x-axis). The panels are arranged so as to reproduce the relative positions of the lines in the apparatus. Crosses are hits in a time window of 3 microseconds around the trigger; full circles are hits passing the trigger condition; open boxes are hits used in the final fit. The symbols are coloured, according to the illustrated code, based on the hit amplitude. The final fit is used to draw the pink lines.}
	\label{fig:neutrinoevent}
\end{figure*}

An example of an energetic upgoing neutrino candidate event, observed on eight ANTARES detector lines is shown in Figure~\ref{fig:neutrinoevent}. For each detector line, a panel shows the vertical position (y-axis) and the arrival time of the hits (x-axis). In this coordinates system, the hits must lie on a hyperbola. A reconstruction algorithm~\cite{bib:bbfit} 
is applied and the curves show the results of the best fit. The ``aperture'' of the asymptotes is related to the angle of the muon with respect to the detector line and the ``summit'' gives the altitude of the closest approach of the muon to the line. 

\begin{figure}[!h]
	\centering
		\includegraphics[width=7.5cm]{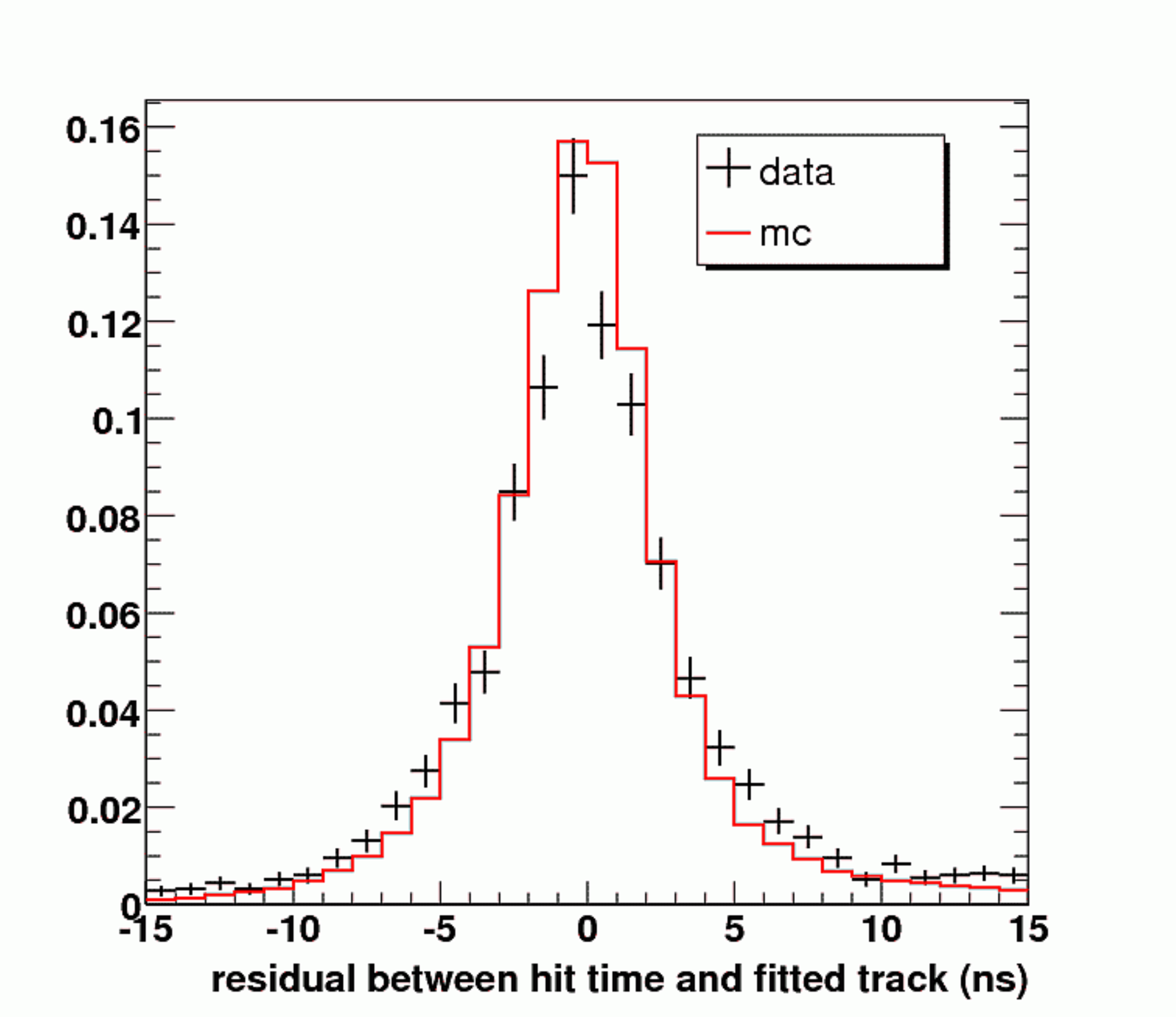}
	\caption{Time residuals for tracks of neutrino candidate events as measured from the data (black crosses) and as expected from the simulation (red histogram).}
	\label{fig:residuals}
\end{figure}

In Figure~\ref{fig:residuals} the time residuals of the track fits for data and Monte Carlo simulation are shown. A
minimum number of 15 hits used in the fit is required. Good agreement can be seen between the data and the simulation with a core timing resolution of 2 ns, obtained by fitting a Gaussian to the data in a range of values between residuals from  -4~ns to 4~ns. The tail of late hits is attributed to light scattering and to the presence of showers in the track sample.

In the absence of a point-like source, demonstration of the experimental angular resolution and absolute pointing of the 
detector can be provided by observation of the moon shadow with cosmic rays. However several years of data taking will be needed. A program is also planned to look for events detected in coincidence by the apparatus and by an array of scintillators floating on the sea surface above \cite{bib:helycon}. 

\begin{figure}[!h]
	\centering
	\includegraphics[width=7.5cm]{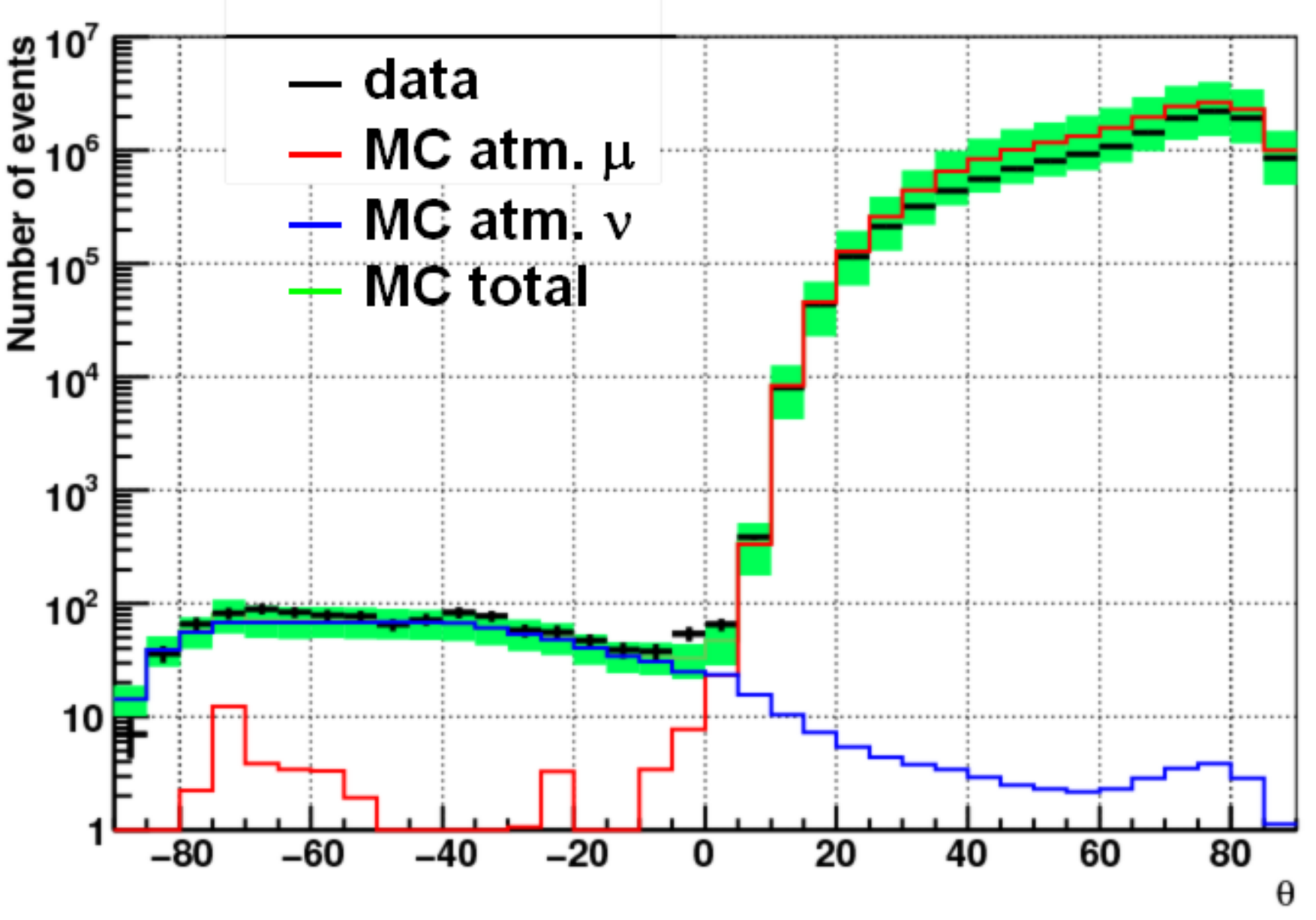}
	\caption{Elevation distribution of events.}
	\label{fig:neutrinodistribution}
\end{figure}

In Figure~\ref{fig:neutrinodistribution} is shown the measured elevation distribution of selected events. Also shown is the corresponding expectation from the Monte Carlo simulation, which takes into account the best measurements and estimates of the contributions to detector efficiency mentioned above. The overall agreement between data and Monte Carlo is well within the estimated systematic uncertainties of about 20~\% for the detector effects and an additional 30~\% of uncertainty on the absolute flux of atmospheric particles.

\begin{figure}[!h]
	\centering
		\includegraphics[width=7.5cm]{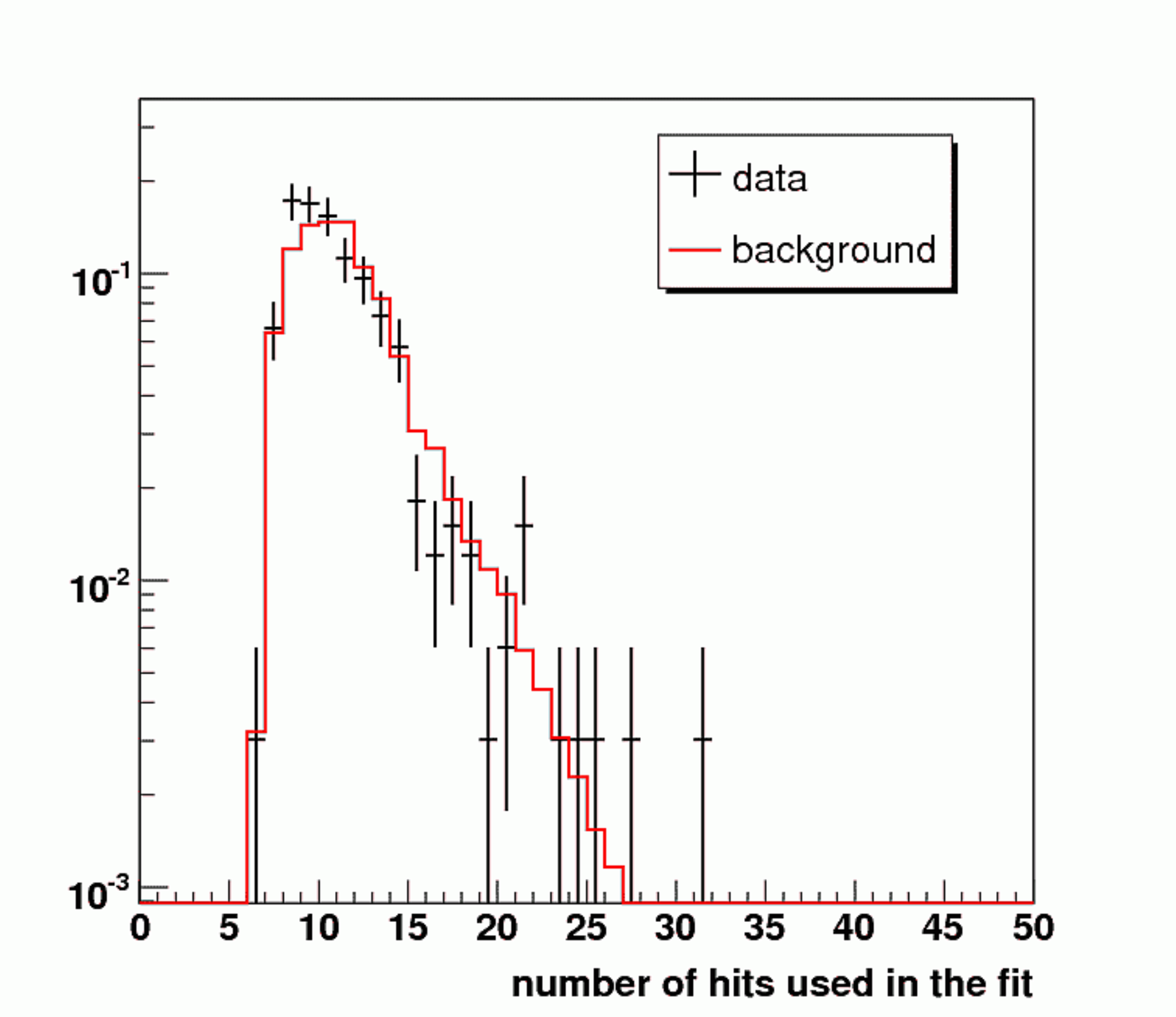}
	\caption{Number of hits in neutrino candidate events as measured from the data (black crosses) and as expected from the simulation (red histogram). The two distributions have been normalized to the same event count.}
	\label{fig:track_hits}
\end{figure}

A further check of the efficiency assumptions in the Monte Carlo, independently of the absolute flux of particles, is shown in Figure~\ref{fig:track_hits}. Here the number of hits associated to the fitted tracks is compared to the Monte Carlo expectation 
for the upward going events of Figure~\ref{fig:neutrinodistribution}. A good agreement between data and Monte Carlo is observed.

\section{Conclusions}
\label {sec:conclusions}
After an extensive period of R\&D and prototyping, the construction of ANTARES has been successfully completed. The high energy neutrino telescope consisting of 12 lines holding optical modules is deployed on the seabed off the Toulon coast at 2475~m depth. Since the deployment of the first line in 2006, data taking has proceeded essentially continuously.

The methods and the procedures to control such a novel detector have been developed including {\it in situ} timing calibration, acoustic positioning of the detector elements and charge calibration. 

All the design goals of the detector have been attained. The measurements of the position of the optical modules is achieved to accuracy better than 20 cm and the expected time resolution of 1 ns is reached. This allows the reconstruction of the events with the desired angular resolution.
 
ANTARES has demonstrated that undersea neutrino telescopes are feasible and manageable from the onshore infrastructure.
The successful operation of ANTARES represents an important step towards a future km$^{3}$-scale high-energy neutrino observatory and marine sciences infrastructure.
\\\
\\\
We dedicate this paper to the memory of our colleague and friend Patrice Payre, who passed away during the preparation of this paper.

\section*{Acknowledgements}
\label {sec:Acknowledgements}
The authors acknowledge the financial support of the funding agencies: Centre National de la Recherche Scientifique (CNRS), Commissariat \`a l'\'energie atomique et aux \'energies alternatives  (CEA), Agence Nationale de la Recherche (ANR), Commission Europ\'eenne (FEDER fund and Marie Curie Program), R\'egion Alsace (contrat CPER), R\'egion Provence-Alpes-C\^ote d'Azur, D\'e\-par\-tement du Var and Ville de La Seyne-sur-Mer, France; Bundesministerium f\"ur Bildung und Forschung 
(BMBF), Germany; Istituto Nazionale di Fisica Nucleare (INFN), Italy; Stichting voor Fundamenteel Onderzoek der Materie (FOM), Nederlandse organisatie voor Wetenschappelijk Onderzoek (NWO), the Netherlands; Council of the President of the Russian Federation for young scientists and leading scientific schools supporting grants, Russia; National Authority for Scientific Research (ANCS), Romania; Ministerio de Ciencia e Innovaci\'on (MICINN), Prometeo of Generalitat Valenciana and MultiDark, 
Spain. We also acknowledge the technical support of Ifremer, AIM and Foselev Marine for the sea operation and the CC-IN2P3 for the computing facilities.


\newpage
\listoffigures
\listoftables

\end{document}